%% file: ScECAL09FNAL_v02-02.tex
\documentclass[11pt, a4]{article}
\usepackage{multirow}
\usepackage{graphicx}
\usepackage{amssymb}
\usepackage{lineno}
\usepackage{braket}
\usepackage{color}
\usepackage{ulem}
\usepackage{hyperref}

\usepackage{cite}

\usepackage{xspace}
\usepackage{caption}

\usepackage[title]{appendix}

\newcommand{\celdeg}{$^{\circ}$C}

\newcommand{\cmipcert}{\ensuremath{c^{\mathrm{MIP}}(T_0)}}
\newcommand{\cmiptwenty}{\ensuremath{c^{\mathrm{MIP}}(T_0=20^{\circ}}C)}
\newcommand{\cmip}{\ensuremath{c^{\mathrm{MIP}}}}
\newcommand{\cphoton}{\ensuremath{c^\mathrm{p.e.}}}
\newcommand{\cinter}{\ensuremath{c^{\mathrm{inter}}}}

\newcommand{\cmipslope}{\ensuremath{\mathrm{d} c^{\mathrm{MIP}}/\mathrm{d} T}}

\newcommand{\ebeam}{\ensuremath{E_{\mathrm{beam}}}}
\newcommand{\gevmip}{\ensuremath{\mathrm{d MIP}/\mathrm{d}\ebeam}}
\newcommand{\Cstoch}{\ensuremath{C_{\mathrm{stoch}}}}
\newcommand{\Cconst}{\ensuremath{C_{\mathrm{const}}}}

\newcommand{\Nfired}{\ensuremath{N_{\mathrm{fired}}}}
\newcommand{\Nin}{\ensuremath{N_{\mathrm{in}}}}
\newcommand{\Npix}{\ensuremath{N_{\mathrm{pix}}}}
\newcommand{\Npixeff}{\ensuremath{N_{\mathrm{pix}}^{\mathrm{eff}}}}

\newcommand{\Aadc}{\ensuremath{A_{i}[\mathrm{ADC]}}}
\newcommand{\AcorrMIPS}{\ensuremath{A_{i}^{\mathrm{corr}}[\mathrm{MIP}]}}
\newcommand{\ADChigh}{\ensuremath{A_i^{\mathrm{high}}}}
\newcommand{\ADClow}{\ensuremath{A_i^{\mathrm{low}}}}

\newcommand{\Ereco}{$E_\mathrm{reco}$}
\newcommand{\sigmaEreco}{$\sigma_E$}

\newcommand{\diff}{\mathrm{d}}
\newcommand{\scmokka}{\textsc{Mokka}}
\newcommand{\scgeant}{{\scshape Geant4}}
\newcommand{\scaleLarge}{$900\,\times\,900$\,mm$^2\,\times$\,90\,layers}
\newcommand{\plm}{$\pm$}

\newcommand{\gev}{\ensuremath{\mathrm{\,Ge\kern -0.1em V}}\xspace}
\newcommand{\mev}{\ensuremath{\mathrm{\,Me\kern -0.1em V}}\xspace}
\newcommand{\kev}{\ensuremath{\mathrm{\,ke\kern -0.1em V}}\xspace}
\newcommand{\ev}{\ensuremath{\mathrm{\,e\kern -0.1em V}}\xspace}
\newcommand{\gevc}{\ensuremath{{\mathrm{\,Ge\kern -0.1em V\!/}c}}\xspace}
\newcommand{\mevc}{\ensuremath{{\mathrm{\,Me\kern -0.1em V\!/}c}}\xspace}
\newcommand{\gevcc}{\ensuremath{{\mathrm{\,Ge\kern -0.1em V\!/}c^2}}\xspace}
\newcommand{\gevgevcccc}{\ensuremath{{\mathrm{\,Ge\kern -0.1em V^2\!/}c^4}}\xspace}
\newcommand{\mevcc}{\ensuremath{{\mathrm{\,Me\kern -0.1em V\!/}c^2}}\xspace}
\newcommand{\chisq}{\ensuremath{\chi^2}\xspace}

\newcommand{\eg}{\mbox{\itshape e.g.}\xspace}
\newcommand{\ie}{\mbox{\itshape i.e.}\xspace}

\newcommand{\vs}{\mbox{\itshape vs.}\xspace}

\newcommand{\textred}{\textcolor{black}}
\newcommand{\textmagenta}{\textcolor{black}}
\newcommand{\textblue}{\textcolor{black}}

\newcommand{\textredSecond}{\textcolor{black}}
\newcommand{\textblueSecond}{\textcolor{black}}
\title{Construction and Response of a Highly Granular Scintillator-based Electromagnetic Calorimeter
 }
\begin{document}

%%\maketitle
%\begin{flushright}
%Draft to CALICE \\
%\today \\
%ver. 06-02-03 %% Nigel eddit til Section 4. Sep 13 version
%\end{flushright}
%\bigskip\bigskip\bigskip\bigskip\bigskip
\begin{center}
\Large\bf
% Construction of the CALICE High Granular  Scintillator Based Electromagnetic Calorimeter Prototype and Its Response to Electrons
Construction and Response of a Highly Granular Scintillator-based Electromagnetic Calorimeter
\end{center}
\begin{center}
%\LARGE
\renewcommand{\thefootnote}{\fnsymbol{footnote}}
{\bf The CALICE Collaboration}\footnote{Corresponding author: Katsushige Kotera, ({\tt coterra@azusa.shinshu-u.ac.jp})}
\end{center}

%%%%%%%%%%%%%%%%%%%%%%%%%%%%%%%%%%%%%%%%%%%
%%%%%%%%%%%%%%%%%%%%%%%%%%%%%%%%%%%%%%%%%%%
%%        Abstract                        %%%%%%%%%%%%%%%%%%%%%%%%%%%%
%%%%%%%%%%%%%%%%%%%%%%%%%%%%%%%%%%%%%%%%%%%
%%%%%%%%%%%%%%%%%%%%%%%%%%%%%%%%%%%%%%%%%%%
\abstract{ A highly granular electromagnetic calorimeter with
  scintillator strip readout is being developed for future linear %%%% NIMA rev2 lepton
  collider experiments. 
  % A prototype of 21.5\,$X_0$ depth, with
  %transverse dimensions %180\,mm\,$\times$\,180\,mm 
  %\textred{180\,$\times$\,180\,mm$^2$} 
  %consisting of 2160 individually read out scintillator strips of size
 %% %10\,mm\,$\times$\,45\,mm\,$\times$\,3\,mm
  %\textred{10\,$\times$\,45\,$\times$\,3\,mm$^3$}, was constructed.  
  %%% 20170511 Nigel
  A prototype of 21.5\,$X_0$ depth and $180\,\times\,180$\,mm$^2$ 
  transverse dimensions was constructed, consisting of 2160 individually 
  read out $10\,\times45\,\times\,3$\,mm$^3$ scintillator strips.
  This prototype was tested using electrons  of 2\,--\,32\,\gev
  at the Fermilab Test Beam
  Facility in 2009.  Deviations from linear energy response were less
  than 1.1\%, and the intrinsic energy resolution was determined to be
  $(12.5\pm0.1(\mathrm{stat.})\pm0.4(\mathrm{syst.}))\%/\sqrt{E[\mathrm{\gev}]}\oplus
  (1.2\pm0.1(\mathrm{stat.})^{+0.6}_{-0.7}(\mathrm{syst.}))\%$, where
  the uncertainties correspond to statistical and systematic sources, 
  respectively.

%\newpage
%\input{CALICEMemberList161116.tex}
\newpage
\input{CALICEMemberList170519.tex}

\newpage
\tableofcontents
\newpage

\renewcommand{\thefootnote}{\arabic{footnote}}
\setcounter{footnote}{0}
%%%%%%%%%%%%%%%%%%%%%%%%%%%%%%%%%%%%%%%%%%%
%%%%%%%%%%%%%%%%%%%%%%%%%%%%%%%%%%%%%%%%%%%
%%  Introduction                 %%%%%%%%%%%%%%%%%%%%%%%%%%%%%%
%%%%%%%%%%%%%%%%%%%%%%%%%%%%%%%%%%%%%%%%%%%
%%%%%%%%%%%%%%%%%%%%%%%%%%%%%%%%%%%%%%%%%%%
\section{Introduction}
Detectors for the International Linear Collider (ILC) are designed to
perform high precision measurements, taking advantage of the
well-defined initial conditions of electron-positron collisions \cite{DBD}.  To
characterise final states that are dominated by the production and
decay of quarks, gauge \textred{bosons and/or} Higgs bosons, the accurate reconstruction of
jets of hadrons is mandatory.  One way to achieve this is by measuring
each particle within a jet individually, and combining information
from calorimeters and tracking detectors.  This method, known as the
particle flow approach (PFA) \cite{JCB,MarkT}, requires highly
granular calorimeters.  
%%%ERIKA In the electromagnetic calorimeter (ECAL),
%%%ERIKA both the lateral and longitudinal segmentation should be around
%%%ERIKA 5--10~mm~\cite{MarkT,LOI}.  
%%%%%%Erica In the electromagnetic calorimeter (ECAL), 5--10\,mm transverse segmentation is required to  separate particle clusters in jets, as well as 20--30 layer segmentation in longitudinal as a consequence of a  sampling calorimeter to prevent the total thickness~\cite{MarkT,LOI}. 
%
%%%% NIMA rev2 The requirement of single particle separation within a jet translates for the electromagnetic calorimeter (ECAL) in a lateral segmentation 
 %%%%better then the Moli\`{e}re radius of Tungsten (9.3\,mm), and a longitudinal sampling at least every $X_0$, resulting in 20-30 layers. 
\textredSecond{
To achieve this single particle separation the
electromagnetic calorimeter (ECAL) must have 
longitudinal sampling at least every $X_0$,
and a lateral segmentation better than the Moli\`{e}re radius of the absorber (\eg\,9.3\,mm for Tungsten).
 Because we require more than 20\,$X_0$ for the total thickness of
 ECAL to prevent energy leakage, the ECAL must therefore have at least
 20--30 layers.
}
At the ILC, 
%PFA also requires the
%intrinsic energy resolution of the ECAL
%\textcolor{red}{
an ideal value %20170511 David  of the 
for the
intrinsic energy resolution of the ECAL is required 
to be less than 15\%$/\sqrt{E\mathrm{[\gev]}}$
by PFA \cite{MarkT}.
%}
 Emerging designs for scintillator-based sampling
calorimeters now have the potential to realise these design criteria.

The previous limiting factors for the segmentation of a
scintillator-based calorimeter were the size and sensitivity of the
readout technology.  
%%%%%%Erica This situation changed drastically with the
%%%%%%Erica introduction of pixelated photon detectors (PPD or SiPM)
This situation changed drastically with the
introduction of \textredSecond{the} silicon photomultiplier (SiPM)
\cite{Bondarenko,BuzhanICFA,Buzhan,AHCAL,Erika}.  Small scintillator
elements can be read out individually using SiPMs without introducing
large dead volumes for the readout systems.
%%%%%%%%%%%%%%%%%%%%%%%%%%%%%%%%%%%%%%%%
%%%%%%% Franks suggestion that we should mention about AHCAL%%%%%
%The CALICE Collaboration (CALICE) introduced this technology to the hadron calorimeter (HCAL) which has ... channels of 30\,$times$\,30\,$\times$,\5\,mm$^3,
%This technology was  introduced to the CAL
%
% KK tried to make a sentence of explanation of great AHCAL. but this breaks the fluent of the story.
%
This technology is used in the scintillator strip %% 20170511 based
electromagnetic calorimeter (ScECAL) being developed by the CALICE
Collaboration. To reduce both the total number of readout channels and
the overall insensitive volume associated with the readout SiPMs,
strips of scintillator, each with a length of 45\,mm and a width of
between 5 and 10\,mm, are used.  Strips in successive layers have an
orthogonal orientation relative to each other \cite{%caliceRepo,
DBD}
and an algorithm has been developed to achieve fine effective
segmentation from such a strip-based design.
\textredSecond{
%For example, a performance of 
\textblueSecond{A study \cite{SSA} of } the invariant mass resolution of 
neutral pions, carried out using a full simulation of a detector for
the ILC, showed that a 
45\,$\times$\,5\,mm$^2$ ScECAL using this algorithm had almost
\textblueSecond{the same} performance \textblueSecond{as} a
5\,$\times$\,5\,mm$^2$ ScECAL.}
%% in full simulation of \textblueSecond{a} detector \textblueSecond{concept} for the ILC.}
%%The energy resolution of up to 250\,GeV jets with  45\,$\times$\,5\,mm$^2$  ScECAL is comparable\,---\,$\sim$\,7\% larger\,---\,with 5\,$\times$\,5\,mm$^2$ tile ScECAL
 %%\cite{SSA}.}  
 
 To achieve
the required longitudinal segmentation, the ScECAL is designed as a
sampling calorimeter using 25--30 tungsten layers of thickness of
2--4\,mm, interleaved with scintillator strip sensor layers.
%%% into previous paragraph 20170511 Nigel
The first CALICE ScECAL prototype\,\cite{desyTB} consisted of 26 sensor
layers, interleaved with 3.5\,mm thick tungsten carbide
%\textcolor{red}{
(WC)
%}
absorber
layers, and had a transverse area of 90\,$\times$\,90\,mm$^2$.
%%%% Frank required to write about absorber.
%%%%%%%%%%%%%%%%%%%%%

%%%% 20170511 Nigel removed: It was tested using positrons with energies between 1 and 6 \gev.  

The current prototype consists of 30 detector layers and has transverse
dimensions of 180\,$\times$ 180\,mm$^2$ and a depth of
%%%% Erika 
 21.5\,$X_0$ (266\,mm),
%%%%%% Frank require to put explicit dimension 
%%%%%%%%%%%%%%%%%%%%%% 
reducing the effect of lateral and longitudinal shower leakage relative to the previous prototype.
%%Nige Watson 20160913 Should be obvious
%%Nige Watson 20160913 The scintillator-PPD unit has been improved using experience gained from the first prototype.
The basic unit was a 45\,$\times$\,10\,$\times$\,3\,mm$^3$
scintillator strip with a central hole of 1.5\,mm diameter running
along its length, hermetically wrapped with reflective foil.  A
wavelength shifting (WLS) fibre inserted into the hole guides light to
a SiPM placed at one of the ends of the scintillator strip.  A LED-based
gain monitoring system was implemented for each strip, an improvement
on the first prototype in which only one LED was provided per layer.
This prototype was tested in conjunction with the CALICE 
analogue hadron calorimeter (AHCAL)\,\cite{AHCAL, EMrespoAHCAL} \footnote{
Electromagnetic response of AHCAL is also available.
}
 and 
tail catcher muon tracker (TCMT)\,\cite{TCMT} prototypes. 

This paper is organised as follows. 
\textred{Details of the prototype design including properties of applied % Erika MPPCs 
SiPMs are given in Section\,\ref{section:ScECAL}.
The test beam experiment at Fermilab is described in
Section~\ref{section:FNALTB}, and the analysis including %calibration procedure 
\textblue{detector calibration} %with determinations of calibration factors 
and results obtained using electron beams are given in Sections~\ref{section:analysis} and \ref{section:results}.
% Section~\ref{section:simulation} compares the
%results given in the previous section with Monte Carlo simulations,
%while Section~\ref{section:discussion} discusses the results presented and Section~\ref{section:summary} draws conclusions.}
}
Section~\ref{section:simulation} compares the analysis results with Monte Carlo simulations, Section~\ref{section:discussion} discusses the results and Section~\ref{section:summary}  draws conclusions.

%%%%%%%%%%%%%%%%%%%%%%%%%%%%%%%%%%%%%%%%%%%
%%%%%%%%%%%%%%%%%%%%%%%%%%%%%%%%%%%%%%%%%%%
%%          ScECAL explanation                %%%%%%%%%%%%%%%%%%%%%%%%
%%%%%%%%%%%%%%%%%%%%%%%%%%%%%%%%%%%%%%%%%%%
%%%%%%%%%%%%%%%%%%%%%%%%%%%%%%%%%%%%%%%%%%%
\textred{\section{Construction}\label{section:ScECAL}
\subsection{Detector}\label{section:construction}}

The prototype, shown in Fig.~\ref{fig:overview} in front of the
CALICE AHCAL, has a total thickness of 266~mm.  It
consists of 30 pairs of alternating 3.5~mm thick tungsten carbide
absorber and scintillator layers, with the first layer being absorber.
%%Nige Watson 20160913 which
%%Nige Watson 20160913 are described in more detail later in this subsection.
%absorber layer, alternately followed by scintillator and absorber layers.
%%%%%%%%%%%%%%%%%%%%%%%%%%%%%%%%%%%%%%%%%%%
%%         Fig flatCable                %%%%%%%%%%%%%%%%%%%%%%%%%%
%%%%%%%%%%%%%%%%%%%%%%%%%%%%%%%%%%%%%%%%%%%
\begin{figure}[tbph]
\centerline{\includegraphics[width=0.5\textwidth]{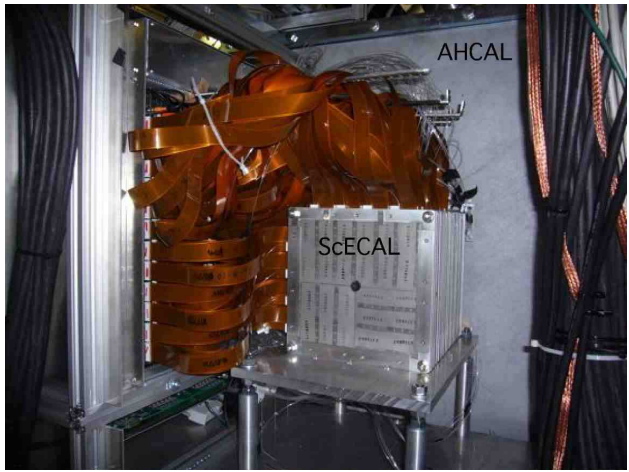}}
\caption{\small  
 %%% 20170511 DavidThe prototype 
 The ScECAL prototype in front of the CALICE AHCAL.}\label{fig:overview}
 \captionsetup{width=.75\textwidth}
\end{figure}
Figure~\ref{fig:a_layer_72holes} shows the design of a scintillator
layer, consisting of four rows of 18 scintillator strips, held in a
rigid steel frame.  Figure~\ref{fig:strip} illustrates the design of a
single polystyrene-based scintillator strip and shows the central hole
for the WLS fibre, manufactured using an extrusion method \cite{KNU}
and cut into strips.  The polystyrene was doped using a mixture of 1\%
2,5--diphenyloxazole and 0.1\%
2,2'--($p$--phenylene)bis(5--phenyloxazole) for fluorescence.  A notch
with a depth of 1.40$\pm$0.05~mm and a width of 4.46$\pm$0.03~mm was
cut mechanically to accommodate the SiPM.  The specific SiPM used was a
multi-pixel photon counter (MPPC), from Hamamatsu K.K.~\cite{HPK}.
The size of the MPPC package was
%1.3\,mm\,$\times$\,(4.2$\pm$0.2)\,mm\,$\times$\,(3.2$\pm$0.2)\,mm.
\textred{1.3\,$\times$\,(4.2$\pm$0.2)\,$\times$\,(3.2$\pm$0.2)\,mm$^3$}.
The four %%%%% Frank's req edges 
%%% 20170511
long 
sides of each strip were polished to 
 control precisely the strip size and
to ensure %% Franks req the 
reflection of the surfaces.

From a randomly chosen sample of 20 strips, the measured mean
values and the sample standard deviations (SD)
%%Nige Watson 20160913 \{$\pm$standard deviation (SD)\}
of the widths, lengths and thicknesses were $9.85\pm0.01$~mm,
$44.71\pm0.04$~mm, and $3.02\pm0.02$~mm, respectively.  A double clad
1~mm diameter Y-11 WLS fibre
%\footnote{Provided by KURARAY Co., Ltd.}
\textred{provided by KURARAY Co., Ltd.\,\cite{kurare}}
with a length of  $43.6\pm0.1$~mm was inserted into the hole of each strip.
Each strip was wrapped with a 57~$\mu$m-thick reflective foil provided
by KIMOTO Co., Ltd~\cite{kimono}.  This foil consists of layers of silver and
aluminium, deposited by evaporation between layers of polyethylene
terephthalate, and has a reflection ratio of 95.2\% for light with a
wavelength of 450~nm. 
%%%%%  come from section 6 according to Jerry's suggestion 
Four out of 2160 channels of the present ScECAL prototype were not operational.
One possible cause is the development of short-circuits between the MPPC electrodes %% g923 occurred 
caused 
by the conductive cut edges of the reflector film.
The CALICE Collaboration has another candidate for the reflector
design that does not have any conductive layer \cite{ThreeM}.
 Each scintillator strip also has
a 2.5~mm diameter hole in the reflective foil to allow the injection
of light from a LED for gain monitoring.

A screen, also made of reflective foil, was used to prevent
scintillation photons impinging directly onto the MPPC, without
passing through the WLS fibre, to ensure uniformity of response 
%%%Ias a function of position 
along the length of the strip. When the screen is
used, the response to single particles at the end of the strip furthest
%away
from the MPPC
%at the end of the strip far from
is (88.3\,$\pm$\,0.4)\% of that directly in front of the MPPC. This is
discussed in more detail in Section~\ref{section:discussion}.  A
photograph of the screen attached to the inside of the scintillator
notch is shown in Fig.~\ref{fig:shade}.
Nine MPPCs were soldered onto a flat polyimide cable, as shown in
Fig.~\ref{fig:flatCable}, and inserted into the corresponding
notches cut into the scintillator strips.
%%%% NIMA rev1
\textredSecond{%%Because the flat readout cables take too large space for the future full scale detector, 
%%the CALICE collaboration is developing a compact and thin baseboard unit inserted directly 
%%between scintillator and absorber requiring minimal space for external interfaces
For \textblueSecond{a} future full scale detector, the CALICE
collaboration is developing a compact,  
thin baseboard unit \textblueSecond{with embedded electronics} inserted directly between scintillator and absorber, requiring 
minimal space for external interfaces instead of the  flat readout cables\,\cite{IEEE2014}.
}

%%%%%%%%%%%%%%%%%%%%%%%%%%%%%%%%%%%%%%%%%%%
%%         Fig  a_layer_72holes                %%%%%%%%%%%%%%%%%%%%%%%%%%
%%%%%%%%%%%%%%%%%%%%%%%%%%%%%%%%%%%%%%%%%%%
\begin{figure}[tbph]
  \captionsetup{width=.75\textwidth}
\centerline{\includegraphics[width=0.6\textwidth]{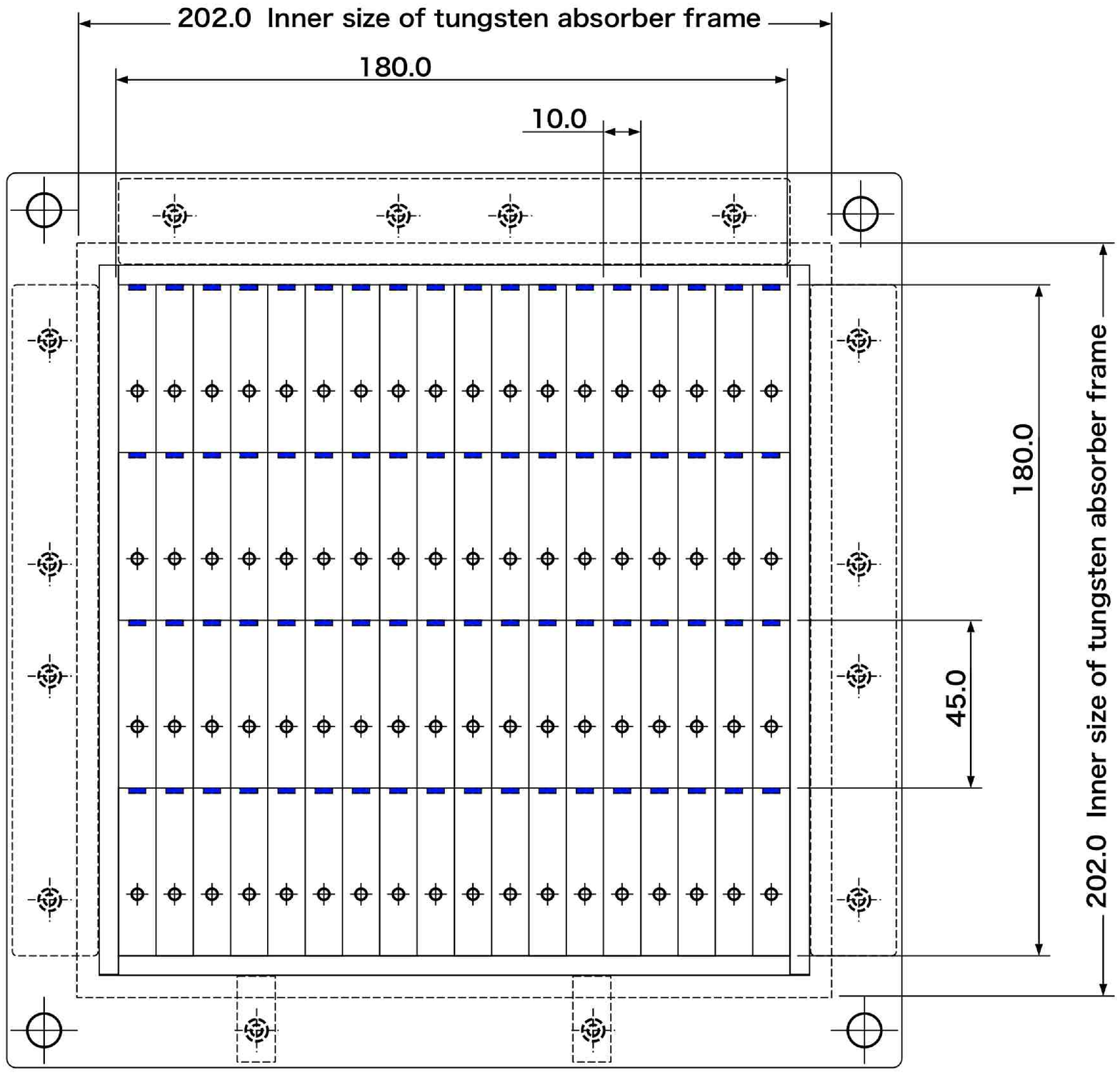}}
\caption{\small The arrangement of 72 strips in a scintillator layer.
  The positions of the MPPC housing notches (blue) are shown, 
  as well as the
  holes in the reflector foil used for the LED calibration.
  All
  dimensions are given in mm.}\label{fig:a_layer_72holes}
\end{figure}

%%%%%%%%%%%%%%%%%%%%%%%%%%%%%%%%%%%%%%%%%%%
%%         Fig strip                 %%%%%%%%%%%%%%%%%%%%%%%%%%
%%%%%%%%%%%%%%%%%%%%%%%%%%%%%%%%%%%%%%%%%%%
\begin{figure}[tbhp]
  \centerline{\includegraphics[width=0.95
    \textwidth]{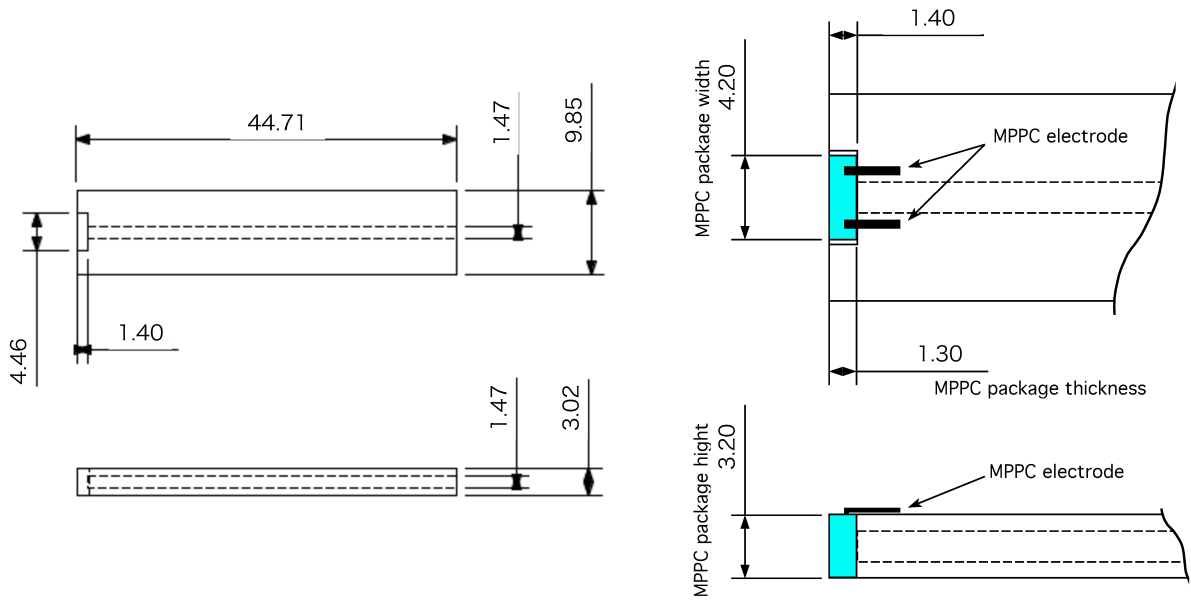} }
%\vspace{-7mm}
\caption{\small Top and side views of a scintillator strip ({\it
    left}) and the notches cut into the strips to accommodate the MPPC
  packages ({\it right}).  All dimensions are given in mm.  }
\label{fig:strip}
\end{figure}
%% According to Frank's suggestion, order of those figures are changed
%%%%%%%%%%%%%%%%%%%%%%%%%%%%%%%%%%%%%%%%%%%
%%         Fig shade                 %%%%%%%%%%%%%%%%%%%%%%%%%%
%%%%%%%%%%%%%%%%%%%%%%%%%%%%%%%%%%%%%%%%%%%
\begin{figure}[tbhp]
 \captionsetup{width=.75\textwidth}
\centerline{\includegraphics[width=0.5 \textwidth]{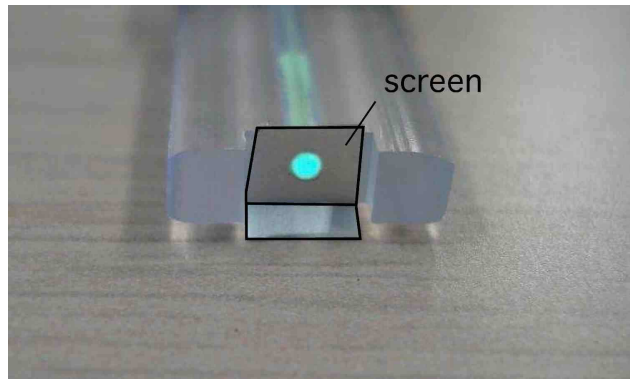}}
\caption{\small The screen used to block direct scintillation
  photons. The bright 
  %% 20170511 magenta circle 
  cyan spot is the transverse section of the
  WLS fibre.}\label{fig:shade}
\end{figure}
%%%%%%%%%%%%%%%%%%%%%%%%%%%%%%%%%%%%%%%%%%%
%%         Fig flatCable                %%%%%%%%%%%%%%%%%%%%%%%%%%
%%%%%%%%%%%%%%%%%%%%%%%%%%%%%%%%%%%%%%%%%%%
\begin{figure}[tbhp]
\captionsetup{width=.75\textwidth}
\begin{center}
\includegraphics[width=0.16 \textwidth]{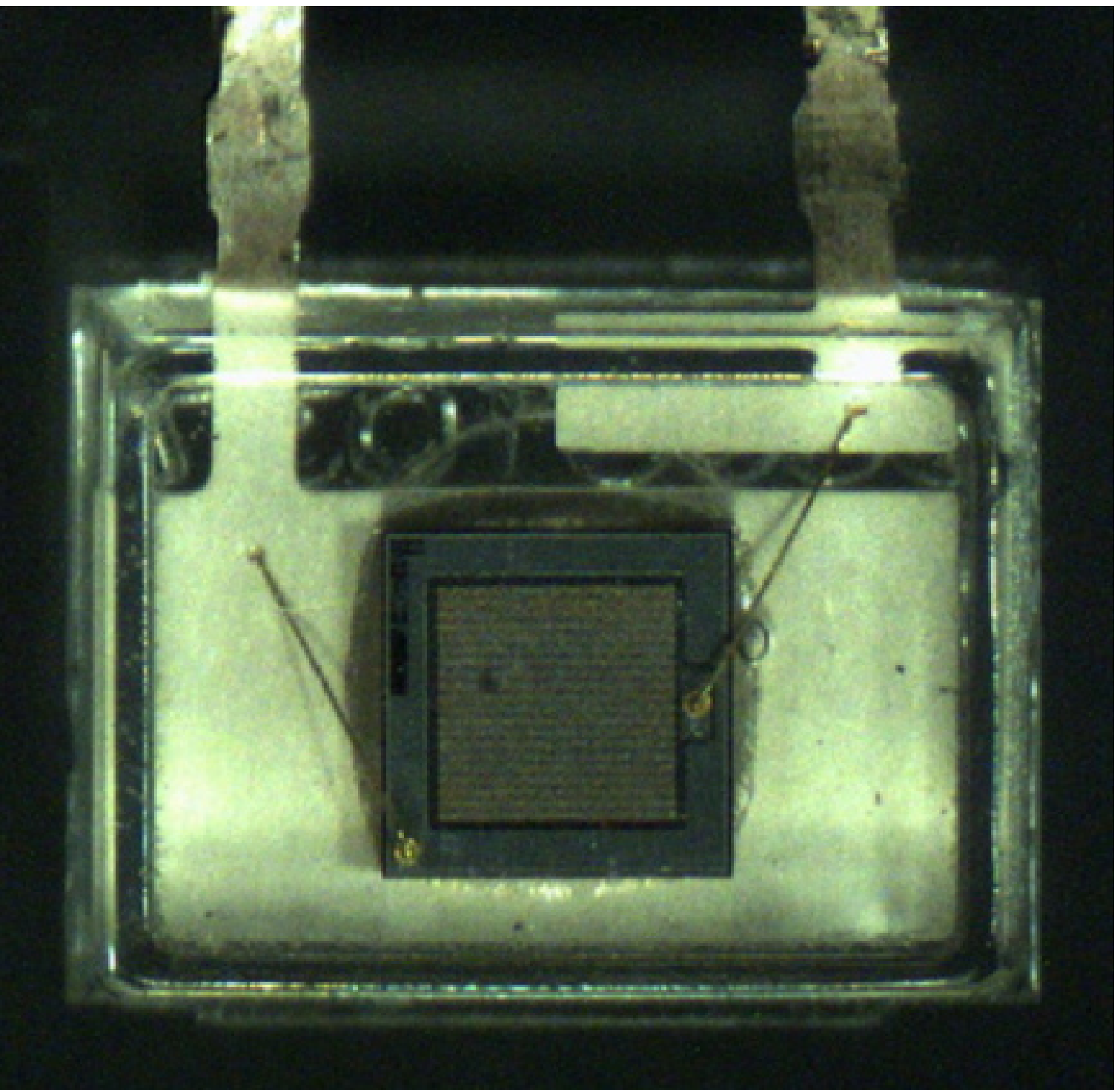}
\includegraphics[width=0.4  \textwidth]{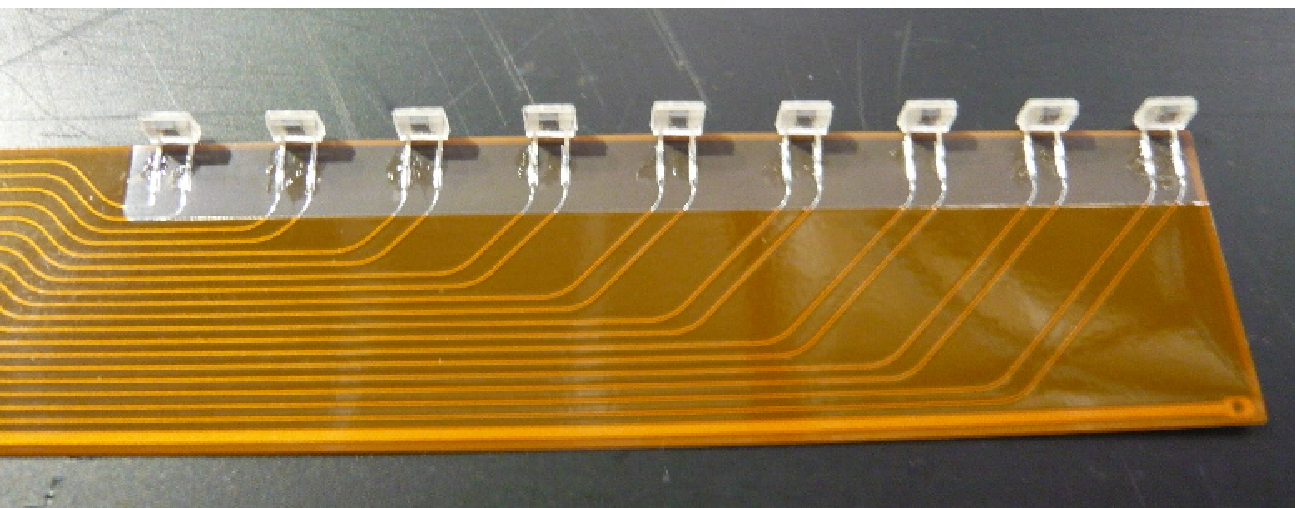}
\caption{\small Photographs of a single MPPC ({\it left}) and nine
  MPPCs soldered onto a flat cable (\it{right}).
}\label{fig:flatCable}
\end{center}
\end{figure}
Each pair of absorber and scintillator layers was held in a rigid
steel frame.  Each frame held four
%100\,mm$\times$\,100\,mm\,$\times$\,(3.49$\pm$0.01)\,mm
\textred{ 100\,$\times$\,100\,$\times$\,(3.49$\pm$0.01)\,mm$^3$ }
tungsten carbide plates aligned to make a %200~mm $\times$ 200~mm
\textred{ 200\,$\times$ 200\,mm$^2$ }
absorber layer in front of the scintillator.  The density of 
%% 20170511 David
the 
absorber plates, based on a sample of eight, was 14.25$\pm$0.04 g/cm$^3$, and
the mass fractions of elemental components were measured using X-ray
diffraction and energy-dispersive X-ray spectroscopy (EDX) to be
(tungsten:carbon:cobalt:chrome) = (0.816:0.055:0.125:0.005).  The
orientation of the scintillator strips in each layer was rotated by
$90^{\circ}$ with respect to that of the previous layer.

%LED monitor
To monitor the stability of response of each MPPC, a LED-based gain
monitoring system was implemented in the prototype.  Each of the 18
strips in a given row within a layer was supplied with LED light via a
clear optical fibre in which notches had been machined at appropriate
positions.  Figure~\ref{fig:clearFiber} shows a photograph of these
fibres, in which light can be seen being emitted at the notches.  The
LED is driven by a dedicated 
%circuit board
electronic circuit
\,\cite{Ivo}.
Details of the calibration procedure are discussed in
Section~\ref{section:calibrationProcedure}.

%The ADC--photo-pixel conversion factor of each MPPC was measured during the test beam experiment by using this LED system. 
%This conversion factor was used to implement the MPPC saturation correction discussed in the next section.
%%%%%%%%%%%%%%%%%%%%%%%%%%%%%%%%%%%%%%%%%%%
%%         Fig  clearFibre                %%%%%%%%%%%%%%%%%%%%%%%%%%
%%%%%%%%%%%%%%%%%%%%%%%%%%%%%%%%%%%%%%%%%%%
\begin{figure}[h!]
\captionsetup{width=.75\textwidth}
\centerline{\includegraphics[width=0.3\textwidth, angle =-90]{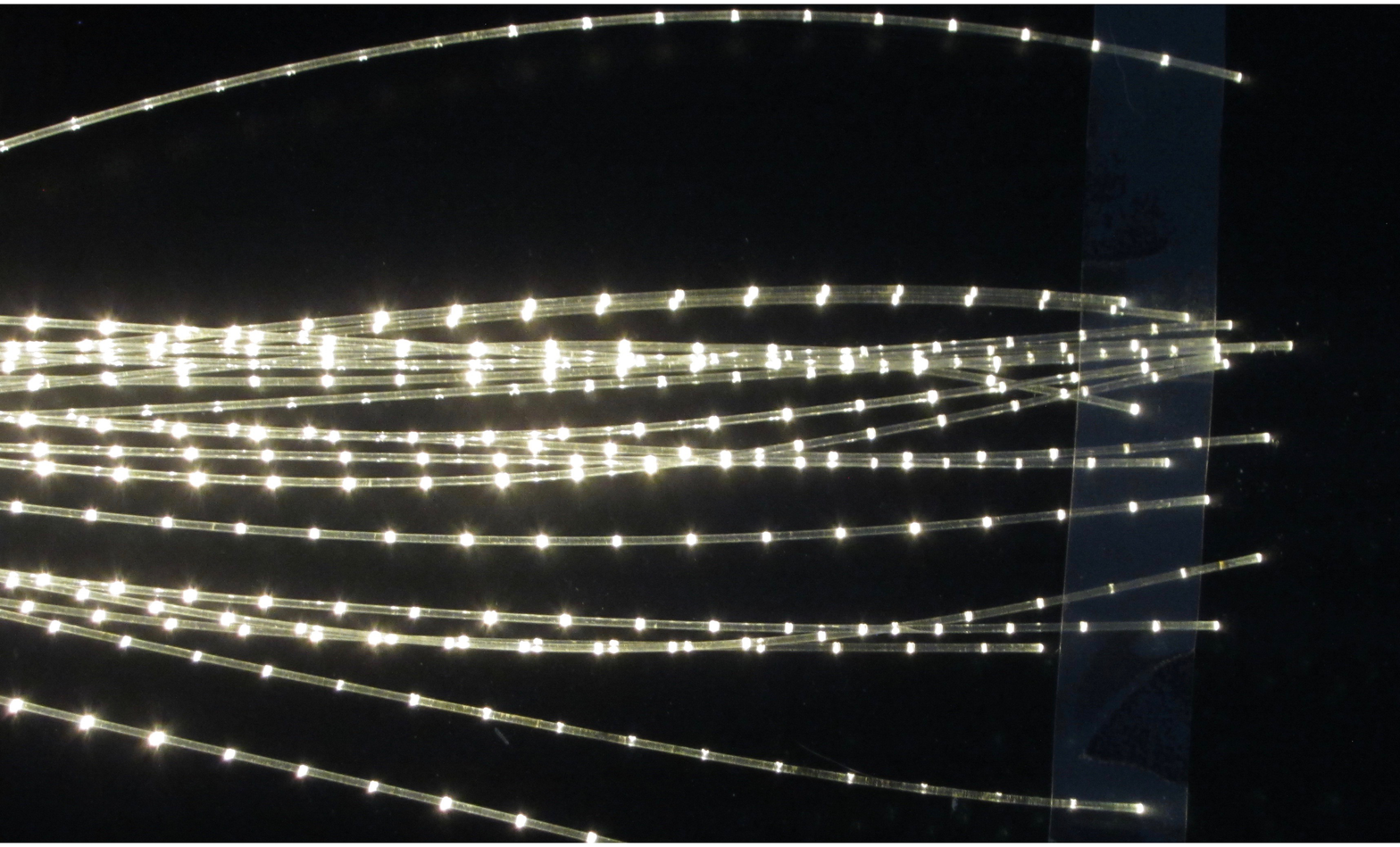}}
\caption{\small A bundle of clear fibres. Each fibre has 18 notches to
  supply the LED light to the 18 strips in a row of scintillators.}
\label{fig:clearFiber}
\end{figure}

%%%%%%%%%%%%%%%%%%%%%%%%%%%%%%%%%%%%%%%%%%%%%%%%%%%%%%%%%%%%%%%%%%%%%
%%%%%%%%%%%%%%%%%%%%%%%%%%%%%%%%%%%%%%%%%%%%%%%%%%%%%%%%%%%%%%%%%%%%%
\subsection{Data acquisition system}\label{section:daq}
%%%%%%%%%%%%%%%%%%%%%%%%%%%%%%%%%%%%%%%%%%%%%%%%%%%%%%%%%%%%%%%%%%%%%
Nine MPPC signal lines and their power supply lines were grouped
together on a flat cable, as noted above, and 12 of these cables were 
connected to a single base board.
%A MPPC signal line together with other eight lines and their power supply lines bundled on a flat cable mentioned
%in previous section\,\ref{section:construction}  were connected on a base board.
The base board contains up to six analogue boards, each of which
contained a single ASIC\,\cite{omega, ASIC}.  Each ASIC controlled 18 MPPCs,
such that 108 MPPCs were controlled by one base board.  The ASIC
performs the following functions:
%consisted of a multiplexed chain of pre-amplifiers, shapers and sample-and-hold circuits \cite{ASIC}:
\begin{itemize}
%%%% 20170511 Nigel
\item fine tuning of %a 
MPPC bias %voltage 
voltages via an % Erika 8-bit DAC 
8-bit DAC over 4.5 V; % Frank Sep. fine bias tuning of MPPC via 8-bit ADC;
\item variable gain charge pre-amplification;
\item variable pulse shaping via  a CR-(RC)$^2$ shaper, and 
\item multiplexing 18 channel signals.
\end{itemize}
This ASIC used a peak-hold method; the hold time was adjusted to give
the largest signal when collecting beam data.
The ASIC provides both a low- and high-gain operation mode; the
low-gain (8.2\,mV/pC) mode was used for the collection of beam runs,
while the high-gain (92\,mV/pC) mode was used to collect the LED
calibration runs used for MPPC gain monitoring. 
 The optimal hold
times were determined separately for both the high-gain  and low-gain
modes.
The spread of gain parameters were 5\,--\,10\%~\cite{AHCAL}.
The analogue outputs from the ASICs were sent to VME-based CALICE
readout cards (CRC), which provided 16 bit ADCs to digitise the MPPC
signals, as well as to perform trigger handling and synchronisation
with the data taken by the AHCAL and TCMT prototype.

%The raw data from the CRC was converted into LCIO \cite{LCIO} format,
%the standard for ILC related studies, and includes information of the
%detector configuration, temperature recordings, voltage settings,
%calibration constants and other specific information associated with
%the run, such as parameters related to the particle beam itself.
%
%\textred{
The raw data from the CRC includes information about the 
detector configuration, temperature recordings, voltage settings,
calibration constants and other specific information associated with
the run in addition to the MPPC signals.
It was stored in the LCIO format, the standard for ILC R\&D. %%% 20170511 Nigel related studies.
%}
%
The actual readout system of the ScECAL prototype was based on that
used for the CALICE AHCAL, as described in Ref.~\cite{AHCAL}.
%%% g702 The readers can find detail figuration of the CALICE data acquisition system in \,\cite{AHCAL}.

%%%%%%%%%%%%%%%%%%%%%%%%%%%%%%%%%%%%
%%%%%%%%%%%%%%%%%%%%%%%%%%%%%%%%%%%%%
%%%%%%%%%%%%%%%%%%%%%%%%%%%%%%%%%%%%%%%%
\subsection{Characterisation of MPPCs and their non-linear response}
\label{section:saturationCorrection}
\label{section:effectiveN}
%%%%%%%%%%%%%%%%%%%%%%%%%%%%%%%%%%%%%%%%%%%%
%%%%%%%%%%%%%%%%%%%%%%%%%%%%%%%%%%%%%%%%%
%%%%%%%%%%%%%%%%%%%%%%%%%%%%%%%%%%%%%%%%
The prototype used 2160 MPPCs.
This subsection discusses how MPPCs were characterised in our laboratory. % before they were mounted.
One of MPPC's characteristics is its non-linear response which is inherent for all SiPMs.
%PPDs are inherently non-linear devices including MPPCs,
%%%%%% after Erica Therefore, we discuss also the way in which corrections for the effects of non-linear response are applied.
The correction for the effects of this non-linear response is described in the calibration procedure in Section \ref{section:calibrationProcedure}.

%%%%%%%%%%%%%%%%%%%%%%%%%%%%%%%%%%%%%%%%%%%%%%%%%%%%%%%%%%%%%%%%%%%%%
\begin{description}
  %%Nige Watson 20161222 \begin{enumerate}
%\subsubsection{MPPC characterization}\label{sec:MPPCchar}
%%Nige Watson 20161222 \item {\bf MPPC characterization}\label{sec:MPPCchar}\\
\item[MPPC characterisation]
  \label{sec:MPPCchar}

The gain $G$ of the MPPC is proportional to the excess voltage applied
above the breakdown voltage (over-voltage), $\Delta$V.  The gain can
therefore be expressed as $G = C\Delta$V, where $C$ is the average
single pixel capacitance of the MPPC.
Two sets of MPPCs were used to %20170511 Nigel instrument 
in 
the prototype: the first 276
pieces were produced in 2007 while the remaining 1884 were produced in
2008 \footnote{ The bespoke model provided by Hamamatsu
  K.K. \cite{HPK} to CALICE was Model MPPC-11-025M,
%%5887
% For ILC-ScECAL  <--- why is this relevant to this paper?
%%%A
%%%A	MPPC-11-025M 5887 is special products for CALICE ScECAL.
%%%A	You cannot find this product name in HPK catalogue. 
%%%A	Therefore, I wrote, "...corresponding to S10362-11-25P"
%%%A
%provided  % <--- or produced?
%%in 2008
%by 
%2008
 corresponding closely to the commercially available device
 S10362-11-25P.}.
 All MPPCs had 1600 pixels in an active area of
 1\,$\times$\,1\,mm$^2$.  The properties of all MPPCs in these two
 sets were measured before constructing the detector prototype.  For
 each MPPC, the gain, noise rate and capacitance were measured as a
 function of the bias voltage.  Figure~\ref{fig:breakdown_capacitance}
 {\it left} shows the distribution of the breakdown voltage of MPPCs,
 and {\it right} shows the extracted single-pixel capacitance $C$ for
 the two sets.  
 %MPPCs from the 2008 set were used in layers 1--20, 29
 %and 30, while those produced in 2007 were installed in the first and
 %the fourth rows of layers 21--28, thereby avoiding the two central
 %rows of the layers where the probability of the energy deposits are
 %highest.
%%% MPPCs from the 2008 set were used in all layers, while those produced in 2007 were 
%%% installed in the first and the fourth rows of layers 21--28, thereby avoiding the two central
%%%% rows of the layers where the probability of the energy deposits are highest.
%%% here by Jerry
\textred{ The MPPCs produced in 2008 \textblue{were used %%% 20170511 Nigel to instrument
 throughout 
} %populated 
most of the prototype, while those produced 
in 2007 were installed only in the first and the fourth rows of layers 21--28.  
By installing the 2007 devices in the layers and rows with low energy deposits, 
possible systematic effects associated with two sets of \textblue{sensors} %devices 
are reduced.
}
%
%The bias voltage on each MPPC at the test beam was determined optimizing  with those results to have the same over-voltage 
%($\Delta$V = -\,3.0\,V) for all channels.
The %%%% 20170511 David: over-voltage (
$\Delta V$
%),
 is tuned to 3.0\,V for all channels in
the test beam experiment.

%%\subsubsection{Number of effective pixels}\label{section:effectiveN}
%%Nige Watson 20161222 \item {\bf Number of effective pixels}
\item[Correction for the non-linear response]

%NKW: just as good to refer to the subsection itself.
%%Nige Watson 20161222 \label{section:effectiveN}\\
%%Nige Watson 20161222
%%%%%%%%%%%%%%%%%%%%%  20170226 move to new section for MPPC ----PPDs are inherently non-linear devices.
%The output of the MPPC in terms of the number of fired pixels
%($N_{\mathrm{fired}}$) can be parametrized as a function of the number
%of incident photons ($N_{\mathrm{in}}$) by the response function:
The non-linear response of MPPCs is approximately described by:
%%%%Equation~\ref{eq:saturationSimple}. 
\begin{eqnarray} \label{eq:saturationSimple} 
%%F(\Nin) := \Nfired = 
  F(\Nin) \equiv \Nfired =
         \Npix\Big\{ 1 - \exp \Big( 
            \frac{ - \epsilon \Nin}{\Npix}
         \Big) \Big\} \,\, ,
\end{eqnarray}
\textredSecond{
where $N_\mathrm{fired}$ is the number of fired MPPC pixels,  \Npix\ is the number of pixels on the MPPC, $\epsilon$ is the
photon detection efficiency, and $N_\mathrm{in}$ is the number of photons incident on the MPPC surface.
For \textblueSecond{a} low light levels, \textblueSecond{the} output
spectrum of \textblueSecond{the} MPPC has clear peaks corresponding to
the fired pixels, allowing the number of ADC counts \textblueSecond{corresponding} to one fired pixel to be determined. For \textblueSecond{higher light levels,}
 where such  discrete peaks are  smeared out, \textblueSecond{the mean signal} is divided by the number of  ADC counts corresponding to one fired pixel to determine the number of fired pixels.}

\if 0 %%%% NIMA rev1
where \Npix\ is the number of pixels on the MPPC, $\epsilon$ is the
photon detection efficiency,
%%%%% NIMRev1 including the normalization factor of PMT, 
and \Nin\ is the number of incident photons on the MPPC surface.
%Erika
\fi
 
However, this function requires
modification to take into account the possibility that a single pixel \textred{may}
fire more than once during a signal pulse event. 
The fact that the 12\,ns decay time of a WLS fibre\,\cite{WLSdecayT} is longer than the 4\,ns~MPPC pixel
recovery time~\cite{Oide}, illustrates  this phenomenon.
%%% This may take place because the decay time of a WLS fibre, $\sim 12$~ns
%%%%\cite{WLSdecayT}, is longer than the MPPC pixel recovery time, which
%%%has been measured to be $\sim 4$~ns.
%
In this study the effective number of pixels, \Npixeff, rather than a
constant number of pixels, \Npix, is used to represent this behaviour.
The parameter \Npixeff\ was determined empirically through
measurements %using 
of 72 channels in layer 30 of the \textred{prototype by} fitting Equation~\ref{eq:saturationSimple} to the signals from these channels.
 The 30$^{\mathrm{th}}$ layer consists entirely of MPPCs produced in 2008.
 %\textred{
 The impact of possible differences in \Npixeff\ between the 2007 and 2008 MPPCs is 
 discussed in Section\,\ref{section:systematicFromNpix}.
 %}
 The signals are collected using a ps pulsed laser, of wavelength 408~nm and  FWHM
 31~ps\footnote{PiL040X (Head)
   + EIG2000DX (Controller) provided by Advanced Laser Diode System
   A.L.S.\ GmbH.}, after the detector had been disassembled into layers and transported to Matsumoto, Japan.  
Figure~\ref{fig:NpixMeasurements} shows a schematic of the setup used to measure the saturating response, 
while Fig.~\ref{fig:fittingNpix} {\it left} shows a typical MPPC %%%signal taken by this measurement.
response, \ie the number of MPPC pixels fired as a function of the
incident photon signal as measured using \textredSecond{a photomultiplier tube (PMT)}. %%%%%%% NIM_rev1
\textredSecond{Therefore, $\epsilon$ in
  Equation\,\ref{eq:saturationSimple} includes the normalisation
  factor relating \textblueSecond{the} PMT \textblueSecond{signal} 
to the number of photons incident \textblueSecond{on} the MPPC.}
Equation~\ref{eq:saturationSimple}, even with \Npixeff, is only applicable within a limited
range, outside of which the response function changes at high photon yields,
%%Nige Watson 20160913 I think this should be 'rate' and not 'ratio'
%%Nige Watson 20160913 because the true recovery ratio of pixels depends on the number of
because the recovery %rate  coterra 20160918
of pixels depends on the number of incident
photons~\cite{saturationCorr}: a constant parameter, \Npixeff,
characterises the behaviour.
%Although study of the reason is under investigation, it has not yet clarified.  
The upper limit on the range over which Equation~\ref{eq:saturationSimple}
is fitted is based on the point at which the data stop exhibiting
exponential behaviour.  Figure~\ref{fig:fittingNpix}~{\it right} shows
%(on a logarithmic scale)
 the \textredSecond{slope} of %%%% NIMA-rev1  gradient of
Fig.~\ref{fig:fittingNpix}~{\it left} with respect to the PMT
response. The plot has two distinct regions of approximately linear
behaviour on a logarithmic scale. These are fitted separately, and the
intersection of these two linear fits is taken as the upper limit of
the fit range for Fig.~\ref{fig:fittingNpix}~{\it left}.
%%Nige Watson 20160913 Not 100% sure I follow this statement? I have
%%updated with my best understanding, as in email.
%%Nige Watson 20160913 The ratio of the number of hit channels having \Nfired\ greater than 2000 for every hit channel is less than one percent with a 32 GeV electron beam.
To put these effects of non-linear response into context, considering only MPPCs
that register a signal during a 32\gev\ electron beam run, fewer than 1\%
of MPPCs have $\Nfired>2000$.
Figure~\ref{fig:distNpix} shows the distribution of \Npixeff, obtained
by fitting the parameters of Equation~\ref{eq:saturationSimple} to
measurements from 72 strips in layer 30. This distribution has a mean
and %standard deviation
a SD of 2428 and 245 pixels, respectively. This mean
value was used to implement the correction for the MPPC non-linear response for all
channels.
%%Nige Watson 20160913 Not necessary to advertise in advance what the SD will be used for.
%%Nige Watson 20160913 , and the  and the SD is later used to 
%%Nige Watson 20160913 estimate the systematic effect originating from the uncertainty of $N_{\mathrm{pix}}^{\mathrm{eff}}$. 

\end{description}
%%Nige Watson 20161222 \end{enumerate}

%%%%%%%%%%%%%%%%%%%%%%%%%%%%%%%%%%%%%%%%%%%
%%         Fig  fig:breakdown_capacitance                %%%%%%%%%%%%%%%%%%%%%%%%%%
%%%%%%%%%%%%%%%%%%%%%%%%%%%%%%%%%%%%%%%%%%%
%%% he:/home/coterra/FNALpaper/MPPCproperties
\begin{figure}[tbp]
\captionsetup{width=0.9\textwidth}
\begin{center}{\includegraphics[width=0.45\textwidth]{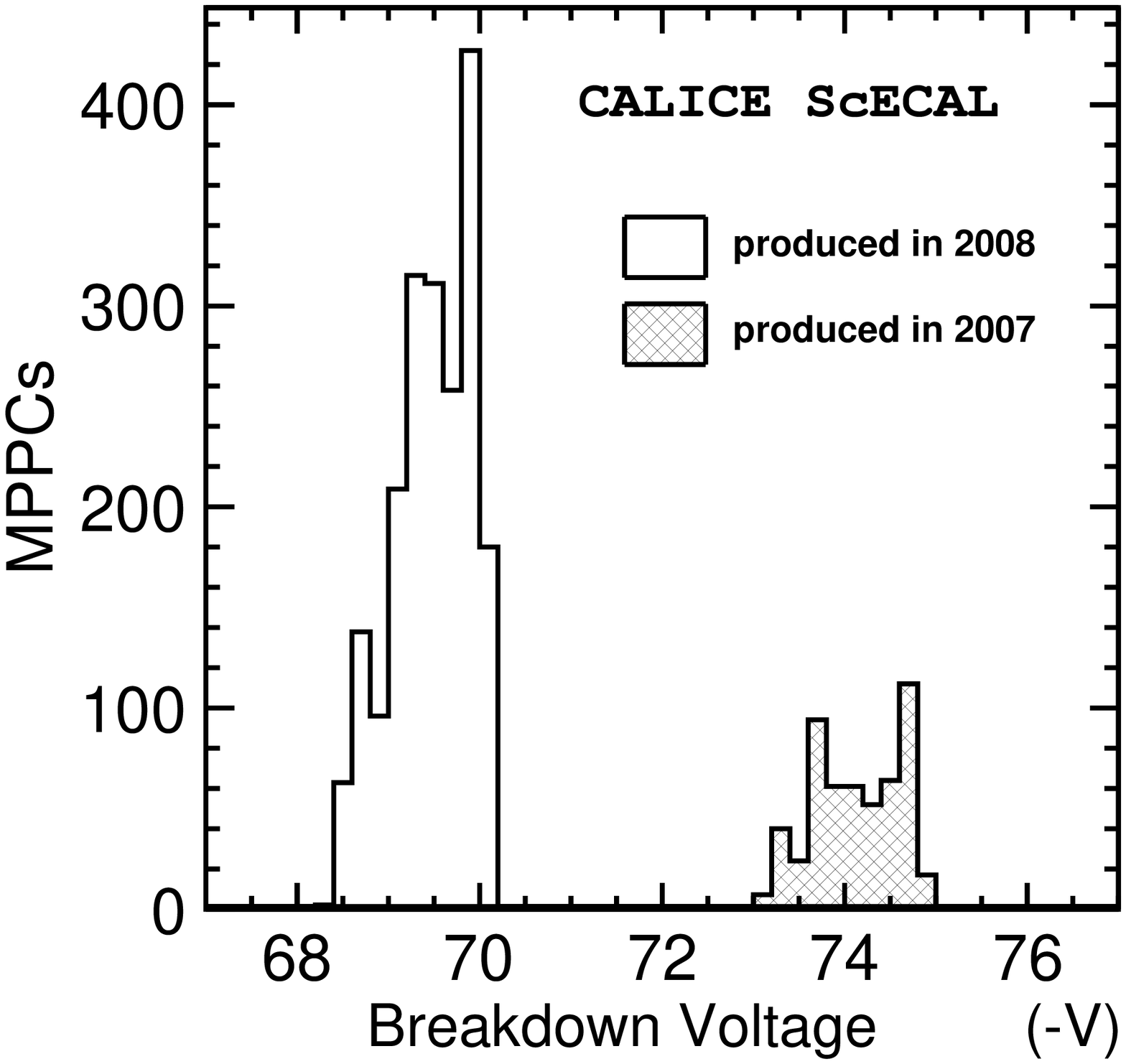}
\includegraphics[width=0.45\textwidth]{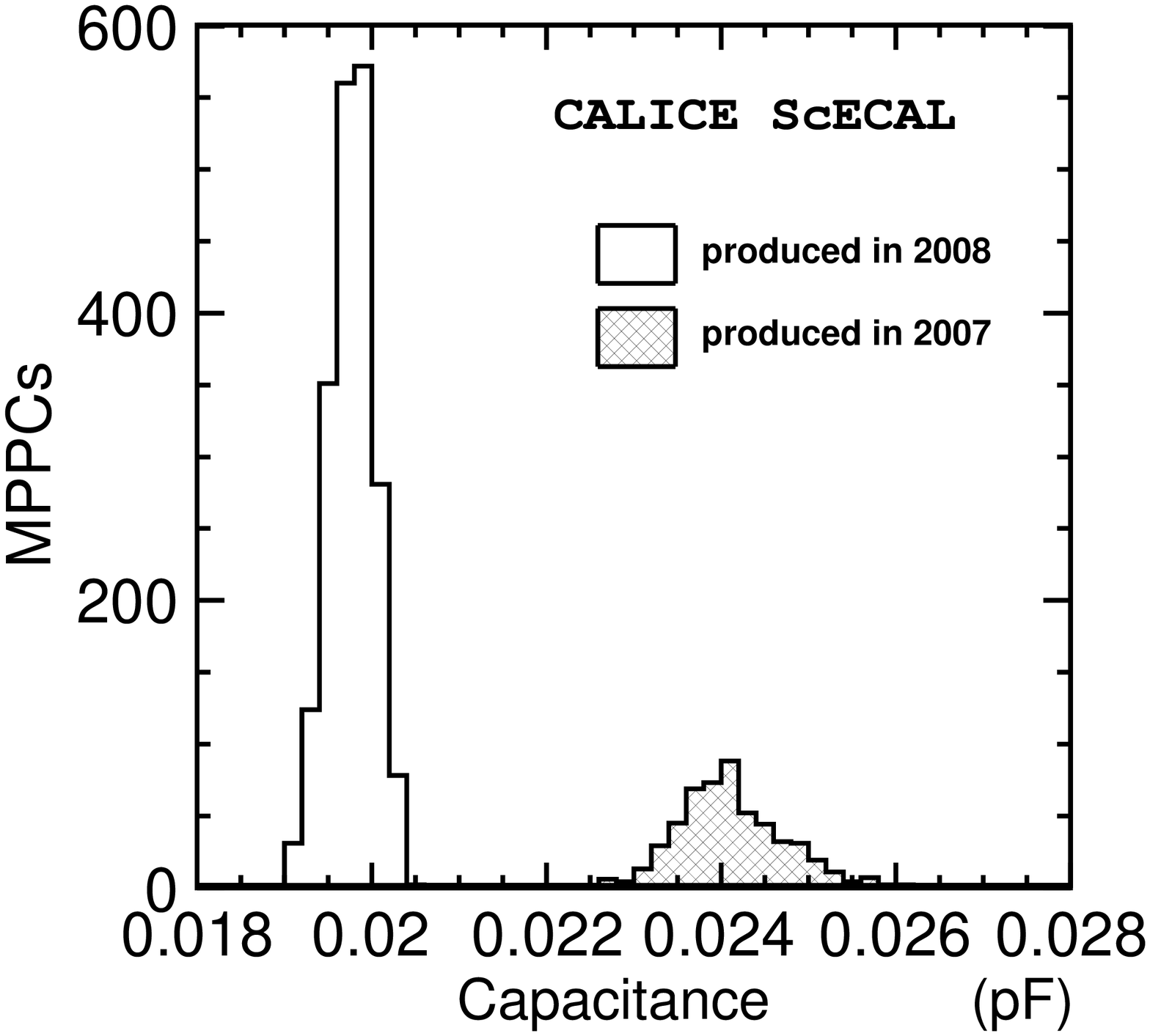}}
\caption{\small Distributions of breakdown voltage ({\it left}) and
  pixel capacitance ({\it right}) of the MPPCs produced in 2007
  (hatched) and 2008 (open).  }\label{fig:breakdown_capacitance}
\end{center}
\end{figure}

%%%%%%%%%%%%%%%%%%%%%%%%%%%%%%%%%%%%%%%%%%%
%%         Fig  NpixMeasurements              %%%%%%%%%%%%%%%%%%%%%%%%%%
%%%%%%%%%%%%%%%%%%%%%%%%%%%%%%%%%%%%%%%%%%%
\begin{figure}[tbp]
\captionsetup{width=0.9\textwidth}
\centerline{\includegraphics[width=0.6\textwidth]{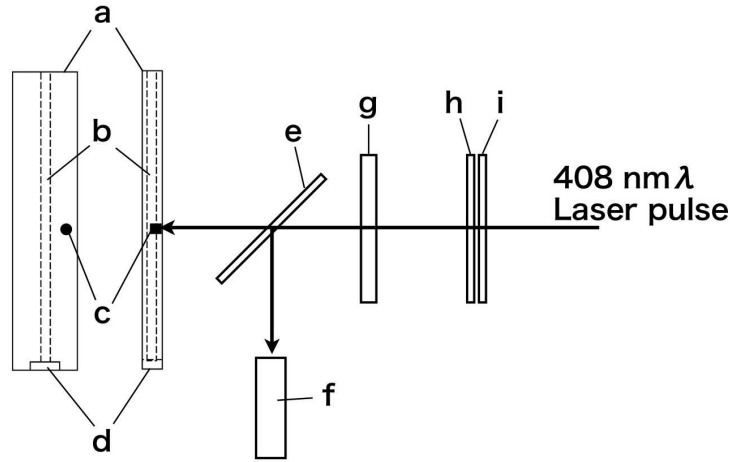}}
\caption{\small Experimental setup for the \Npixeff\ measurement: a)
  target scintillator wrapped in reflective foil (front-view and side-view); %%%{\textit left} topview, {\it right} side view);
   b) WLS fibre; c) irradiation
position with a small hole in reflector; d) MPPC; e) semi-transparent
mirror; f) photomultiplier tube; g) lens; h) polarising plate (fixed);
and i) polarising plate (rotatable).}\label{fig:NpixMeasurements}
\end{figure}
%%%%%%%%%%%%%%%%%%%%%%%%%%%%%%%%%%%%%%%%%%%
%%         Fig  fittingNpix                %%%%%%%%%%%%%%%%%%%%%%%%%%
%%%%%%%%%%%%%%%%%%%%%%%%%%%%%%%%%%%%%%%%%%%
%%%% he:/home/coterra/FNALpaper/FitGetNpixChoi/useInPaper	
\begin{figure}[tbhp]								
\begin{center}\includegraphics[width=0.45\textwidth]{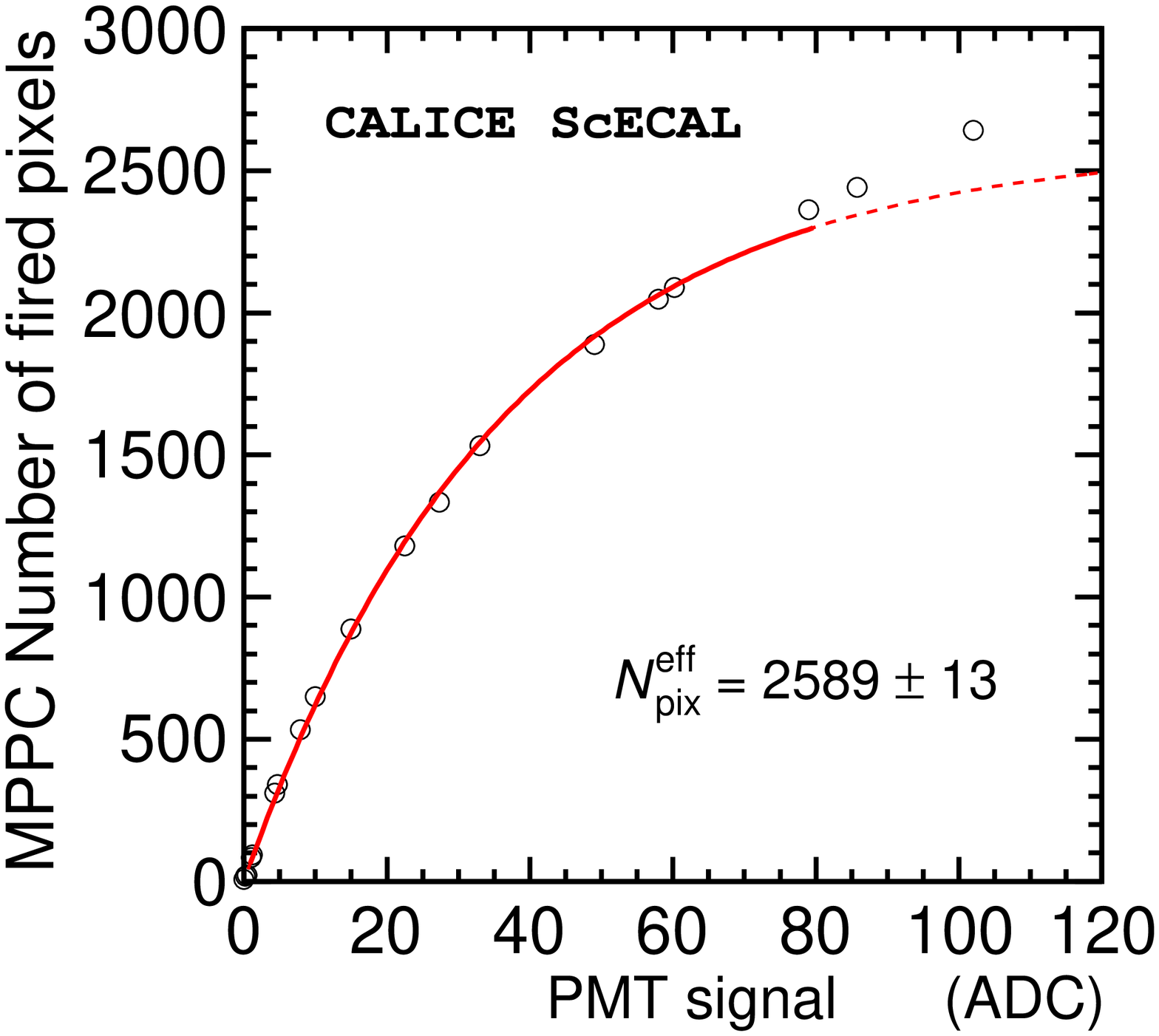}
\includegraphics[width=0.45\textwidth]{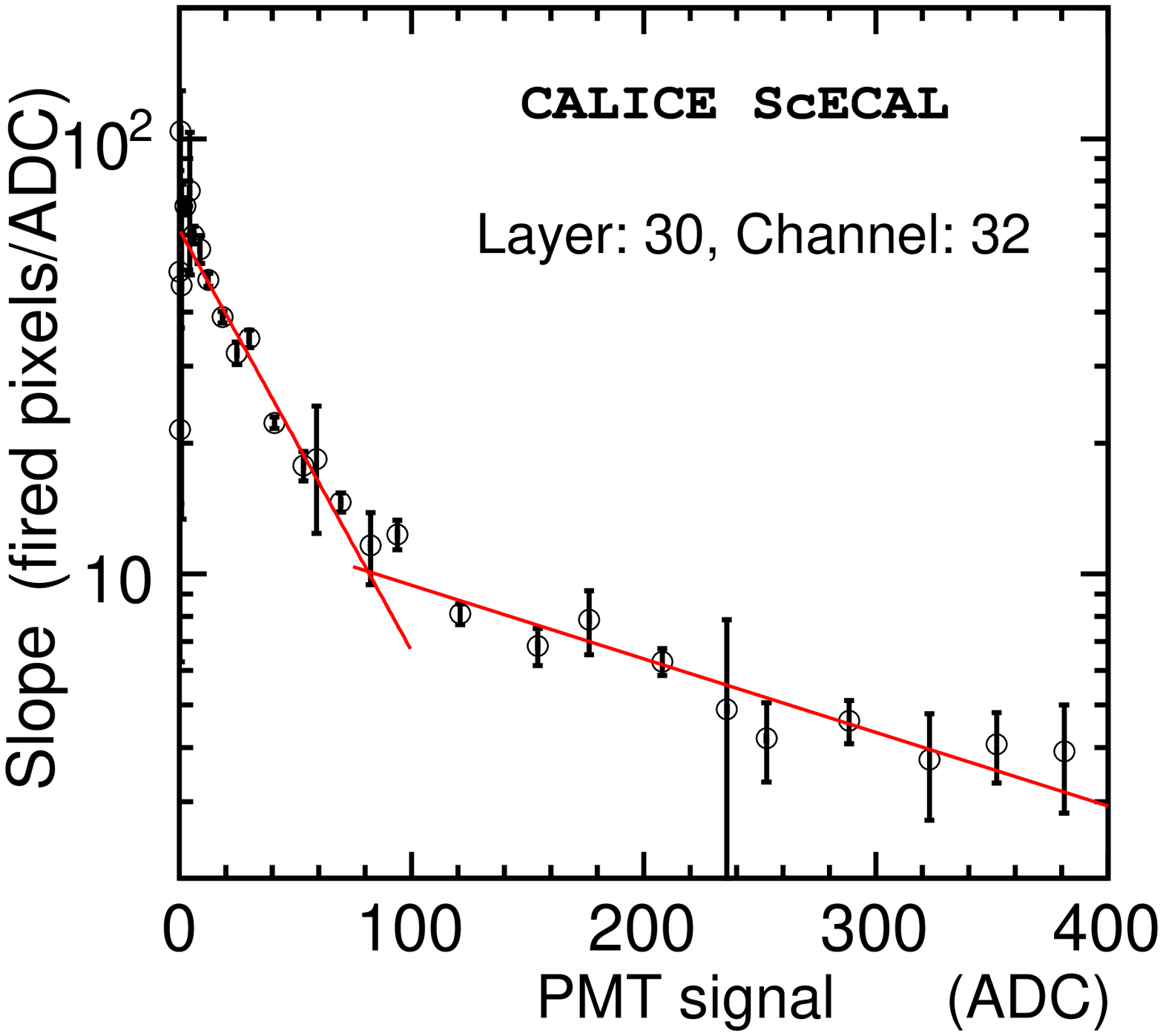}
\caption{\small  {\it left} 
Number of MPPC pixels fired as a function of the incident photon
signal measured using a PMT. The curve shows the results of a
fit using Equation~\ref{eq:saturationSimple}: the solid line indicates
the region over which the fit is performed; the dashed curve shows
the extrapolation of the fit results outside this range.
{\it right} %%%% NIMA-rev1 Gradient 
\textredSecond{Slope} of {\it left} plot vs. %PMT response, 
PMT signal showing
fits in two different regions and their intersection, from which 
the fitting range of the {\it left} plot was determined, see text for
details.
%%Nige Watson 20160913 %%Nige Watson 20160913 Do not think we need to explain what gradient is, just a detail.
%%Nige Watson 20160913 %%Nige Watson 20160913 Gradient\{$(A_{\mathrm{PMT},i+1} + A_{\mathrm{PMT},i})/2\} =$ (MPPC~response$_{i+1} -$ MPPC~response$_{i} )/ (A_{\mathrm{PMT},i+1} + A_{\mathrm{PMT},i})$.
%%Nige Watson 20160913 The dashed vertical line shows the upper range of ADC count considered
%%Nige Watson 20160913 on the {\textit left} plot.
}\label{fig:fittingNpix}
\end{center} 
\end{figure}

%%%%%%%%%%%%%%%%%%%%%%%%%%%%%%%%%%%%%%%%%%%
%%         Fig  fig:distNpix                %%%%%%%%%%%%%%%%%%%%%%%%%%
%%%%%%%%%%%%%%%%%%%%%%%%%%%%%%%%%%%%%%%%%%%
%%% he:/home/coterra/FNALpaper/Npix
\begin{figure}[tbhp]
\captionsetup{width=0.9\textwidth}
\centerline{\includegraphics[width=0.6 \textwidth]{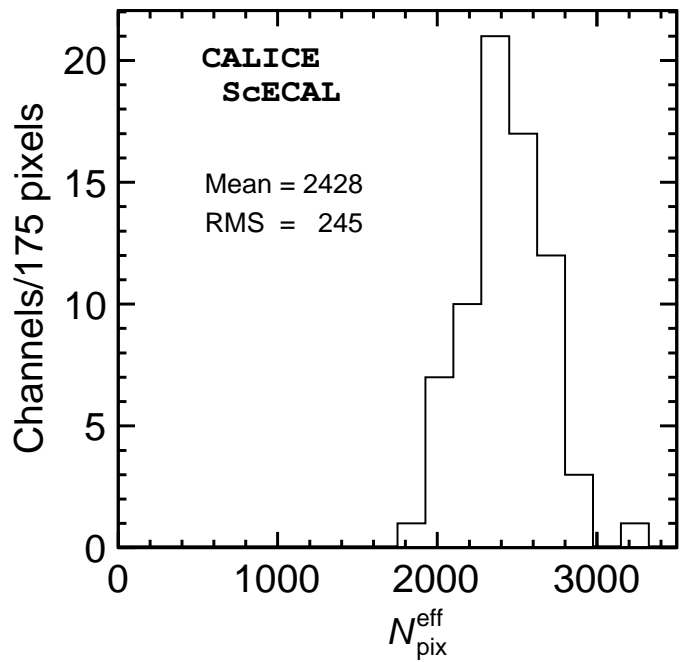}}
\caption{\small Distribution of the number of effective pixels, \Npixeff, measured in 72 strips.}\label{fig:distNpix}
\end{figure}

%%%%%%%%%%%%%%%%%%%%%%%%%%%%%%%%%%%%%%%%%%%%%%%%%%%%%%%%%%%%%%%%%%%%%
%%%%%%%%%%%%%%%%%%%%%%%%%%%%%%%%%%%%%%%%%%%%%%%%%%%%%%%%%%%%%%%%%%%%%

%%%%%%%%%%%%%%%%%%%%%%%%%%%%%%%%%%%%%%%%%%%
%%%%%%%%%%%%%%%%%%%%%%%%%%%%%%%%%%%%%%%%%%%
%%          TEST Beam                %%%%%%%%%%%%%%%%%%%%%%%%
%%%%%%%%%%%%%%%%%%%%%%%%%%%%%%%%%%%%%%%%%%%
%%%%%%%%%%%%%%%%%%%%%%%%%%%%%%%%%%%%%%%%%%%
\section{Test beam at FNAL}\label{section:FNALTB}
%%%%%%%%%%%%%%%%%%%%%%%%%%%%%%%%%%%%%%%%%%%
%%%%%%%%%%%%%%%%%%%%%%%%%%%%%%%%%%%%%%%%%%%
\subsection{Beams and setup}
%%%%%%%%%%%%%%%%%%%%%%%%%%%%%%%%%%%%%%%%%%%
%MT6 and its particle kinds
The prototype described in Section~\ref{section:construction} was
exposed to particle beams of varying type and energy at the Meson Test
Beam Facility number 6 (MT6) at Fermilab: electrons between 1 and
32\gev\ to study the electromagnetic response of the detector; 32\gev\
muons for detector calibration; charged pions between 1 and 32\gev\ to
study the hadronic response in combination with the AHCAL and TCMT. %,
%%and 60 GeV charged pions\footnote{\textcolor{red}{
%%A study indicated that the beam dominated with protons \cite{ThesisOskar}.}
%%}
%% with an iron %%% Frank's suggestion Sep. 18 target
%%target 1.8~m upstream of the
%%prototype, creating neutral pions to evaluate two-cluster separations.
%The muon beams were generated by introducing another iron absorber into
%the 32\gev\ pion beam, upstream of MT6 by the Fermilab beam control
%centre.  
%\textred{
%%%%%The muon beams were generated by introducing another iron absorber, upstream of MT6, into the 32\gev beams which contain the secondary particles of initial proton beams on a 300\,mm aluminium target collision.
The time structure of the beams %in September 2008 and May 2009 
was one 4\,s spill per minute in MT6. %}
This paper reports the response of the prototype to the electron beam data collected in May
2009 at energies between 2\gev\ and 32\gev.

\textred{
The setup of the beam line is shown in Fig.~\ref{fig:setup}.  A
\v{C}erenkov counter \cite{NilsFeege} placed upstream of the experimental area was used
for triggering, together with various combinations of plastic
scintillators.  %A 20\,cm $\times$ 20\,cm %
\textred{A 200\,$\times$\,200\,mm$^2$ }
counter provided the trigger
signals for muon runs, while a pair of %\mbox{10\,cm $\times$10\,cm}
\textred{100\,$\times$100\,mm$^2$} 
counters provided the trigger signals for pion and electron runs: a
coincidence signal from two counters separated by 2.5~m along the beam
direction was required.
%Additionally, a signal from a 1\,$\times$\,1\,m$^2$ veto counter with a 20\,cm $\times$ 20\,cm hole at the center rejected 
%events having such a signal to be recorded. 
Additionally, a 1\,$\times$\,1\,m$^2$ counter with a %20~cm $\times$20~cm
200\,$\times$\,200\,mm$^2$
 hole at its centre was used as a veto counter.  The combinations
of trigger counter and the pressure of the \v{C}erenkov counter
nitrogen gas for the electron and muon runs are listed in
Table~\ref{table:trigger}.
%
\if 0 DANIEL req. 2017.04
This \v{C}erenkov counter is a differential one: an outer mirror counter requires a
\v{C}erenkov cone opening angle larger than a threshold angle and an inner mirror counter requires the cone angle less 
than the threshold angle \cite{NilsFeege}. 
The threshold angle is determined \textblue{by the} %with a 
geometrical relation of a hole at the centre of the outer mirror and the position of the mirror. 
%The opening cone angle is proportional to the difference between mass threshold which available to make \v{C}erenkov light and the particle 
%mass: electrons can make larger angle than other particles. 
The opening cone angle is proportional to the difference between mass of the particle and the lower limit of mass which available to make 
\v{C}erenov light. 
Therefore, the light mass of electron makes larger angle than other particles.
During the electron runs, 
we set the \v{C}erenkov pressure listed in Table~\ref{table:trigger}, so that only electrons hit the outer mirror counter.
%}
\fi
For beam energies $E= 8$\gev and 12\gev,
two different
\v{C}erenkov counter pressures were used.
The effects of these differences are small and are included
in estimating systematic uncertainties.
The %20~cm $\times$ 20~cm 
\textred{200\,$\times$\,200\,mm$^2$} 
counter also served as a multiplicity
counter to distinguish multi-particle events from single particle
events: % Frank's suggestion to v04-02 but Nigel didn't :  signal amplitudes of this counter were used to remove the
the signal amplitude of this counter was used to remove %the
 multi-particle events in the off-line analysis.
}

%%Nige Watson 20160913 Not used.
%%Nige Watson 20161203 \if 0
%%Nige Watson 20161203 MTest \v{C}erenkov counter has two light cone counters: one is called
%%Nige Watson 20161203 ``inner'' which detects narrow cone coming from the particles having
%%Nige Watson 20161203 %%%20140523   just over the threshold pressure of the particles, 
%%Nige Watson 20161203 speed which is only a little over the threshold,
%%Nige Watson 20161203 the other is  
%%Nige Watson 20161203 called ``outer'' which detects light having enough speed making large cone angle. 
%%Nige Watson 20161203 Electron has enough low pressure threshold than other particles. Therefore, the  \v{C}erenkov gas pressure was set keeping more than 100 hPa lower  threshold pressure of pion up to 15~GeV electron,
%%Nige Watson 20161203 while it was set to close threshold of pion for 20--32~GeV electron.  
%%Nige Watson 20161203 \fi
%%Nige Watson 20160913 Not used.

%%%%%%%%%%%%%%%%%%%%%%%%%%%%%%%%%%%%%%%%%%%
%%         Fig  setup                %%%%%%%%%%%%%%%%%%%%%%%%%%
%%%%%%%%%%%%%%%%%%%%%%%%%%%%%%%%%%%%%%%%%%%
 \begin{figure}[tbp]	
 \captionsetup{width=0.9\textwidth}								
\begin{center}\includegraphics[width=14cm]{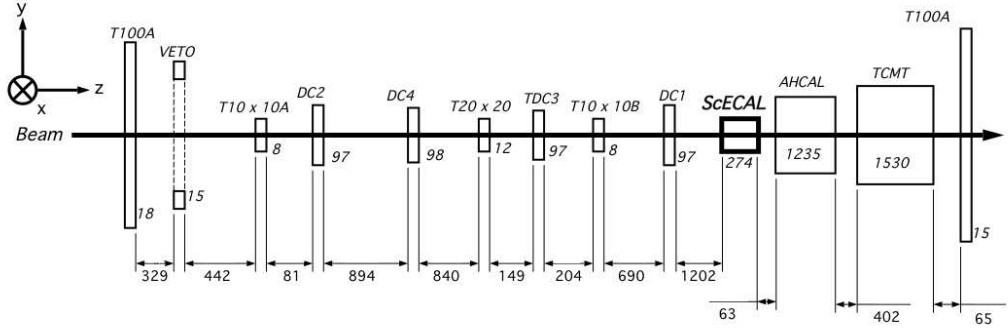}
\caption{\label{fig:setup} \small{ {Configuration of detectors on the
      MT6 beam-line at the MTBF. \textredSecond{Schematic is not to scale.}  A %%%%% NIMA rev1
      right-handed coordinate system is shown.  Italic numbers at
      bottom/right of detector elements show their thicknesses. All
      dimensions are in mm.}}}
\end{center} 
\end{figure}

\begin{table}[tbhp]%[!]
\captionsetup{width=0.8\textwidth}
\begin{center}
\caption{\label{table:trigger}{Trigger systems used for different
    particles and energies.  The pressure of the \v{C}erenkov counter
    used for each trigger configuration is also indicated.}
%%Nige Watson 20160913 This comment is better in the text not the
%%table, and it is clear from the table that two pressures were used.
%%Nige Watson 20160913 For beam momenta $p$(GeV/$c$)= 8\,GeV and 12\,GeV, two different
%%Nige Watson 20160913 \v{C}erenkov counter pressures were used.
%%Nige Watson 20160913 The effects of those differences are not significant. %and they are included in the systematic uncertainties of run by run. 
}
\begin{tabular}{lrcc}
\hline \hline
&&\vspace{-3mm}\\
Particle & $E$[\gev] &Trigger  & \v{C}erenkov pressure (hPA) \\
\hline %\hline
&&\vspace{-3mm}\\
muon & 32 & 200\,$\times$200\,mm$^2$& - \\
electron & 2& 100\,$\times$100\,mm$^2$& 345 \\   
electron & 4& 100\,$\times$100\,mm$^2$& 345 \\
electron & 8&  100\,$\times$100\,mm$^2$& 282, 158 \\
electron & 12& 100\,$\times$100\,mm$^2$& 158, 138 \\    
electron & 15& 100\,$\times$100\,mm$^2$& 138 \\
electron & 20& 100\,$\times$100\,mm$^2$& 138 \\
electron & 30& 100\,$\times$100\,mm$^2$& 103 \\   
electron & 32& 100\,$\times$100\,mm$^2$& 103 \\            
\hline
\end{tabular}
\end{center}
\end{table}

\subsection{Temperature measurement}
\textred{The temperature of the prototype was measured using two thermocouples\textblue{,} %; 
one located on the top of the first ScECAL layer and the other at the bottom of the
last layer.}
%\textcolor{red}{
Figure~\ref{fig:runs_temperature}
%} 
shows the temperature of
data acquisition periods, averaged over each run with a 1~Hz data
recording rate and
%averaging over the 1 Hz temperature data taking rate and
 over the two sensors. 
%The temperature of the prototype varied between 19$^{\circ}$C and 27.5$^{\circ}$C. 
%Frank's requirement
Data were recorded in runs with durations varying between 16 and 85
minutes, and the average temperature of the prototype within a given
run was stable to within 0.24$^{\circ}$C. The temperatures %between runs 
recorded varied between 19.0$^{\circ}$C and 27.5$^{\circ}$C.  %This
Although this
 large variation was caused in part by a malfunction of the air
conditioning of the experimental hall over a period of two days at the
beginning of the data-taking period, 
%The performance of the prototype presented in this paper was exposed to rather severe conditions ($\Delta T = 8^{\circ}$C).
%This provided a good test of tolerance of ScECAL toward the temperature variation, since the sensitivity of MPPC varies with %changes in temperature.
%%These large variations in temperature provided 
this allowed 
a robust test of the
sensitivity of the ScECAL  to be performed %($\Delta T = 8^{\circ}$C) 
%\textcolor{red}{
and %} 
 confirms the %tolerance 
resilience  of the prototype. %%%to large variations in temperature.

% is presented in this paper.
%%%%%%%%%%%%%%%%%%%%%%%%%%%%%%%%%%%%%%%%%%%
%%         Fig  runs_temperature                %%%%%%%%%%%%%%%%%%%%%%%%%%
%%%%%%%%%%%%%%%%%%%%%%%%%%%%%%%%%%%%%%%%%%%
\begin{figure}[tb]		
\captionsetup{width=0.9\textwidth}							
\begin{center}\includegraphics[width=12cm]{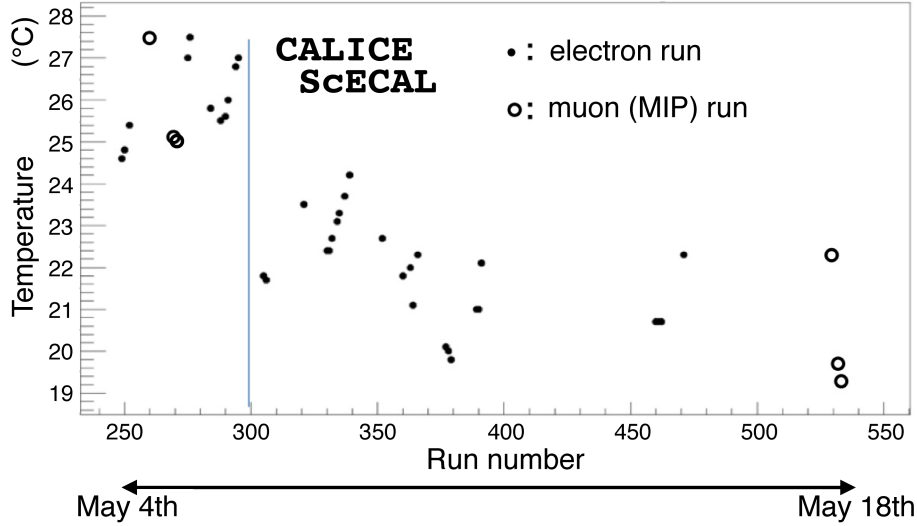}
\caption{\label{fig:runs_temperature} \small{ {Temperature of the ScECAL
      prototype during the muon and electron runs collected in 2009.
      The air conditioning system of the experimental hall was not
      operational in the period to the left of the vertical line.  }}}
\end{center} 
\end{figure}

\section{Reconstruction procedure}\label{section:analysis}

%%%%This section summarises the calibration procedures used and the way in
%%%%which the energy spectra of electrons are measured using the
%%%%%calibrated data. 
\textred{This section %%% 20170511 Nigel: represents 
gives an overview of the calibration procedure,
%%% 20170511 Nigel: and 
the determination of the calibration factors,
and the subsequent measurement of the
%%% 20170511 Nigel: \textmagenta{Following the calibration, the extraction of the measured electron 
 energy spectra. %%% are given.}
%TODO move to somewhere \textblue{
%%The signal pedestals} %The pedestal values of the electronics
% %were monitored by recording 500 randomly triggered events in the period
%%between beam spills.  The means of these pedestal events were
%%calculated separately for each channel, and \textblue{subtracted from signals collected during the subsequent set of beam events. 
%%then subtracted from the
%subsequent set of beam events.
}
}

%%%%%%%%%%%%%%%%%%%%%%%%%%%%%%%%%%%%%%%%%%%%%%%%%%%%%%%%%%%%%%%%%%%%%
%%%%%%%%%%%%%%%%%%%%%%%%%%%%%%%%%%%%%%%%%%%%%%%%%%%%%%%%%%%%%%%%%%%%%
\subsection{Calibration procedure}\label{section:calibrationProcedure}
%%%%%%%%%%%%%%%%%%%%%%%%%%%%%%%%%%%%%%%%%%%%%%%%%%%%%%%%%%%%%%%%%%%%%
%%%% g702 The ScECAL takes three step calibration procedures:
The ScECAL calibration is performed in three steps:
\begin{enumerate}
\item relative calibration of cells, to ensure uniform cell-to-cell
  response;
\item gain calibration (in ADC counts), to determine the signal amplitude
  corresponding to a single fired pixel, and 
\item calibration to an absolute energy scale, using electromagnetic
  showers.
\end{enumerate}
For the first step, the cell-to-cell response of cells is normalised
using the response of each cell to beams of muons, which approximate
%to 
minimum ionising particles (MIP).  The most probable value (MPV) of
the signal distribution obtained using muons and measured in ADC
counts, \cmip, is the calibration factor of this procedure.  
%With 
After this
calibration, the visible energy in the detector is %given as 
 %explained 
expressed  in units of MIPs.
%%ADC-MIP conversion factor (\cmip).

A second calibration step %% 20170511 Nigel: is %necessary to account for the fact that
%MPPCs are inherently non-linear devices. 
is performed to correct for the non-linear response of MPPCs.
%%%for the saturation correction of MPPCs.
%%This formula does not include the effect from the cross-talk phenomenon between pixels and the after-pulse.
The inverse of Equation~\ref{eq:saturationSimple} unfolds %, with the effective number of pixels, \Npixeff, 
the effects of non-linear response as discussed in Section\,\ref{section:effectiveN}. 
However, as $F^{-1}$ is a function of  the number of fired pixels,
% and the input of $F^{-1}$ is the number of fired pixels.
the amplitude of the signal must be converted accordingly.  
The relevant ADC--photoelectron conversion factor (\cphoton) is
determined in situ for each channel, 
%%%% 20170628
%%% 20171128 where a photoelectron corresponds to an  electron-hole pair triggering a geiger discharge of a pixel, 
where a photoelectron corresponds to an electron-hole pair in the SiPM triggering a geiger discharge of a pixel, 
\textredSecond{which also corresponds to a fired SiPM pixel.}   %%%%% NIMA rev1 added for 
%%%
This is an essential role for the
LED-based gain monitoring system discussed in
Section~\ref{section:construction}.

%For the calibration of each MPPC, one additional calibration
%\textred{
The second step---the calibration of each MPPC---includes
one additional calibration coefficient, 
%}
%coefficient is required 
because \cphoton\ is measured using a high-gain amplifier %for enough 
to achieve a sufficient separation of photoelectron peaks \cite{AHCAL}, %photoElecSep}, 
whereas the signals in physics data taking are
acquired using a lower gain due to the wider dynamic range required.
This calibration coefficient, referred to as inter-calibration
coefficient (\cinter), is measured for each channel as the ratio of
the amplitudes of the response to LED light with the high-gain to the
low-gain settings.
%%% to answer Erika's request.
Therefore, a \cinter\ includes not only the ratio of amplifier in electronics but also the effect from 
the %%%% NIM rev1
difference of pulse shape time between the high-gain and low-gain modes.

These calibration constants, namely the ADC-MIP conversion factor, the
ADC--photoelectron conversion factor and the inter-calibration
coefficient for each channel, were determined in situ and %therefore
are discussed in Section~\ref{AnalysisFlow}.
%%their specific values of them are discussed in Section~\ref{section:analysis}.
%%%%%%%%%%%%%%%%%%The details of the saturation correction via the ADC--photo-electron
%%%%%%%%%%%%%%%%%%%conversion factor are discussed in the following subsection.
%
With these constants, a signal in channel $i$ for the
physics study can be written as:
\begin{eqnarray} \label{eq:unfold} 
 %%%% Erika2  \AcorrMIPS = F^{-1}\Big(
  %%%% Erika2  \Aadc(T) \frac{ \cinter_{i} }{ \cphoton_{i}(T)}\Big)\ 
  %%%% Erika2           \frac{\cphoton_{i}(T)}{\cinter_{i}\cdot\cmip_{i}(T)}   \,\, ,
  \AcorrMIPS = F^{-1}\Big(
  \Aadc(T) \frac{ \cinter_{i} }{ \cphoton_{i}(T)}\Big)\ 
           \frac{\cphoton_{i}(T)}{\cinter_{i}\cdot\cmip_{i}(T)}   \,\, ,
\end{eqnarray}
where $\Aadc(T)$ is the uncorrected signal of the cell in ADC counts
for a detector of temperature $T$, and 
%%%%$F^{-1}$ has a parameter \Npixeff\ instead of \Npix\ as discussed in Section\,\ref{section:effectiveN}.
$F^{-1}$ has a parameter \Npixeff\ instead of \Npix\ as  discussed in 
Section\,\ref{section:effectiveN}, and $\epsilon$ %%%%%%%%%%becomes unit because \cmip\ and $\Aadc$ are also values 
%%%%%%%%%% affected by the same PDE.
is cancelled in Equation \ref{eq:unfold} because both a $F^{-1}$ and a \cmip\ are inversely proportional to the $\epsilon$.
 
Each calibration factor is
determined as a function of temperature.  The sum of these signals
represents the energy of an event in a physics run, in units of MIPs,
and is given by
\begin{eqnarray}\label{signalMIP}
		E_\mathrm{reco}[\mathrm{MIP}] = \sum_{\mathrm{all\, strips}}  \AcorrMIPS \;\;\; .
\end{eqnarray}
%	is the energy of the event in a physics run.

The mean of $E_\mathrm{reco}$ as a function of the incident beam energy represents
the calibration of the ScECAL to an energy scale in \gev\ as required for
the third calibration step.  The demonstration of this calibration is
one of the primary goals of the test beam activity reported in this
article.  Detailed results are discussed in
Section~\ref{section:results}.

%%%%%%%%%%%%%%%%%%%%%%%%%%%%%%%%%%%%%%%%%%%
%%%%%%%%%%%%%%%%%%%%%%%%%%%%%%%%%%%%%%%%%%%
\subsection{Calibration runs \textmagenta{and pedestal measurements}}\label{calibrationRuns}
%%%%%%%%%%%%%%%%%%%%%%%%%%%%%%%%%%%%%%%%%%%

To determine the \cmip, six muon runs were
recorded over a wide range of temperatures, allowing the temperature
dependence to be quantified.
%
%
%The ADC--photo-electron
%conversion factors are required to convert ADC counts to a number of photon-electrons for the 
%saturation correction as discussed in section \ref{section:effectiveN}.
To determine the \cphoton, several LED
calibration runs were typically recorded per day.
%%Nige Watson 20160913 In these runs, events were collected with LED light pulses provided via the clear fibres discussed in section \ref{section:construction}. 
%responded to Frank's question 
%During each run, 50,000 events were collected in total  of eleven LED power settings, 
During each run %with a total 
of 50\,000 events, the LED power was
changed in \textred{eleven} steps to ensure that some events with a suitable photon
yield were present in all channels.
%Since the ADC--photo-electron conversion factor depends on the temperature, a temperature-dependent conversion function was derived.

Inter-calibration runs were also taken in the LED calibration
runs. For these, intermediate intensity LED light was injected into
each channel and the signal in ADC counts was measured in both the
low-gain and high-gain modes.  In each such
run, 50\,000 events were taken for each of the \textred{eleven} different LED power
settings.

\textmagenta{The signal pedestals
 were monitored by recording 500 randomly triggered events in the period
between beam spills.  The mean values of these pedestal events were
calculated separately for each channel, and \textblue{subtracted from signals collected during the subsequent set of beam events. }
The widths of the pedestals were also calculated as the RMS for each channel.
}

\subsection{Determination of calibration constants}
\label{AnalysisFlow}
Three calibration factors, \cmip, \cphoton\, and \cinter, 
discussed in Section\,\ref{section:calibrationProcedure}
are determined in this subsection.
%are required for each cell: MIP calibration using
%muon run data; LED calibration to monitor the MPPC gain used in the
%MPPC saturation correction, and inter-calibration to convert between
%the calibration and physics gain modes. 
The methods through which these are also
evaluated are described.

\subsubsection{%\cmip}%
ADC-MIP conversion factor}% as a function of temperature}
\begin{figure}[tbp]			
\captionsetup{width=0.8\textwidth}						
\begin{center}\includegraphics[width=0.6\textwidth]{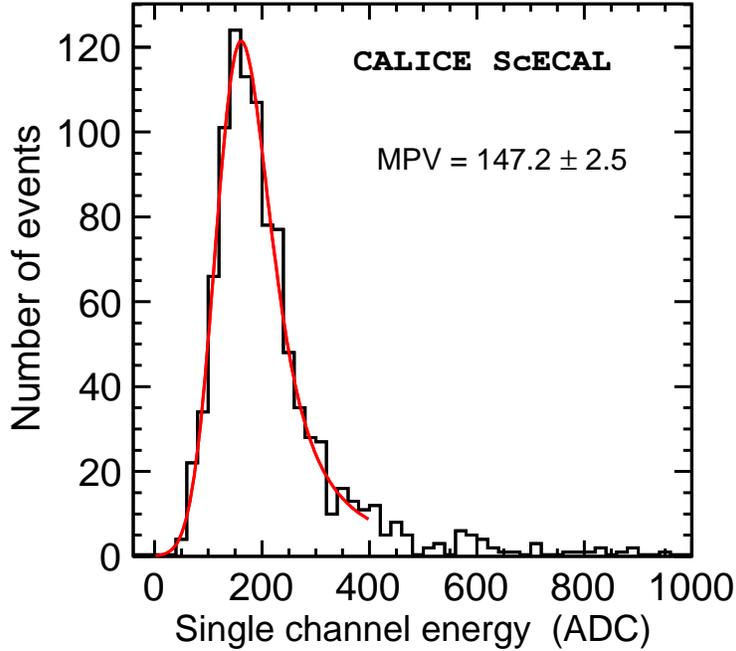}
\caption{\label{fig:mip} \small{ {Distribution 
%\textcolor{red}{
of
%}
pedestal-subtracted
      energy deposits from MIP-like particles in a single (typical)
      channel. The solid line shows the result of a fit using a
      Gaussian-convoluted Landau function.}}}
\end{center} 
\end{figure}

To select muon events
%%Nige Watson 20161031 for a strip in an $x$-oriented ($y$-oriented) layer,
%%%% 20170511 Nigel: we required that hits were 
hits were required to be
present in at least ten layers in the same lateral
%strip position of the $x$-oriented ($y$-oriented) layers.
strip position of the same oriented layers of $y$---having detail segmentation in $x$---($x$ layer) or $x$---having detail segmentation in $y$---($y$ layer).
 A strip was defined to have been hit if the recorded signal value was more than three
times the width of a Gaussian function fitted to the corresponding pedestal distribution.
%To select muon events, at least ten layers were required to have a hit in the same position strip of the X-oriented (Y-oriented) layers for a strip in a X-oriented (Y-oriented) layer.
As an example, Fig.~\ref{fig:mip} shows the distribution of signal
recorded in a typical single strip for muon events, 
fitted with a Gaussian-convoluted Landau function.
%%Nige Watson 20161031 NKW: Suggest that we do not have to give this level of detail in the paper.
%%Nige Watson 20161031 \begin{equation}
%%Nige Watson 20161031 	L(x) = \int^{\infty}_{-\infty}dx'
%%Nige Watson 20161031         \mathcal{L}(x-x')\mathcal{G}(x') \,\, ,
%%Nige Watson 20161031 \end{equation}
%%Nige Watson 20161031 where $\mathcal{L}$ and $\mathcal{G}$ represent Landau and Gaussian
%%Nige Watson 20161031 functions, and $x$ and $x'$ are measured in ADC units.
The MPV of the function is taken to be the \cmip, 
and the mean uncertainty % 20170511 David: from 
of 
all channels on the fitted MPV was ($1.8\pm0.7$)\%.
%%%%%%%
%%%
%%%  data: he: /home/coterra/frmDESYlc4/coterra/scecalana/work/MIPLandauRangeFrm1
%%%	mipAll110824.eps, mipErr110824.eps,  ref. ./resultData/
%%%	checked /home/coterra/frmDESYlc4/coterra/scecalana/calice_reco/raw2calohit/src/ScECALMIPCalibrationProcessor.cc
%%%
%%%%%%%

%%%%%%%%%%%%%%%%%%%%%%%%%%%%%%%%%%%%%%%%%%%
%%         Fig  mip                %%%%%%%%%%%%%%%%%%%%%%%%%%
%%%%%%%%%%%%%%%%%%%%%%%%%%%%%%%%%%%%%%%%%%%
%%% he:/home/coterra/FNALpaper/MIPFit

\begin{figure}[tbhp]					
\captionsetup{width=0.8\textwidth}					
\begin{center}\includegraphics[width=0.6\textwidth]{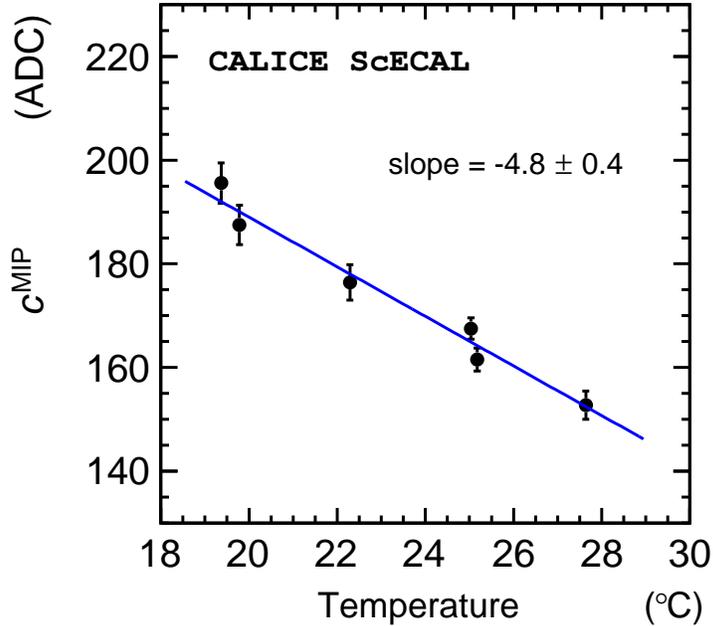}
\caption{\label{fig:mip_temp} \small{
{Measured \cmip\ 
of a typical channel
as a function of the average detector temperature
  during a data taking run.
%%% Frank v04-02 Solid 
The line shows the result of a linear fit. 
This fitting gives \cmip(20$^\circ$C) = 189\,$\pm$\,14.
 This channel used a MPPC
from the 2008 production.}}}
\end{center} 
\end{figure}

\begin{figure}[tbhp]								
\begin{center}\includegraphics[width=0.45\textwidth]{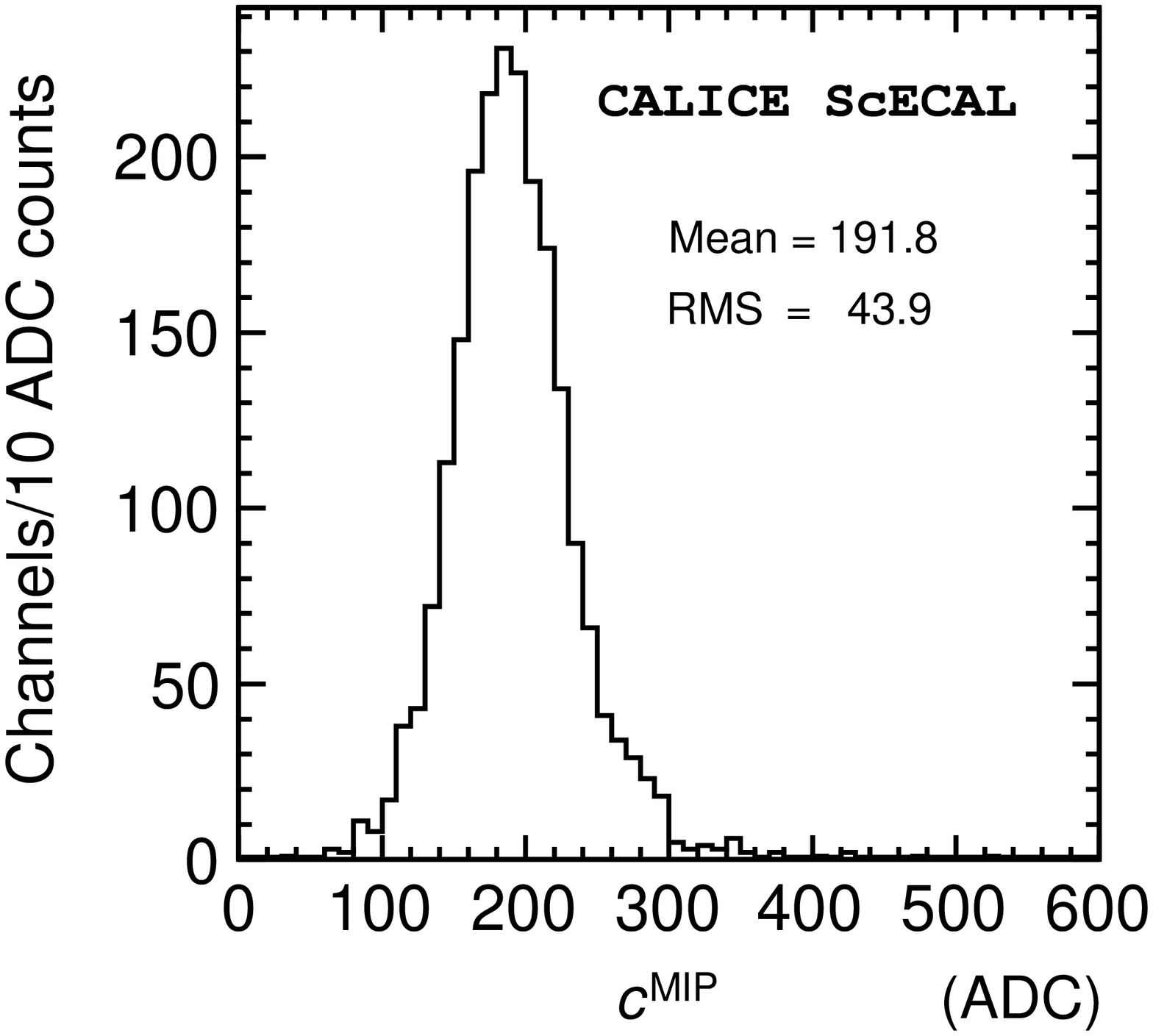}
\includegraphics[width=0.45\textwidth]{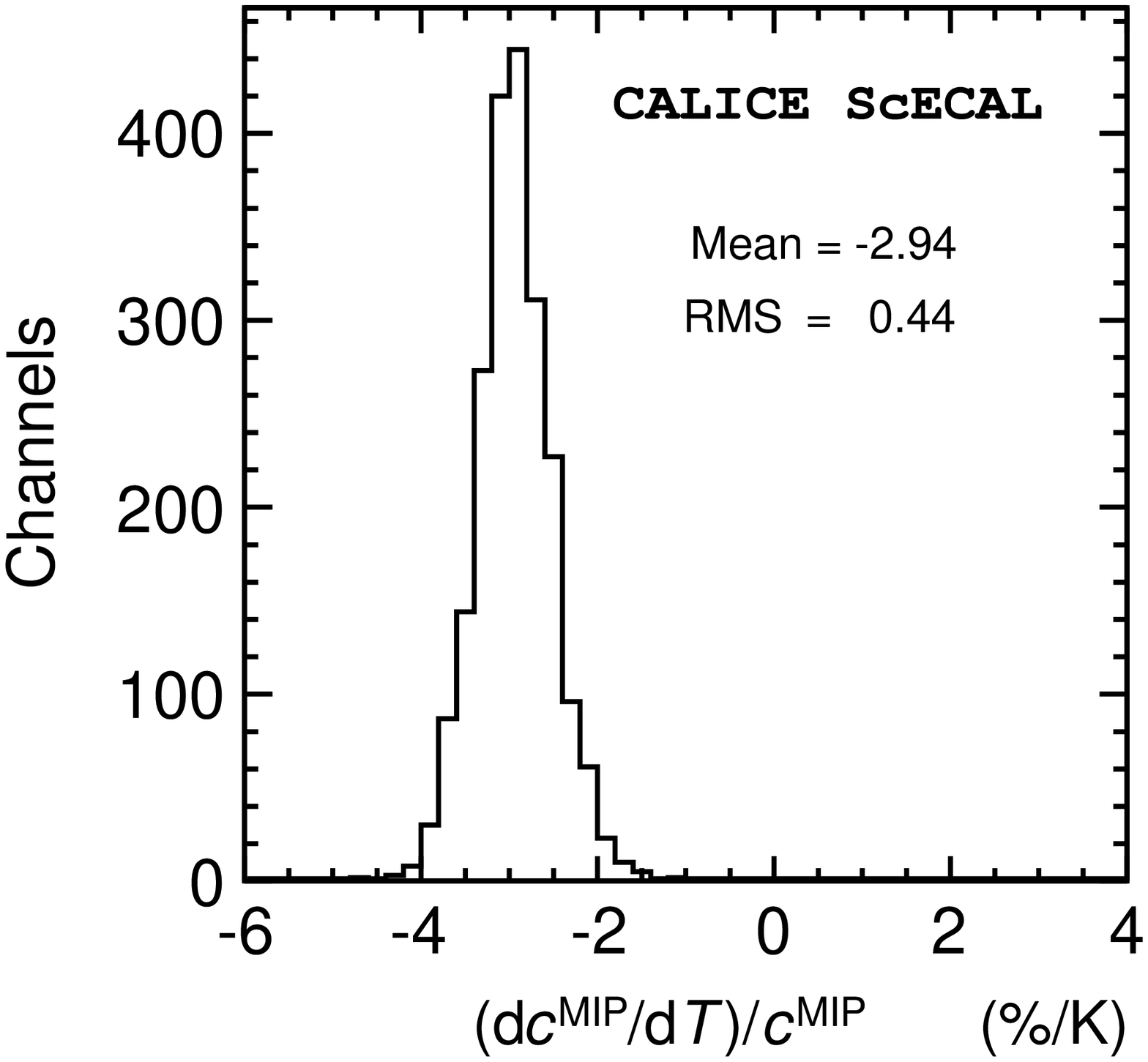}
\caption{\label{fig:adc_mip_distribution} \small{
{{\it left} Distribution of $c^{\mathrm{MIP}}(T_0=20^{\circ}$C). {\it right} Distribution of the slope of \cmip, 
%$(\diff c^{\mathrm{MIP}}/c^{\mathrm{MIP}}$)$/\diff T$.
$(\diff\cmip/\diff T)/\cmip$.
}}}
\end{center} 
\end{figure}

The MPV of each channel was measured in six dedicated runs at various
temperatures, allowing the temperature dependence of the MIP response to
be determined.  This is illustrated for a typical channel in
Fig.~\ref{fig:mip_temp}, showing a linear dependence of the Landau
MPV on the average temperature during a run.
%%Nige Watson 20161031 NKW: Is it necessary to give the equation for a linear function in the paper?
%%Nige Watson 20161031 The conversion factor \cmip\ is linear with temperature, as shown
%%Nige Watson 20161031 the line in Figure~\ref{fig:mip_temp} shows the result of such a linear fit. 
The %ADC-MIP conversion factor for each channel
 \cmip\ is therefore expressed as
\begin{equation}\label{eq:cmip}
\cmip(T) = \cmip(T_0) + \frac{\diff\cmip }{\diff T} (T - T_0)\,\, ,
\end{equation}
where $T$ is the temperature at which the measurement was taken, and
$T_0$ is a reference temperature.  The parameters
$\cmip(T_0)$ and $\diff\cmip/\diff T$ were
determined for each channel and
%to correct the estimated energy of electron beam data to the reference temperature.
account for the effect of temperature on the energy deposit as
measured in each channel.
%%%%%%%%%%%%%%%%%%%%%%%%%%%%%%%%%%%%%%%%%%%
%%         Fig  mip_temp                %%%%%%%%%%%%%%%%%%%%%%%%%%
%%%%%%%%%%%%%%%%%%%%%%%%%%%%%%%%%%%%%%%%%%%
%%% he:/home/coterra/FNALpaper/MIP_temp
%
%
%%Nige Watson 20161031 Figure~\ref{fig:adc_mip_distribution} {\it left} shows the
%%Nige Watson 20161031 distribution of the ADC-MIP conversion factors estimated at
%%Nige Watson 20161031 20$^{\circ}$C \{$c^{\mathrm{MIP}}(T=20^{\circ}$C)\}.  The RMS of the
%%Nige Watson 20161031 distribution is 19\% of its mean.
%%Nige Watson 20161031 Figure~\ref{fig:adc_mip_distribution} {\it right} shows the
%%Nige Watson 20161031 distribution of $(\diff c^{\mathrm{MIP}}/c^{\mathrm{MIP}}$)$/\diff T$.
%%Nige Watson 20161031 The mean of this distribution with the RMS is (-2.95$\pm$0.44\%/K).
Figure~\ref{fig:adc_mip_distribution} shows the
distributions of the %ADC-MIP conversion factors,
\cmip, estimated at
20$^{\circ}$C, and $(\diff\cmip/\diff T)/\cmip$.
%%%%%%%%%%%%%%%%%%%%%%%%%%%%%%%%%%%%%%%%%%%
%%         Fig  adc_mip_distribution                %%%%%%%%%%%%%%%%%%%%%%%%%%
%%%%%%%%%%%%%%%%%%%%%%%%%%%%%%%%%%%%%%%%%%%
%%% he:/home/coterra/FNALpaper/MIP_dist
%%% he:/home/coterra/FNALpaper/MIP_slope

\subsubsection{%\cphoton}%
ADC--photoelectron conversion factor} % as a function of temperature}
\label{section:adcphoton}

\begin{figure}[tbp]						
\captionsetup{width=0.8\textwidth}				
\begin{center}\includegraphics[width=0.6\textwidth]{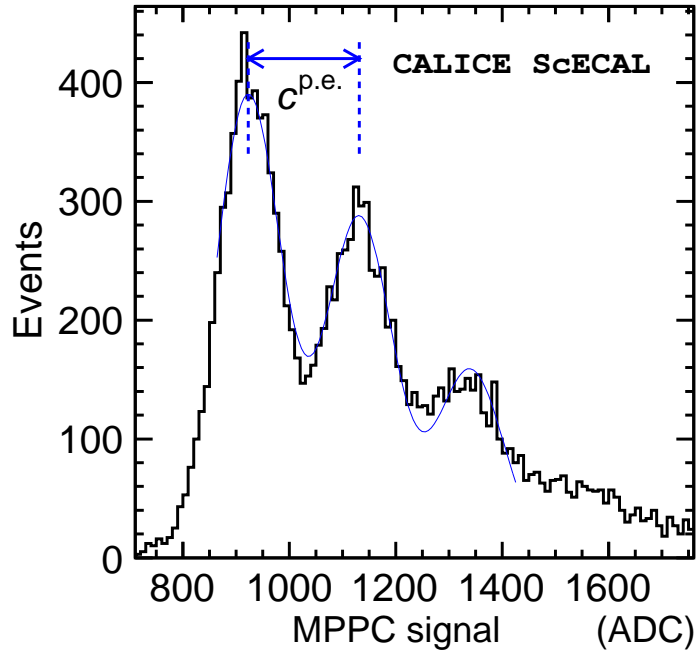}
\caption{\label{fig:adc_photon} \small{
{Typical spectrum of a LED run for a single channel, with the results
  of a  three-Gaussian function fit overlaid. The arrow indicates \cphoton\ for this channel.}}}
\end{center} 
\end{figure}

The \cphoton\  was determined by measuring signal distributions
consisting of a few peaks of photoelectrons induced by LED light during the
dedicated runs discussed in Section~\ref{calibrationRuns}.  %The number of ADC counts corresponding to a single photo-electron for a given channel is \cphoton.  
%The temperature dependence of $c^{\mathrm{p.e.}}$ is affected only by gain variations.
Figure~\ref{fig:adc_photon} shows an example of an MPPC signal
distribution, for one of several LED intensities used during a
calibration run.
The pedestal and first two peaks of photoelectrons are easily
distinguished.  Three Gaussian functions are used to fit this
distribution, 
%%%%Erica2 with free parameters: %% 20170511 David: of 
%%%%Erica2 the amplitudes of the three
%%%%Erica2 Gaussians; the peak position of one of three Gaussians; a common SD, 
%%%%Erica2 and a shared separation between adjacent peaks.
with six free parameters: the amplitudes of the three Gaussian functions; the peak position of the first Gaussian function; 
a sigma equal for all three functions, and an equal distance between adjacent peaks, corresponding to the MPPC gain.
The latter parameter is the \cphoton\ of this channel.
%
%The error-weighted average of $c^{\mathrm{p.e.}}$ was taken with regard to the variation of LED intensity: multiple LED intensities allowed
% success of fitting.
Where successful fits were obtained for more than one of the LED
intensities used during the calibration runs, a weighted average of
\cphoton\ was used. The mean fractional uncertainty on \cphoton,
averaged over all channels and temperatures, is (0.7\,$\pm\,$0.3)\%.
%% using
%%%he:/home/coterra/frmDESYlc4/coterra/scecalana/work/Gain/resultsGainR560xxx.dat
%
%
%%%%%%%%%%%%%%%%%%%%%%%%%%%%%%%%%%%%%%%%%%%
%%         Fig  adc_photon                %%%%%%%%%%%%%%%%%%%%%%%%%%
%%%%%%%%%%%%%%%%%%%%%%%%%%%%%%%%%%%%%%%%%%%
%%% he:/home/coterra/FNALpaper/GainFit

\begin{figure}[tbhp]	
\begin{center}\includegraphics[width=0.45\textwidth]{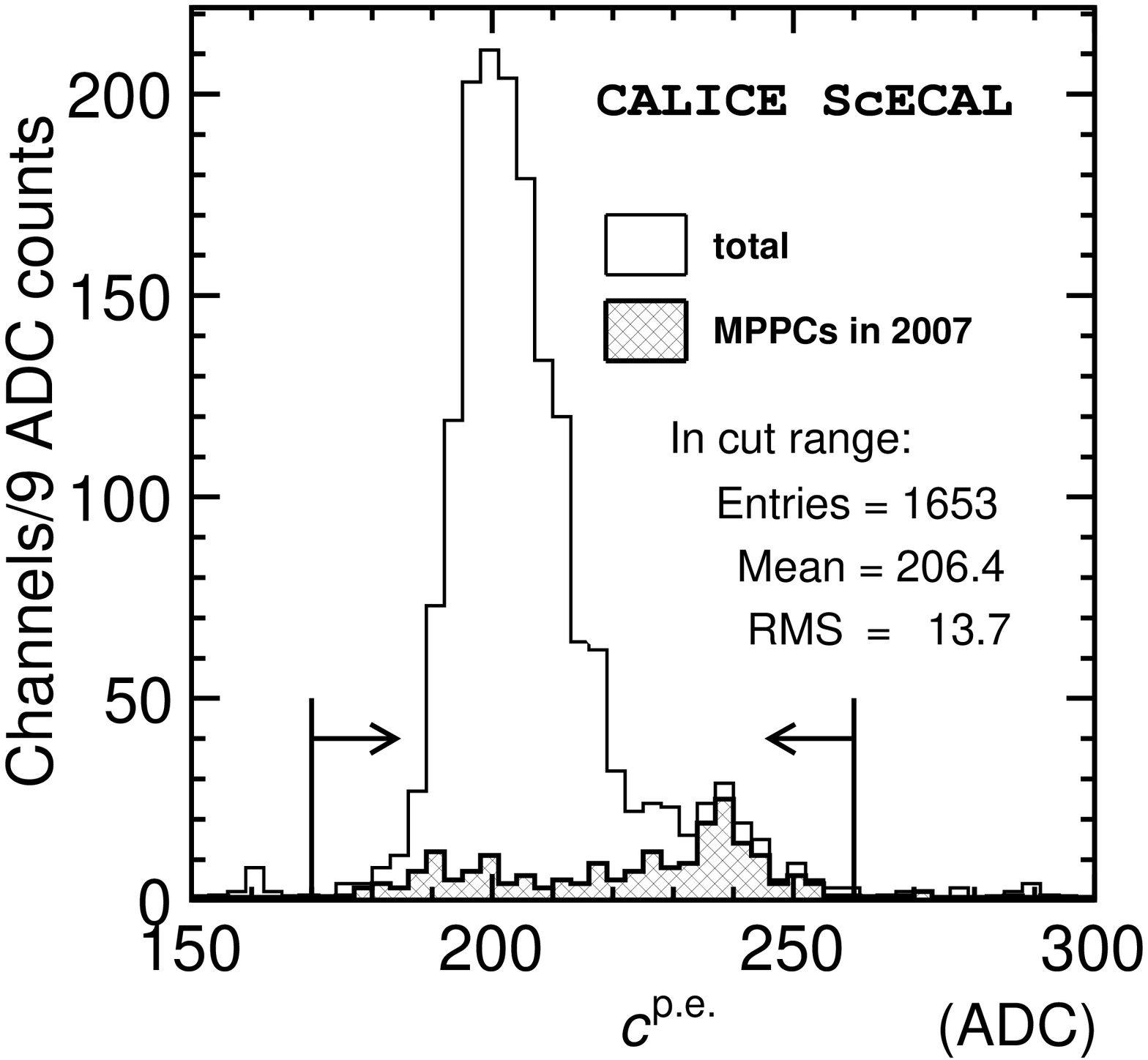}
\includegraphics[width=0.45\textwidth]{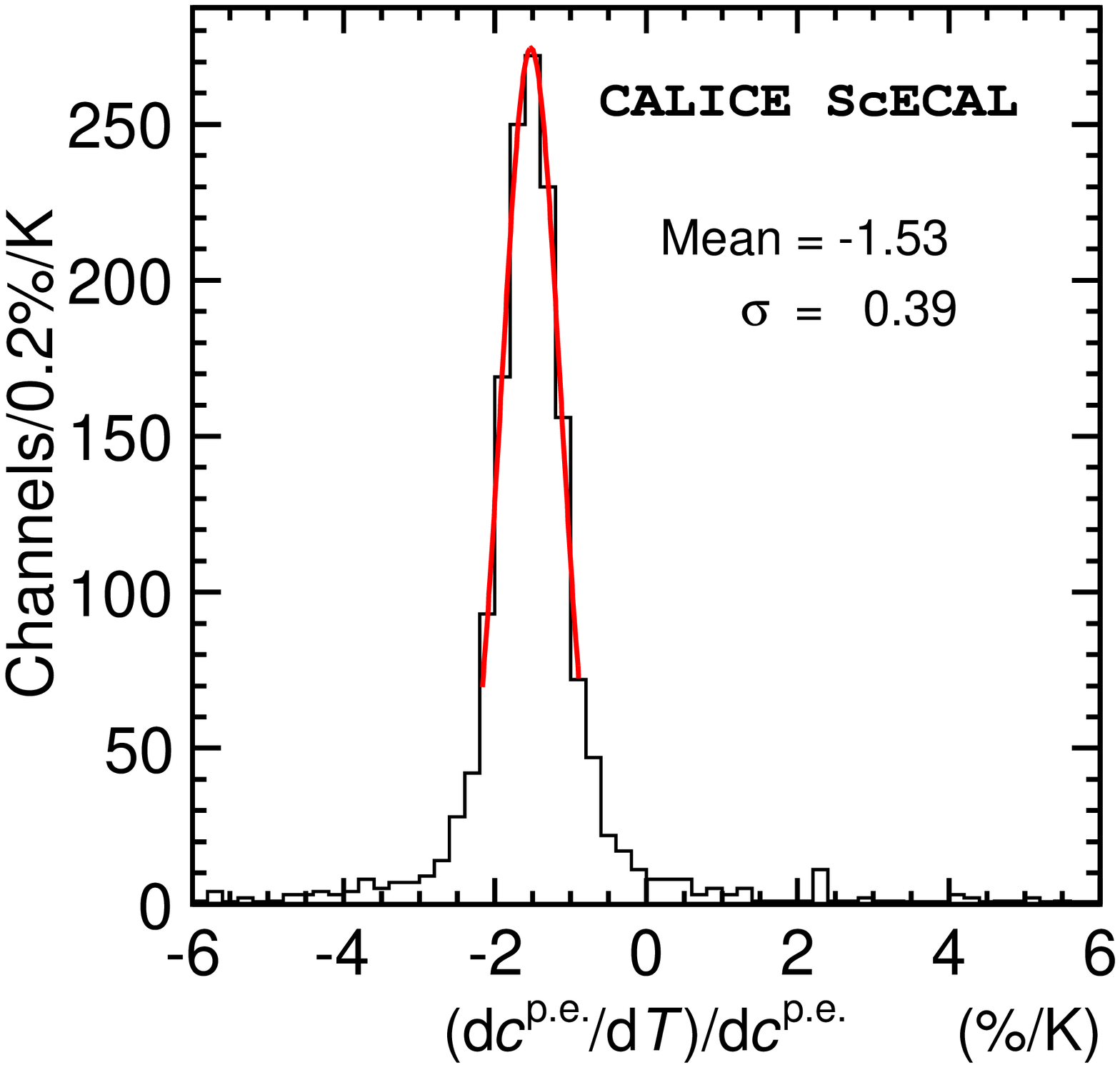}
\caption{\label{fig:adc_photon_distribution} \small{ {{\it left}
      Distribution of $\cphoton(T_0=20^{\circ}$C).  MPPCs produced in
      2007 and 2008 have different characteristics
      (cf.\ Fig.~\ref{fig:breakdown_capacitance}).  {\it right}
      Distribution of 
      %$(\diff\cphoton/\cphoton)/\diff T$. 
 $(\diff\cphoton/\diff T)/\cphoton$.    
      The curve
      shows the result of a Gaussian fit used to extract the mean %%%value
      and %%%the 
      SD.}}}
\end{center} 
\end{figure}

The LED data were collected in nine runs and the variation in
conditions between these runs allowed the temperature dependence of
\cphoton\ to be determined.
The %ADC--photo-electron factor
 \cphoton\ was parametrised in the same
way as \cmip, assuming a linear dependence with temperature.
%%Nige Watson 20161031 NKW: no reference to the factors elsewhere, so can simplify by not
%%Nige Watson 20161031 including this equation.
%%Nige Watson 20161031 \begin{eqnarray} \label{eq:adcPhotonConvFactor} 
%%Nige Watson 20161031 \cphoton(T) = c_{ls}^{\mathrm{p.e.}}(T_0) + \frac{\diff c_{ls}^{\mathrm{p.e.}} }{\diff T} (T - T_0).
%%Nige Watson 20161031 \end{eqnarray}
% Frank to v04-02; Eighty\% of channels were successfully calibrated using the LED system. 
Approximately 80\% of all channels were calibrated using the LED
system.  In the remaining channels, either the pedestals had two peaks
because of noise in the LED circuit or the peaks of photoelectrons were
not sufficiently distinct as to be separable. The majority of these
were concentrated within a few rows of channels, mostly located
on the first or fourth (outermost) rows of layers.  The two-peak
pedestals were not observed during physics runs.
Figure~\ref{fig:adc_photon_distribution} shows the distributions of
\cphoton\ (at $20^{\circ}\mathrm{C})$ and
$(\diff\cphoton/\cphoton)/\diff T$ for completeness.
The temperature dependence of \cphoton\ is only
affected by gain variations, whereas the dependence of \cmip\ on
temperature includes contributions from both variations in gain and
variations in the photon detection efficiency.
%%%%%%%%%%%%%%%%%%%%%%%%%%%%%%%%%%%%%%%%%%%
%%         Fig  adc_photon _distribution               %%%%%%%%%%%%%%%%%%%%
%%%%%%%%%%%%%%%%%%%%%%%%%%%%%%%%%%%%%%%%%%%
%%% he:/home/coterra/FNALpaper/Gain_dist_slope
%%% he:/home/coterra/Gain_dist_slope

%%% replaced 
%%Nige Watson 20161203 \if0 These $c^{\mathrm{p.e.}}$ factors were used to apply the MPPC
%%Nige Watson 20161203 saturation correction to electron data.  For channels in which
%%Nige Watson 20161203 $c^{\mathrm{p.e.}}$ was successfully extracted, besides in which $170
%%Nige Watson 20161203 < c^{\mathrm{p.e.}}(T_0=20^{\circ}\mathrm{C}) < 260$\,ADC
%%Nige Watson 20161203 counts/photo-electron and its fit uncertainty between 0.2 and 50 ADC
%%Nige Watson 20161203 counts/photon-electron, channel-by-channel values of
%%Nige Watson 20161203 $c^{\mathrm{p.e.}}$ were used.  For channels that did not satisfy
%%Nige Watson 20161203 these requirements, the average of successful channels'
%%Nige Watson 20161203 $c^{\mathrm{p.e.}}(T_0=20^{\circ}\mathrm{C})$\,--\,77\% of
%%Nige Watson 20161203 all\,--\,was used. A uniform temperature dependence, given by the
%%Nige Watson 20161203 fitted mean of the $\diff c^{\mathrm{p.e.}}/\diff T$ distribution, was
%%Nige Watson 20161203 assumed for all channels.  \fi

Where they were available, these $\cphoton(T)$ values were used to apply the correction of MPPC non-linear response
to electron data at temperature $T$, channel-by-channel.
%%Nige Watson 20161031 Where fits were successful,Regarding $\cphoton(T_0 =
%%Nige Watson 20161031 20^{\circ}$C), one of two parameters, channel-by-channel values were
%%Nige Watson 20161031 used if fitting was succeeded.
The following criteria were also required: $170 < \cphoton(T_0 =
20^{\circ}\mathrm{C}) < 260$ ADC counts/photoelectron; the
corresponding fit uncertainty between 0.2 and 50 ADC
counts/photoelectron.  For channels where successful fits were not
obtained, the average value of successfully fitted channels was used:
77\% of all channels have individual \cphoton$(T_0 = 20^{\circ}$C). 
 A single value for  $(\diff\cphoton/\diff T)/\cphoton$,
  taken from the
mean of the Gaussian fit as shown in
Fig.~\ref{fig:adc_photon_distribution}, is used for all channels.

\subsubsection{%\cinter}% 
Inter-calibration constant}
\label{interCal}

\begin{figure}[tbhp]		
\begin{center}\includegraphics[width=0.45\textwidth]{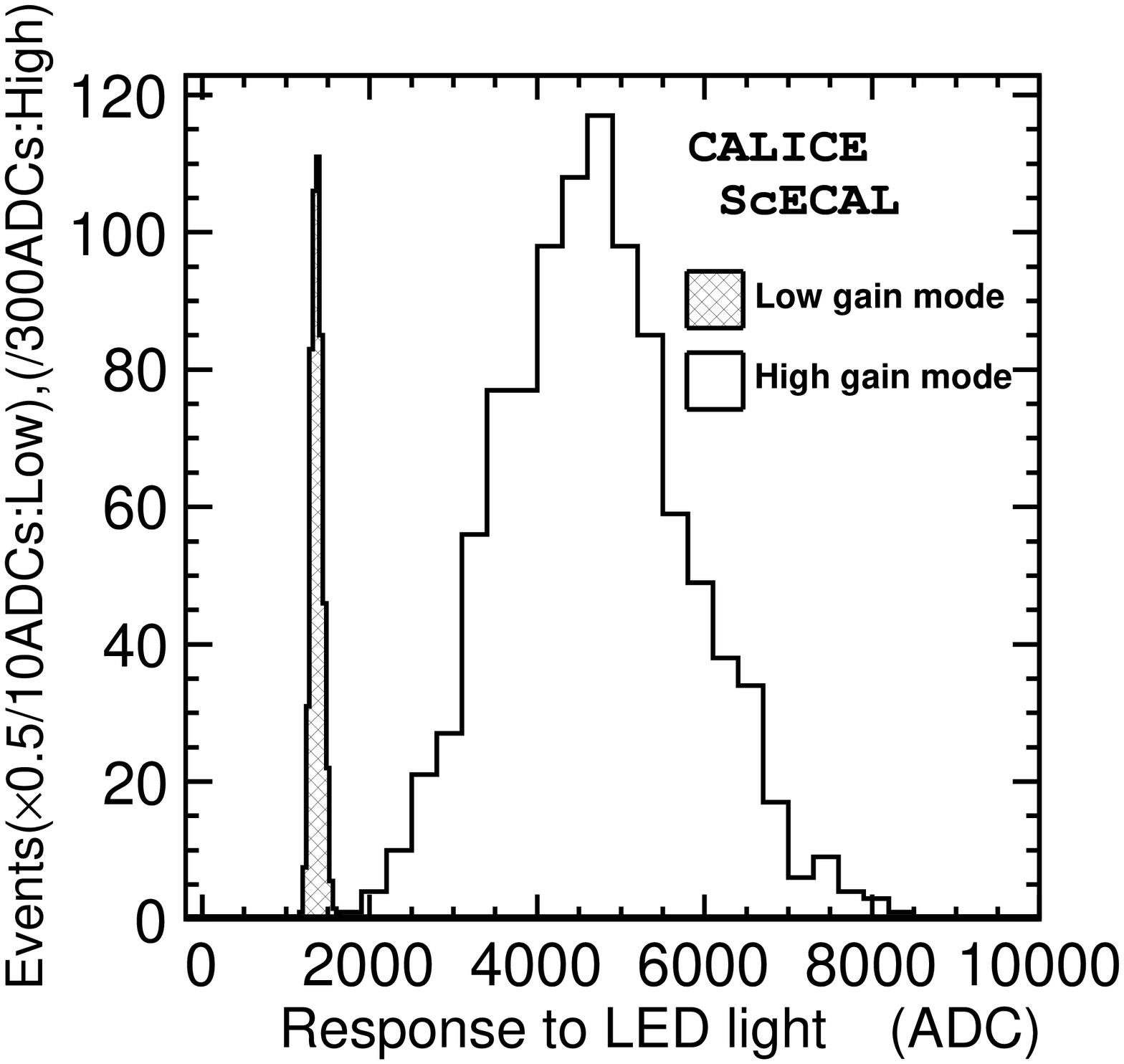}
\includegraphics[width=0.45\textwidth]{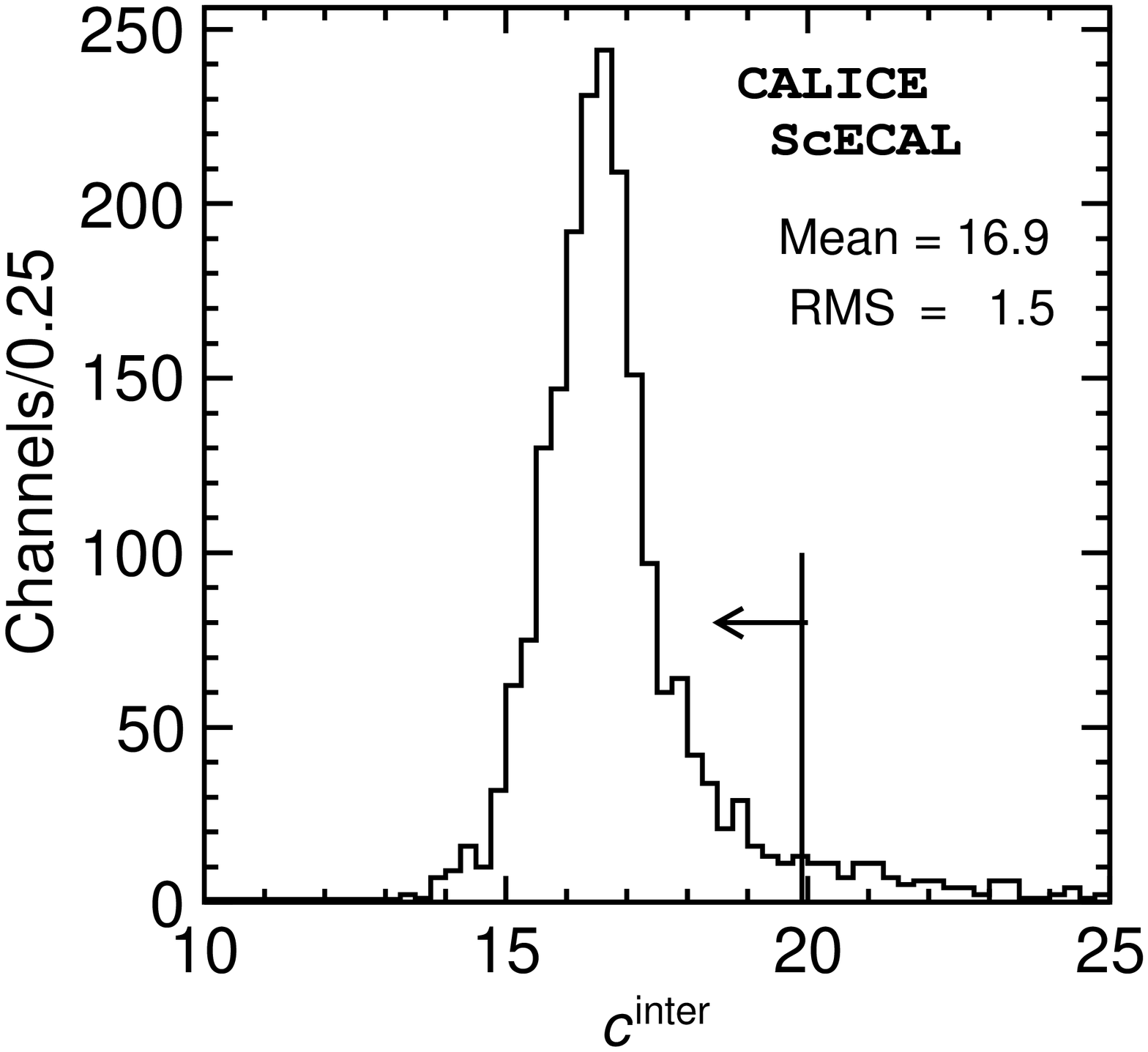}							
\caption{\label{fig:intercalib_signals} \small{{ \it left}: 
Distributions of the MPPC response to LED light in an inter-calibration run
before pedestal subtraction, using the high-gain mode (open)
and the low-gain mode (hatched).  The same LED power was supplied in both cases. \ {\it right}: 
Distribution of the inter-calibration constant \cinter.
%after pedestal subtraction.
The arrow shows the boundary above which an average value of
\cinter\ is used, to avoid excessively large
\cinter\ values, see text for details.}}
\end{center} 
\end{figure}

The dedicated inter-calibration runs used LED light of higher
intensity, which could be measured in both high-gain and 
low-gain modes of the ASIC.
Figure~\ref{fig:intercalib_signals} {\it left} compares MPPC response
in the two operating modes for the same LED power.
%
%The inter-calibration constant
A \cinter\  
for each channel was %%% 20170511 David: measured 
determined 
as
\begin{eqnarray} \label{eq:intercalib} 
\cinter = \frac{\langle\ADChigh\rangle}{\langle\ADClow\rangle} \ ,
\end{eqnarray}
where \ADChigh\ and \ADClow\ are the signal amplitudes in the
high-gain and low-gain modes of preamplifiers, respectively.
%%%%%%%%%%%%%%%%%%%%%%%%%%%%%%%%%%%%%%%%%%%
%%         Fig   fig:intercalib_signals              %%%%%%%%%%%%%%%%%%%%
%%%%%%%%%%%%%%%%%%%%%%%%%%%%%%%%%%%%%%%%%%%
%%% he:/home/coterra/InterCalib
%%% he:/home/coterra/InterCalib
%%%%%%%%%%%%%%%%%%%%%%% Important %%%%%%%%%%%%%%%
%%Nige Watson 20161203 This figure seems confusing if you do not emphasise that the right
%%Nige Watson 20161203 plot is the ratio *after* pedestal subtraction, otherwise the
%%Nige Watson 20161203 approx. value of the ratio of high/low gain modes is ~3 not ~17.
%%Nige Watson 20161203 Also, need to check consistent use of high gain/low gain <=>
%%Nige Watson 20161203 calibration/physics modes.
%%%%%%%%%%%%%%%%%%%%%%% Important %%%%%%%%%%%%%%%

%%Nige Watson 20161203 Just mentioned above, so do not repeat again.
%%Nige Watson 20161203 As discussed in Section~\ref{section:adcphoton}, problems of noise induced by the LED system
%%Nige Watson 20161203 decreased the measurable channels of this ratio to 90\% of all channels.
%%%%%%%%%%%%%%%%%%%%%%%%
%%%%%%%%%%%%%%%%%%%%%%%%%%
\textred{As shown in Fig.~\ref{fig:intercalib_signals} {\it right}, the
distribution of \cinter\  has a tail 
extending to large values. }
\textmagenta{
These channels have unexpectedly small  \ADChigh\ and \ADClow\  values
due to insufficient light supplied by the LED even at its highest power setting.}
\textblue{
Most of these channels were located at the far end of
the fibres distributing the LED light.
An additional contribution to large \cinter\ values in such cases is a possible downward pedestal shift
during LED runs due to large} \textmagenta{power consumption of other highly illuminated channels.}
\textred{
The impact of this effect should %drastically become strong
be more pronounced 
 for small values of \ADClow.
%We 
Having confirmed 
\footnote{\textmagenta{
Exchanging the electronics of the DAQ for those channels with normal channels confirmed that
the large \cinter\ was not due to a genuine change of  \ADChigh\ / \ADClow.
%Note that \cinter\ is a parameter belonging to an electronics function.
This behavior on those channels was uncorrelated with \cphoton. %which was measured with 
% lower LED power  and we know the real pedestal position for each channel as represented in Fig.\,\ref{fig:adc_photon}.
The \cmip\ on those channels %%also has uncorrelated behaviour with the \cinter.}}
shows no correlation with \cinter.}}
that the large \cinter\ does not represent real \cinter}, 
\textred{%As a solution from those fact, 
 we replaced the \cinter\ value of
 all channels that are more than %2\,$\sigma$ 
 2\,RMS
above the mean of the
 entire distribution by the mean calculated using only channels
 that are below this boundary.
 The sensitivity of the energy resolution to the choice of this
boundary is taken into account as a potential systematic uncertainty.}
\subsection{Reconstruction  of electron energy spectra}\label{electronSpectra}
%%%%%%%%%%%%%%%%%%%%%%\subsubsection{Reconstruction}

The energy $E$ of events recorded during electron runs were
reconstructed according to Equations~\ref{eq:unfold} and
\ref{signalMIP},
%%Nige Watson 20161222 with $F$ replaced by $F'$ of \Npixeff,
as discussed in Section~\ref{section:calibrationProcedure}.
In this reconstruction procedure, signals that are \textred{less than three \textmagenta{RMS} % sigma of pedestal width
\textblue{ above}} the mean were rejected in both data and the simulation. 
%
%%Nige Watson 20161203 \if 0
%%Nige Watson 20161203 The energy $E$ of events collected during electron runs was reconstructed as:
%%Nige Watson 20161203 \begin{eqnarray} \label{eq:intercalib} 
%%Nige Watson 20161203 	E = \sum_{\mathrm{all strips}}  A_{i}^{\mathrm{corr.MIPs}},
%%Nige Watson 20161203 \end{eqnarray}
%%Nige Watson 20161203 where $A_{i}^{\mathrm{corr.MIPs}}$ is the temperature- and saturation-corrected MPPC signal converted into MIP units:
%
%%Nige Watson 20161203 \begin{eqnarray} \label{eq:intercalib} 
%%Nige Watson 20161203 	 A_{i}^{\mathrm{corr.MIPs}} = F^{-1}\Big\{ A_{i}^{\mathrm{ADC}}(T)\ \frac{c_{i}^{\mathrm{inter}}}{c_{i}^{\mathrm{p.e.}}(T)}\Big\}\ 
%%Nige Watson 20161203 	 						 \frac{c_{i}^{\mathrm{p.e.}}(T)}{ c_{i}^{\mathrm{inter}} }\ c_{i}^{\mathrm{MIP}}(T),
%%Nige Watson 20161203 \end{eqnarray}
%%Nige Watson 20161203 where the $F^{-1}$ is the inverse of the function defined in Equation~\ref{eq:saturation}, 
%%Nige Watson 20161203 $A_{i}^{\mathrm{ADC}}(T)$ is a strip's uncorrected signal in ADC counts,
%%Nige Watson 20161203 and $T$ is the temperature of the detector at which the event was collected.
%%Nige Watson 20161203 Note that \AcorrMIPS\ has been corrected for the temperature dependence of both the MIP and single photo-electron response.
%%Nige Watson 20161203 \fi
%%%%%%%%%%%%%%%%\subsubsection{The energy spectrum after event selections}\label{section:6cuts}
Events recorded in electron runs that are triggered by the
\v{C}erenkov counter still include contamination from pions, muons,
and multi-particle events. To further enhance the purity of the
electron sample and to select events that are contained within the
fiducial volume of the ScECAL prototype, the following selection
criteria %are required:
were applied:
\begin{enumerate}
\item the layer with maximum reconstructed energy must be within the first 20 ScECAL layers;
\item the reconstructed energy in this layer must exceed a beam energy
  dependent threshold, as given in  Table~\ref{tab:cutoffs};
%%Nige Watson 20161203 NKW: Moved numerical values to table for easier reading.
%%Nige Watson 20161203 10 MIPs for 1 GeV/$c$;
%%Nige Watson 20161203 15 MIPs for 2 GeV/$c$; \\
%%Nige Watson 20161203 27 MIPs for 4 GeV/$c$; \\
%%Nige Watson 20161203 54 MIPs for 8 GeV/$c$; \\
%%Nige Watson 20161203 80 MIPs for 12 GeV/$c$; \\
%%Nige Watson 20161203 95 MIPs for 15 GeV/$c$; \\
%%Nige Watson 20161203 125 MIPs for 20 GeV/$c$; \\
%%Nige Watson 20161203 200 MIPs for 30 GeV/$c$; and \\
%%Nige Watson 20161203 200 MIPs for 32 GeV/$c$;
\item the reconstructed energy of the highest energy AHCAL layer must
  be less than 20 MIPs;
\item the reconstructed energy in the most downstream layer of the AHCAL must be less than 0.4 MIPs; 
\item\label{item:xcut} the energy-weighted mean position---measured using only $x$ layers---of ScECAL hits must be
    within 40\,mm of the detector centre in the $x$ direction;
\item equivalent of criterion~\ref{item:xcut} in $y$; and 
    %%Nige Watson 20161203 NKW: perhaps easier to not disturb item numbering?
  %%Nige Watson 20161203 \item the energy-weighted mean position of ScECAL hits must be within 4~cm of the detector centre in the $x$ ($y$) direction; and  \setcounter{enumi}{6}
\item %the multiplicity counter output must be less than an energy
  %equivalent to 
  the multiplicity counter signal should correspond to less than
  %$\sim$
   1.4 MIP \cite{NilsFeege},
\end{enumerate}
%%%% added according to Lei's suggetsion%%%%%
where the first three criteria reduce contamination from both pions
and  muons, the fourth further reduces that of muons,
the fifth and sixth define the lateral fiducial area and the seventh reduces the
selection of multi-particle events.
\begin{table}[tbhp]
\captionsetup{width=0.7\textwidth}	
\begin{center}
\caption{\label{tab:cutoffs}\small \textred{%Requirements of energy thresholds for layer with maximum reconstructed energy
\textblue{Energy thresholds required for the layer with maximum reconstructed energy.}
}}
 \begin{tabular}{cc}
\hline \hline
%%NKW: \vspace{-2mm}&\\
Beam energy & Minimum energy \\
 $[$\gev$]$    & reconstructed [MIP]([GeV])\\ \hline
  % 1   & 10 \\
   2   & 15 (0.12) \\
   4   & 27 (0.21)\\
   8   & 54 (0.42)\\
   12  & 80 (0.62)\\
   15  & 95 (0.73)\\
   20  & 125 (0.96)\\
   30  & 200 (1.54)\\
   32  & 200 (1.54)\\
%%%Beam energy & Minimum energy \\  NIMA rev 1
%%% $[$\gev$]$    & reconstructed [MIP]\\ \hline
  % 1   & 10 \\
 %%%  2   & 15 \\
 %%%  4   & 27 \\
  %%% 8   & 54 \\
 %%%  12  & 80 \\
%%%   15  & 95 \\
%%%   20  & 125 \\
%%%   30  & 200 \\
%%%   32  & 200 \\
\hline
\end{tabular}
\end{center}
\end{table}

%%%%%%%%%%%%%%%%%%%%%%%%%%%%%%%%%%%%%%%%%%%
%%         Fig   spectrum32GeV_2              %%%%%%%%%%%%%%%%%%%%
%%%%%%%%%%%%%%%%%%%%%%%%%%%%%%%%%%%%%%%%%%%
%%% he:/home/coterra/Fnal2009Anaf830start/2009MultiEventsCut_intcal2sigAverage/02GeV/All10x10_C2_MultA3800_20GeV4runs
%%% he:/home/coterra/Fnal2009Anaf830start/2009MultiEventsCut_intcal2sigAverage/32GeV/All10x10_C2_MultA3800_20GeV4runs
\begin{figure}[btph]							
\begin{center}\hspace{-3mm}\includegraphics[width=1.05\textwidth]{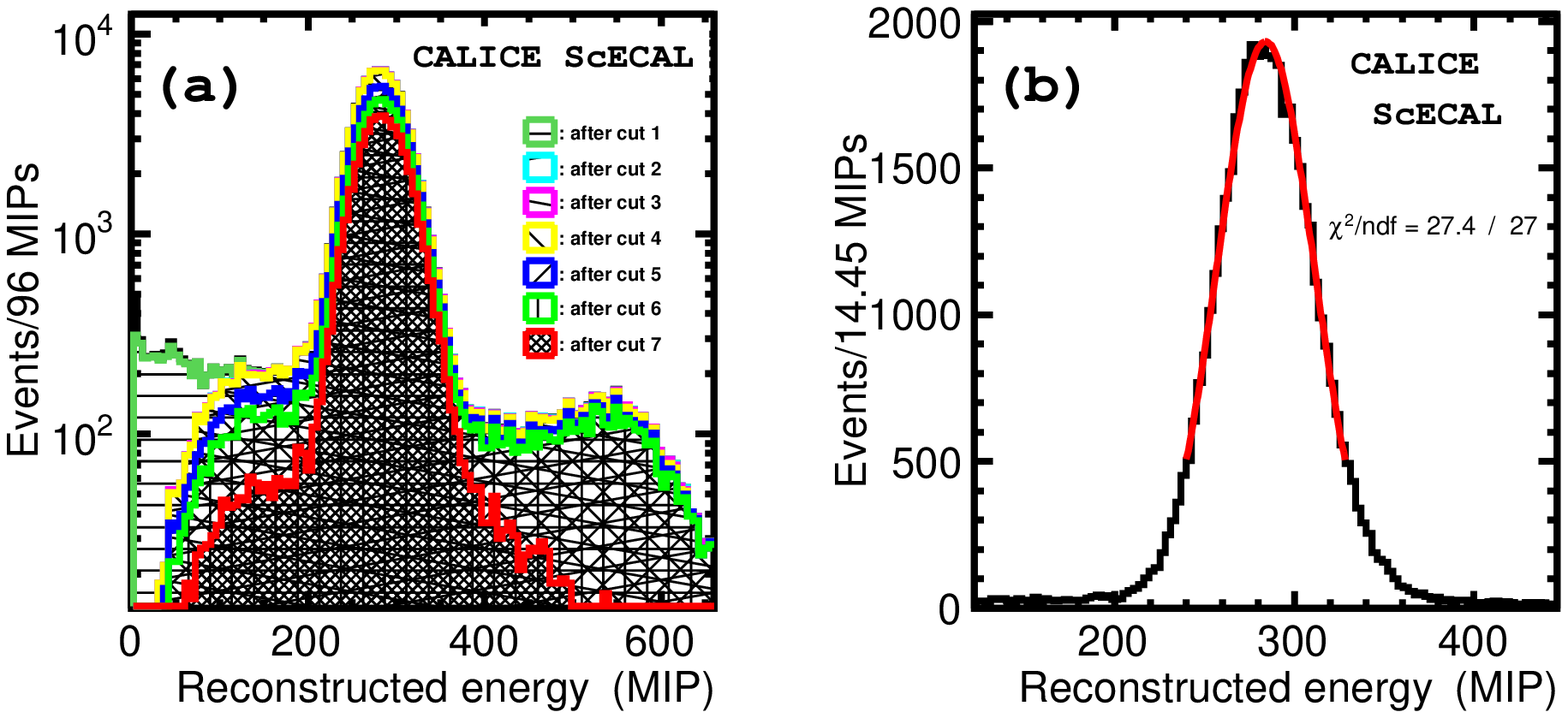}	
		\hspace{-3mm}\includegraphics[width=1.05\textwidth]{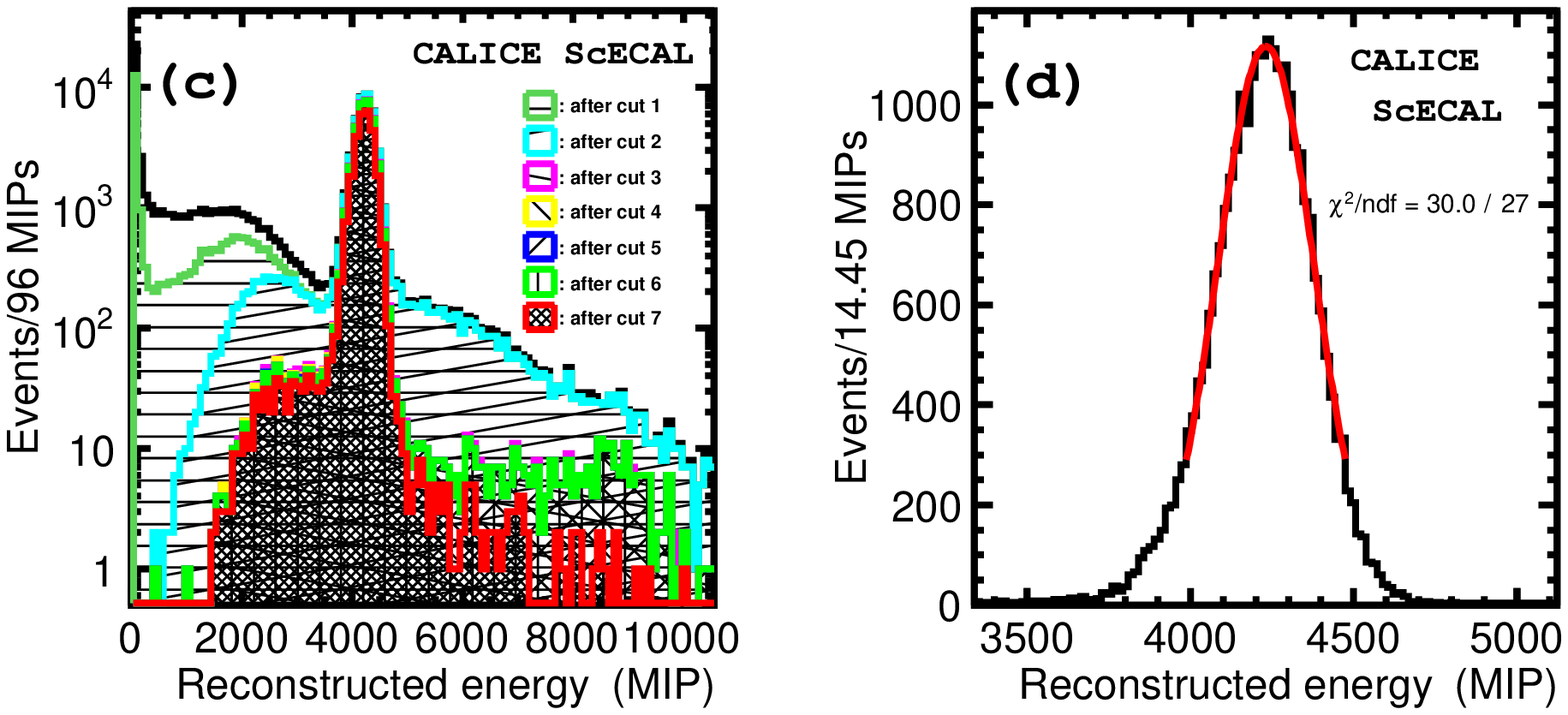}
\caption{\label{fig:spectrum32GeV_2} \small{
%\textcolor{red}{
Energy spectra of events collected in a 2\gev\ electron run ({\it top}: a, b), and 32\gev\ electron run  ({\it bottom}: c, d).
{\it Left}: (a, c) show the effects of the sequential application of
selection criteria 1--7, see text for details.
{\it Right}: (b, d) show the spectrum after all cuts. 
A solid curve in (b) and (d) is the
result of a Gaussian fit in a range that contains 90\% of selected events.
%}
}}
\end{center} 
\end{figure}

Figure~\ref{fig:spectrum32GeV_2} shows the reconstructed energy
spectrum of events recorded in 
a single 2\gev electron run and a single 32\gev electron run
after the sequential application of these selection cuts, and the energy
spectrum after all cuts.  These criteria remove almost all
two-particle events, which are identified as having higher energies
than the main electron peak.  
%The residual contamination from
%particles other than electrons, associated with energies below the
%main peak, is small.
%\textcolor{red}{ 
The residual contamination form particles other than electrons, 
associated with energies below the main peak, is less than between 0.1\% (32\gev) and 1\% (2\gev). 
%}
%%%The final spectrum is described well by a
%%%Gaussian function in a range\textred{
%%%---$\pm\,1.65\,\sigma$---}
%%%that contains 90\% of all entries.
The final spectrum is described well by a
Gaussian function in a range of $\pm\,1.65\,\sigma$.
%%Nige Watson 20161203 NKW: This is the 90%/1.64 sigma, not specific to this fit? %% (1.65\,$\sigma$).
The reduced \chisq\ of the fits to the spectra collected at all
energies were between 0.9 and 1.2.
%
%%%The energy response and resolution were defined as the mean of the
%%%fitted Gaussian and the ratio of the Gaussian width to its mean,
%%%respectively.
%
The mean reconstructed energy, \Ereco, and resolution, \sigmaEreco, were obtained from 
the mean and width of the Gaussian function fitted to the reconstructed energy spectra.
The relative resolution is calculated as the ratio \sigmaEreco/\Ereco.
%%%% NIMA rev-1
\textredSecond{\textblueSecond{The} systematic uncertainty \textblueSecond{originating}  from the  \textblueSecond{restricted} fitting range is discussed in Section\,\ref{section:systematics}.}
%%
%
% merge the following Data from ... into the previous paragraph following Jerry's suggestion.
%%Nige Watson 20161024 Do not think this needs to be written out in a paper.
%%% Frank to v04-02 in\subsection{Combining the reconstructed energies means among runs}
%%Nige Watson 20161024 \subsection{Combining the reconstructed mean energies among runs}
%%%% Frank to v04-02 Means and the SDs, resolutions, of energy spectra among runs at the same nominal beam momentum 
\textred{%Data from different runs that have 
\textblue{Data from runs with} 
the same nominal beam
energy were combined, %%% 20170511 Nigel with weights according to their statistical uncertainties alone.}
weighted by their statistical uncertainties.}
%%Nige Watson 20161203 were combined by the the error-weighted mean method of the single run results:
%%Nige Watson 20161203 \begin{eqnarray} \label{eq:intercalib} 
%%Nige Watson 20161203 	\overline{x} \pm \delta\overline{x} = \frac{\sum_i w_i x_i}{\sum_i w_i} \pm (\sum_i w_i)^{-1/2},
%%Nige Watson 20161203 \end{eqnarray}
%%Nige Watson 20161203 where $x_i$ is the mean response or resolution of a single run and $w_i$ is $1/(\delta x_i)^2$.
%%Nige Watson 20161203 %

\section{Performance of the prototype}\label{section:results}

\subsection{Mean and resolution with statistical uncertainties}
\label{section:meanResolutionWstatUncertainties}
\begin{figure}[hbt]
\captionsetup{width=0.7\textwidth}	
\begin{center}\includegraphics[width=0.5\textwidth]{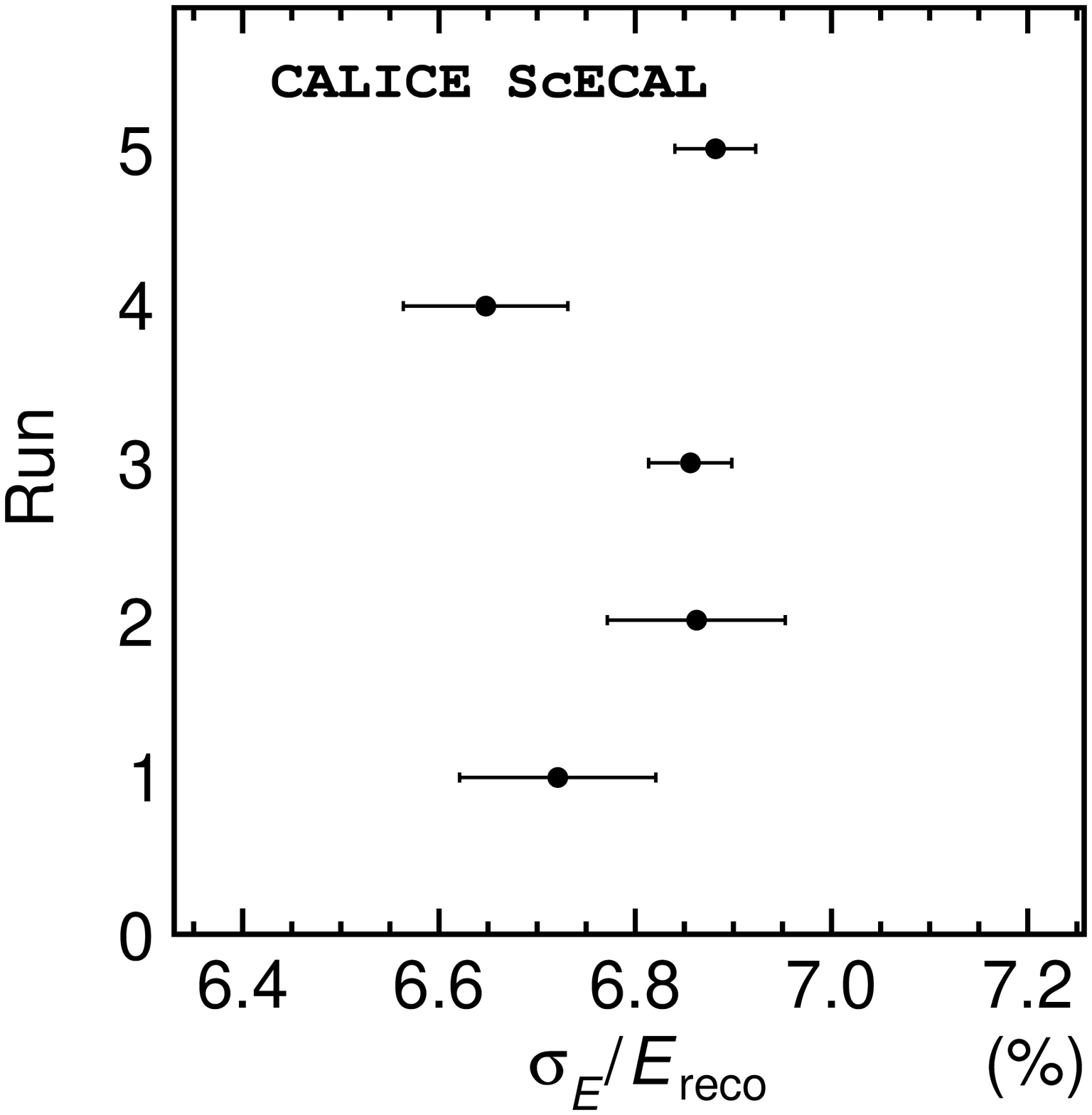}									
\caption{\label{fig:runDiff4GeV} \small{
The energy resolution obtained from the five electron runs collected
at a beam energy of 4\gev.
The uncertainties show only statistical uncertainties.
}}
\end{center} 
\end{figure}

Table~\ref{table:meanResolution_eachBean} summarises the mean energy
response and resolution for each beam energy, together with their
statistical uncertainties.
%The systematic uncertainties of these figures are discussed in the next section. 
%% removed following Jerry's suggestion The linearity of the response and the parameterization of the energy
%%resolution are discussed in Section~\ref{section:results_linearity_resolution}.
%
%The variations in the energy resolution measured in different runs at the same nominal energy are consistent within their uncertainties, 
%as demonstrated in
%
Figure~\ref{fig:runDiff4GeV} shows the energy resolutions of the five runs collected at 4\gev.
% Frank: g923 The variations measured in different runs at the same nominal energy are consistent within their statistical uncertainties as well as those observed at other beam momenta.
%% To answer the reviewr 1. 
%% The variations measured in different runs at the same nominal energy
%%are consistent within their statistical %uncertainties as well as with those observed at other beam energies.
%%uncertainties, as are those at other beam energies.
%%%%%%%\textredSecond{The variations measured in different runs at the same nominal energy are not \textblueSecond{significantly} larger %%%%%%%than their statistical uncertainties: 
%%%%%%%the average of the $\chi^2$ probabilities of all energies is 0.13.}
\textredSecond{The variations measured in different runs at the same nominal energy are all smaller than the uncertainty of the beam energy spread which is discussed in the following subsection.}
\begin{table}[bth]
\captionsetup{width=0.65\textwidth}	
\begin{center}
%\textcolor{red}{
\caption{\label{table:meanResolution_eachBean}\small %Mean response and energy resolution at each beam energy.
Mean reconstructed energy and relative resolution for the combined data sets.
The resolution includes the intrinsic energy spread of the beam.
%%Nige Watson 20161222   , which is discussed in term 6 of the following next
%%Nige Watson 20161222 section.%%\,\ref{section:systematics|}. 
The uncertainties are statistical only.}
%}
\begin{tabular}{ccc}
\hline \hline
\vspace{-2mm}&&\\
Beam energy [\gev]&\Ereco[MIP]&\sigmaEreco/\Ereco%$^{\dag}$ 
(\%)\\ 
\hline %\aline
2   &\ \,281.53$\pm$0.08&9.633$\pm$0.035\\
4   &\ \,545.10$\pm$0.12&6.855$\pm$0.026\\
8   &1076.52$\pm$0.14&5.049$\pm$0.015 \\
12  &1588.43$\pm$0.22&4.388$\pm$0.016\\
15  &1966.31$\pm$0.23&4.222$\pm$0.014\\
20  &2589.30$\pm$0.29&3.791$\pm$0.013\\
30  &3910.4\ \,$\pm$0.6\ \,&3.445$\pm$0.017\\
32  &4201.5\ \,$\pm$0.7\ \,&3.425$\pm$0.020\\
\hline
\end{tabular}
\end{center}
\end{table}
%
%
%%%%%%%%%%%%%%%%%%%%%%%%%%%%%%%%%%%%%%%%%%%
%%         Fig   runDiff4GeV              %%%%%%%%%%%%%%%%%%%%
%%%%%%%%%%%%%%%%%%%%%%%%%%%%%%%%%%%%%%%%%%%

\begin{figure}[t]		
\captionsetup{width=0.75\textwidth}								
\begin{center}\includegraphics[width=0.6\textwidth]{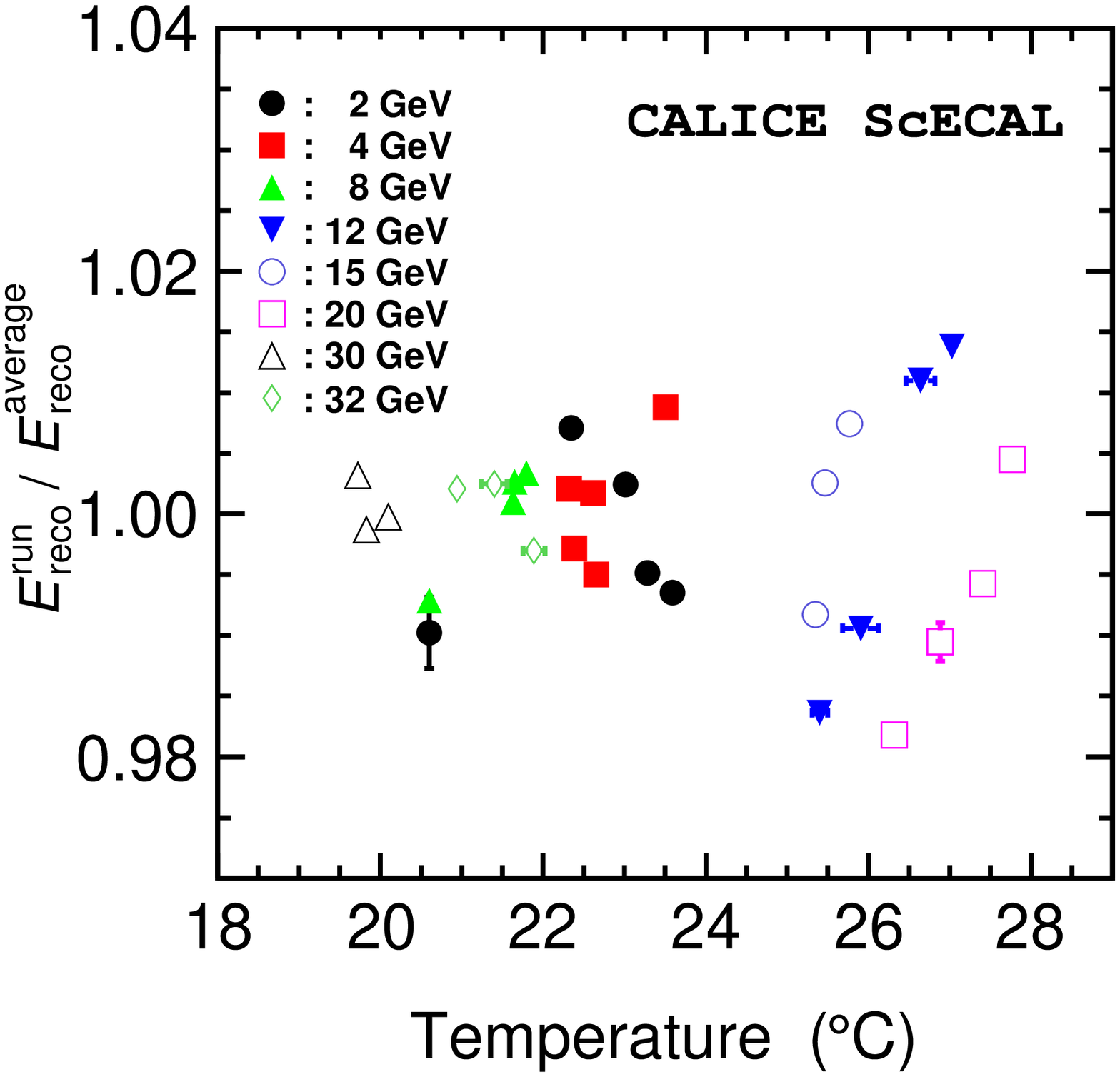}
\caption{\label{meanDepOnTemp} \small{
{\textblueSecond{
Ratio of the
      reconstructed mean of individual runs to their average
      \vs\ temperature during data taking. }%% 20170511 David  at that time.
}}}
\end{center} 
\end{figure}

In contrast, the \textred{mean} reconstructed energies measured in other runs show
variations that are beyond what is expected from their statistical
uncertainties, as seen in Fig.~\ref{meanDepOnTemp}.
% shows such a plot indicating non correlation between means of reconstructed energy  and the temperature except 8\,GeV/$c$, 12\,GeV/$c$, and 20\,GeV/$c$. 
%%% Frank g923 Although, the imperfect of over corrections on
%%%temperature variation considered, Figure~\ref{meanDepOnTemp} shows
%%%non correlation between means of reconstructed energy and the
%%%temperature except 12\,GeV/$c$ and 20\,GeV/$c$.
Imperfections in the correction for temperature variation were
considered as a possible explanation for this difference.
However, Fig.~\ref{meanDepOnTemp} shows that the correlation
between 
the reconstructed energy of individual \textred{runs} 
%the mean reconstructed energy 
\textred{and the temperature is only
%%% 20170511 Nigel clear 
apparent 
for runs taken at 8\gev, 12\gev, and 20\gev.}
The following subsections discuss investigations into potential
sources of systematic effects that may account for these differences.
%%%%%%%%%%%%%%%%%%%%%%%%%%%%%%%%%%%%%%%%%%%
%%         Fig   runDiff: temperrature vs mean (1)             %%%%%%%%%%%%%%%
%%%%%%%%%%%%%%%%%%%%%%%%%%%%%%%%%%%%%%%%%%%
%%% he:/home/coterra/FNALpaper/temper_mean_20GeV4runs

\subsection{Systematic uncertainties}
\label{section:systematics}
%The systematic uncertainties on the mans and the resolutions of the measured energy in Table\,\ref{table:meanResolution_eachBean} are estimated in this section,
%regarding the difference of the measured energy among runs, comes from the event selection criteria, calibration constants, MPPC saturation corrections, and beam momentum 
%fluctuation.
%%% According to Frank's suggestion:

%%%The systematic uncertainties estimated on the mean and the resolutions of the measured energy in Table 2 are discussed in this subsection.
 We consider sources of potential systematic uncertainty from the
 event 
 selection
 criteria, the calibration factors/constants, correction of the MPPC
 non-linear response and the beam energy spread.
%and  difference of the measured energy among runs.
 Table~\ref{table:summaryUncertaintyIndividual} lists 
 the contributions from different sources to the overall uncertainty for each beam energy.
%%%%%%%%%%%%%%%%%%%%%%%%%%%%%%%%%%%%%%%%%%%%%%%%%%
%%%%%%%%%%%%%%%%%%%%%%%%%%% EVENT SELECTIONS
\begin{description}
%%\subsubsection{Event selection}
%%Nige Watson 20161222 \item {\bf Event selection}\\
\item[Event selection]~

As discussed in Section~\ref{electronSpectra},
  seven cuts were used to select well-contained electron events.  To
  estimate the systematic uncertainties associated with these
  selection criteria, the impact of varying the applied cuts was
  evaluated.
%The resulting of change by each criterion in the mean energy response were less than 0.3 \%, except in the case of the requirement on 
The largest contribution to the energy mean arises from the fiducial
volume cut in the $x$ direction, due to the larger beam spread
in $x$ than in $y$. %%% Marina requires Table~\ref{table:summaryUncertaintyIndividual}
%%lists the estimated uncertainties arising from this and other requirements for each beam energy. % variations.
%%%%%%%%%%%%%%%%%
%%%%%%%%%%%%%%%%%% fiducially should be replaced other word???
%fiducial window (cuts 5 and 6). 
%The variation by tightening the fiducial window from 40 to 20~mm was adopted as a systematic uncertainty.
%%%%The effect of these cut variations on the energy resolution was less than 0.05\%.
%negligibly small ($<0.005$).

%%\subsubsection{ADC-MIP conversion factor}\label{section:ADC-MIPsystematic}
%%Nige Watson 20161222 \item {\bf ADC-MIP conversion factor}
\item[ADC-MIP conversion factor]
\label{section:ADC-MIPsystematic}~ 

Systematic uncertainties
  on the ScECAL performance originating from the statistical
  uncertainty in the extraction of \cmip\ were estimated.  The assumed
  values of \cmipcert\ and \cmipslope\ were randomly fluctuated around
  their central values using a Gaussian probability distribution
  function (PDF) with a width equal to the statistical uncertainty of
  these parameters extracted for each channel. The data were
  re-analysed twenty times using such fluctuated calibrations.  The
  absolute value of the systematic uncertainty from \cmipcert
  (\cmipslope ) on the reconstructed energy mean is less than 0.3\%
  (0.06\%) for all beam energies.
%
%%\subsubsection{ADC--photo-electron conversion factor}
%%Nige Watson 20161222 \item {\bf ADC--photo-electron conversion factor}
\item[ADC--photoelectron conversion factor]~

%Regarding the  ADC--photo-electron conversion factors we investigated with the same way for the ADC-MIP conversion factor discussed above.
A similar method is applied to systematic uncertainties originating from
%ADC--photo-electron conversion factors,
 \cphoton, and effects on the
reconstructed %% 20170511 Nigel: energy means 
mean energies 
were found to be negligible.

%%\subsubsection{Inter-calibration constant}\label{section:InterCalibSystematic}
%%Nige Watson 20161222 \item {\bf Inter-calibration constant}
\item[Inter-calibration constant]
\label{section:InterCalibSystematic}~

  Systematic effects
  arising %%% 20170511 David: with 
  from 
  uncertainties in the inter-calibration constants are
  also studied using a pseudo-experiment method.  In the case of
  channels with a successfully measured inter-calibration constant,
  the constant is varied according to a Gaussian PDF, whose width is the
  uncertainty of the inter-calibration constant of the channel under
  consideration.
  In the case of channels where the measurement was
  not successful, the SD of inter-calibration
  constants for all measured channels was used as the width of the Gaussian
  PDF.  On the basis of twenty such pseudo-experiments, changes in performance were
  negligibly small.
As discussed in Section~\ref{interCal}, %all channels having
%\cinter\ greater than 2~$\sigma$ from the mean of
%\cinter\ were replaced with the mean.  
%%% 20170511 Nigel: \cinter\ was replaced with the mean of them for all channels having \cinter\ larger than 2~$\sigma$ from the mean.
the value of \cinter in any channel that was more than 2~$\sigma$ above the mean of all channels, was replaced by the mean value itself.
To investigate the
effect of this procedure, the criterion of the \cinter\
cut was changed from 1~$\sigma$ to 3~$\sigma$ and also for the case of
all measured \cinter.  The relative shifts found in the  mean and 
the resolution of energy with respect to the default case were less
than 0.01\% when changing the criterion from 1 to 3~$\sigma$ for all
energies, and less than 0.1\% when all measured \cinter\
were used.
%%% g928 Therefore, the systematic uncertainties from this procedure do not appear.
Therefore, the systematic uncertainties from this procedure are also
considered %%% 20170511 Nigel: as 
to be 
negligible.

%%\subsubsection{The number of effective pixels of the MPPC}\label{section:systematicFromNpix}
%%Nige Watson 20161222 \item {\bf The number of effective pixels of the  MPPC}
\item[The number of effective pixels of the  MPPC]
  \label{section:systematicFromNpix}~
  
  The number of effective
  MPPC pixels, \Npixeff, was measured in 72
  strips. The mean of these 72 measurements was used when applying
 correction of the MPPC non-linear response to all strips of the prototype.
  Pseudo-experiments in 
  which \Npixeff\ 
  of each strip was varied with a Gaussian PDF were
  performed to study the impact of the uncertainties of this quantity.
  The width of the Gaussian PDF was taken as the SD of
  the 72 measurements.  Effects on calorimeter performance were rather
  small: the absolute value of the systematic uncertainty from the
  uncertainty of the number of effective pixels is less than 0.13\%
  for all beam energies.
%
%\textred{
The 72 MPPC samples are all from the 2008 production.
A  \Npixeff\  of the 2007 production is estimated\footnote{This is after correcting for known differences in the 2007 production due to absence of a photon %%% 20170511 coterra shield
screen  
and use of WLS fibre rather than direct coupling to the MPPC.} to be 2185, which is 
within one SD of %%% 20170511  that obtained for 
of the
2008 products.
We estimated this value using data from the first prototype where all MPPCs were 2007 products \cite{desyTB}. %20170511 ;
%% 20170511 taking from the  \Npixeff\  of the WLS fibre coupling model and correcting the effect from that 
%% 20170511 the first prototype had no screen to avoid photons directly detected via scintillator
%% 20170511 by comparing with the \Npixeff\  of the MPPC-scintillator direct coupling model.
%
Additionally, the 2007 products only represent 13\% of all MPPCs in
the prototype and these are all located in peripheral
regions. Therefore, we ignore the effect of differences between the  2008
and 2007 devices.
%}

%\textred{
\item[\textred{Response dependence on hit position} \textblue{along the strip length}]~%along the length of strip]
  \label{section:position_response}~

\textred{
\textblue{A previous ScECAL prototype using extruded scintillator strips demonstrated a
significant dependence of the response on the hit position along the scintillator strips~\cite{CAN06}. 
This response non-uniformity results in a significant degradation of the energy resolution.
}
%A study with the previous ScECAL prototype clarified that our scintillator by the extrusion method gave an undeniable response dependence on 
%hit position along scintillator strip~\cite{CAN06}. 
%
%This non-uniformity of response makes a large degradation of energy resolution. %}
Applying a screen in front of the MPPC (shown in Fig.~\ref{fig:shade}), together
with higher scintillator quality, has demonstrated significant improvements. % with respect to the previous prototype.
Figure~\ref{fig:uniformity} shows 
\textblue{the MIP response of a channel }
%
%the response of a channel 
as a function of the distance from the MPPC, and the distribution of
the ratio of response at the far end side to the MPPC side for all
channels in the prototype (with the exception of four dead channels).
A response ratio for each channel was determined from the result of a  single
exponential  function: \textblue{
the measured position dependence was fitted by a single exponential function, and a response ratio defined as the ratio of this function at the two strip ends.
}
\if 0
$[a_0\mathrm{exp}\{a_1\times x (=45\,\mathrm{mm})\}]/[a_0\mathrm{exp}\{a_1\times x (=45\,\mathrm{mm})\}]$
}
%
%$\mathrm{exp}\{a_0 + a_1\times x (=
%45\,\mathrm{mm})\}/\mathrm{exp}\{a_0 + a_1\times x (=
%0\,\mathrm{mm})\}$, 
%
where $a_0$ and $a_1$ are fitting parameters.
\fi
\textblue{The mean and RMS of the measured uniformity are  $(88.3\pm4.3)$\%.}
%The mean of measured uniformity with an RMS is $(88.3\pm4.3)$\%.
%characterized by the
 %difference in response for the two ends of a strip, is sufficient for the response linearity and the energy resolution %% g923 for an 
 %required of the ScECAL.
%
This uniformity of the response within each strip has been measured
using muon events by reconstructing the %approximate
 position within a strip using information from layers with different orientation. 
%The distribution of the measured relative differences in response between
%the two ends of a strip has a mean of 11.7\% and a standard deviation
%of 4.2\%.  
%
%The solid headstand triangles  in  Fig.\,\ref{fig:mcresults} {\it right} show the energy resolution in the case where 
%this non-uniformity and fluctuations of $N_{\mathrm{pix}}^{\mathrm{eft}}$ have been implemented in the simulation.
%
\textblue{Simulation studies %A simulation study
 with and without 
 a description of this non-uniformity
 demonstrated that the maximum degradation of the energy resolution due to non-uniformity is
 $\Delta(\sigma_E/E) =+0.04\%$ at 2\,\gev.}
 %
% this non-uniformity according to the distribution in Fig.\,\ref{fig:uniformity} {\it right},
%we found that the maximum shift of the energy resolution is $\Delta(\sigma_E/E) =+0.04\%$.
Details  of the simulation study are given in Section~\ref{section:MC_model}.
\textmagenta{
%We also confirmed that no significant $\sigma_E/E$ exists depending on the beam position 
%originated by this non-uniformity within the uncertainties.
Within  uncertainties, there is no significant change of $\sigma_E/E$ as a function of beam position associated with the non-uniformity.
}
%We also confirmed that no significant $\sigma_E/E$ dependence on the beam position originated by this non-uniformity exists unless we keep
%this uniformity of response by comparing $\sigma_E/E$  for the beam at $(x, y) = $( 0\,mm, 0\,mm) and =  (22.5\,mm, 22.5\,mm).
}
%%%%%%%%%%%%%%%%%%%%%%%%%%%%%%%%%%%%%%%%%%%
%%         Fig  fig:uniformity                %%%%%%%%%%%%%%%%%%%%%%%%%%
%%%%%%%%%%%%%%%%%%%%%%%%%%%%%%%%%%%%%%%%%%%
%%% he:/home/coterra/FNALpape/uniformity
%%% he:/home/coterra/FNALpape/uniformity
\begin{figure}[htbp]									
\begin{center}\includegraphics[width=0.45\textwidth]{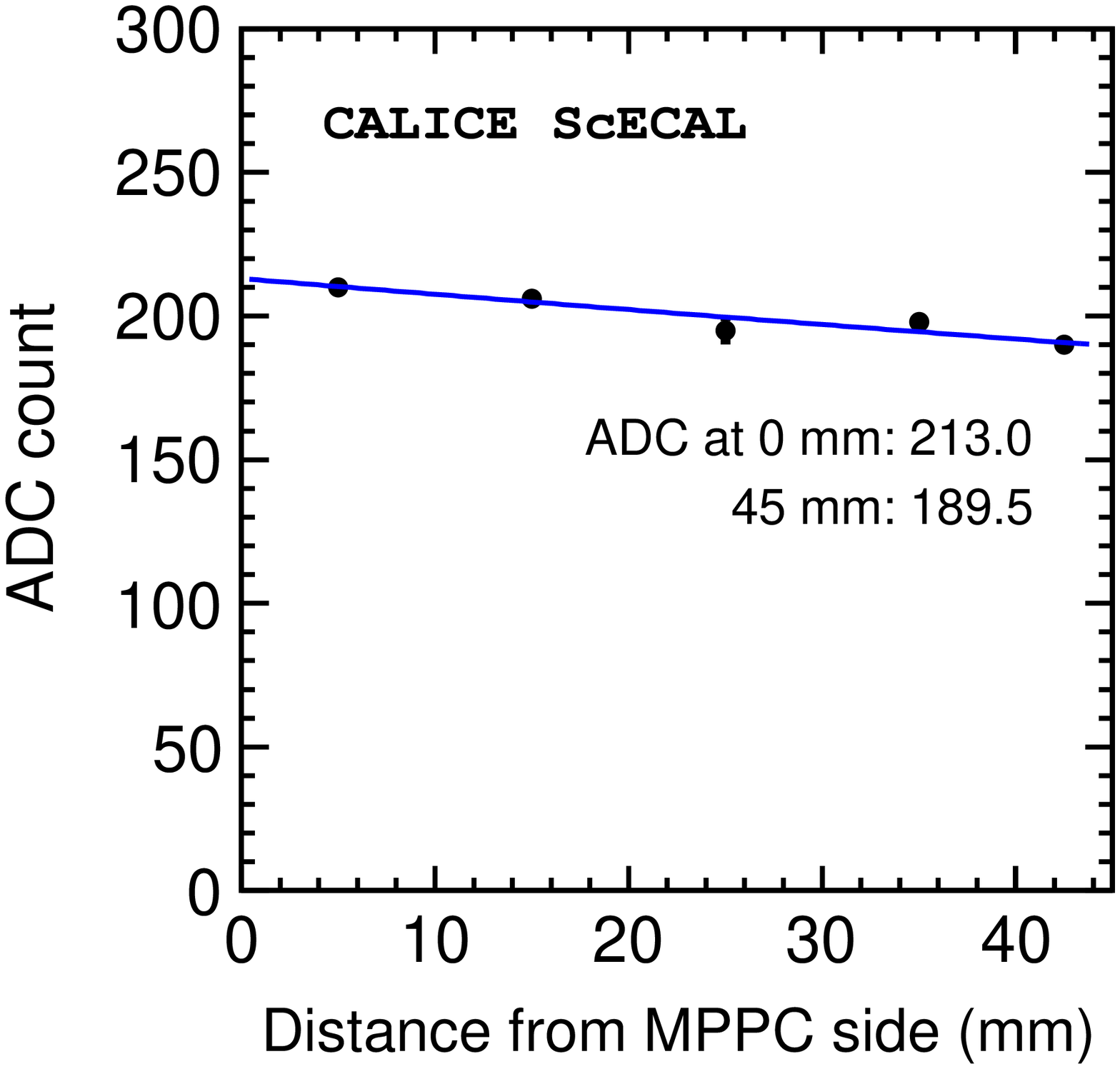}
\includegraphics[width=0.45\textwidth]{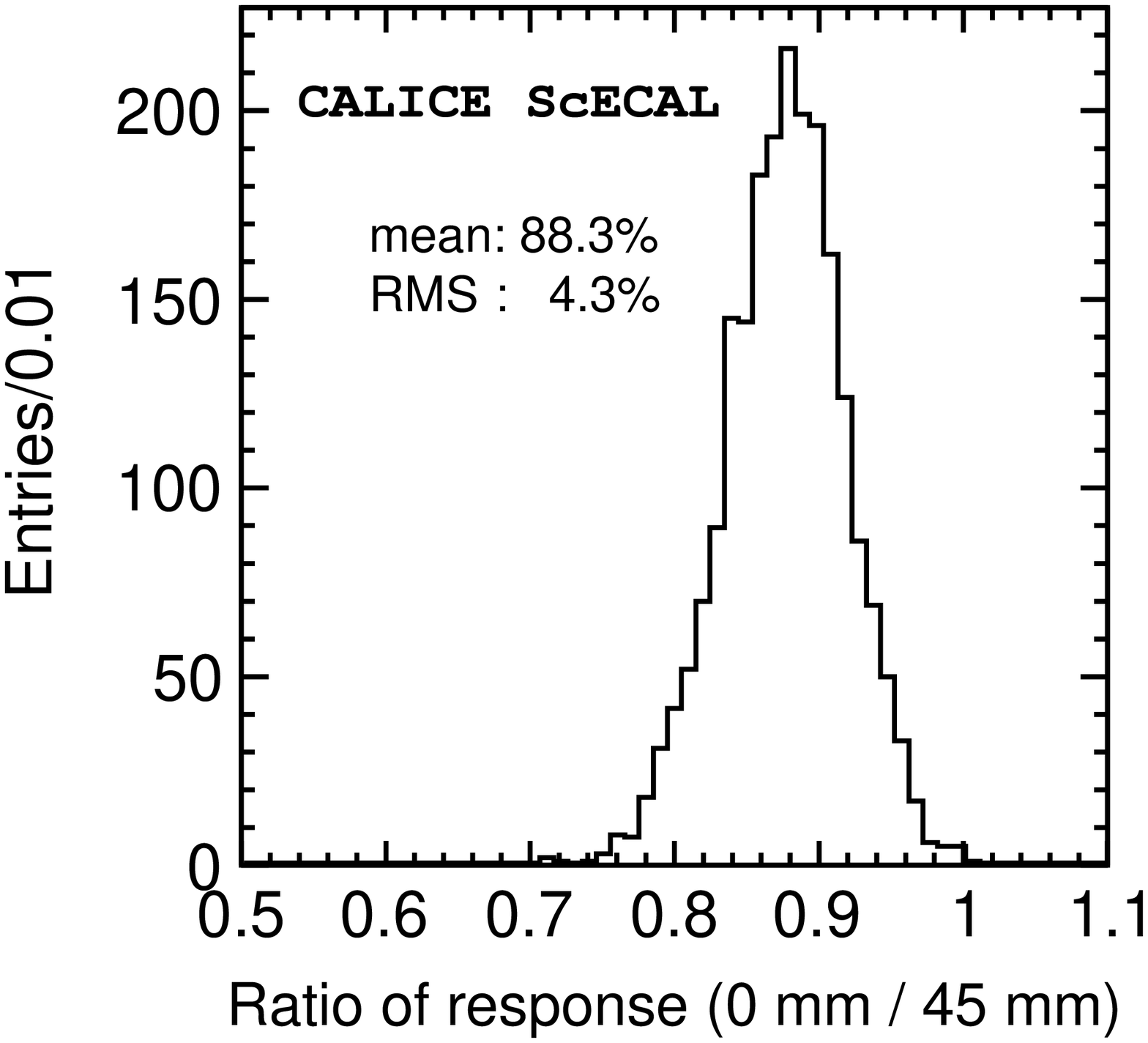}
\caption{\label{fig:uniformity} \small{ An example of the MIP response
    as a function of the distance from the MPPC side ({\it left}), and
    the distribution of the ratio of response at the far end side to
    the MPPC side, %%% 20170511 David: measured 
    determined 
    by fitting with an exponential function
    ({\it right}).}}
\end{center} 
\end{figure}

%%\subsubsection{Beam momentum spread}\label{section:systematicFromBeamMomFlactuation}
%%Nige Watson 20161222 \item {\bf Beam momentum spread}
\item[Beam energy spread]
\label{section:systematicFromBeamMomFlactuation}~

  The beam provided at MT6 \textred{has a %designed 
  relative beam energy spread} $\Delta
  E/E = 2\%$ for beam energies in the range
  1---60\gev\ \cite{MTestDesign}.  Tests of a Pb/glass
  calorimeter performed at the same beam-line led to an estimate of the
  relative beam energy spread of 2.7\,$\pm$\,0.3\% for beam energies
  in the range 1--4\gev~\cite{T1005}.  Another experiment measured
  a relative spread of 2.3\% for 8\gev~\cite{T1018} by using the
  same Pb/glass calorimeter.  A third study has also estimated a
  energy spread of 2.3\% in the range 1.5--3.5\gev \cite{g-2}.
Using these measurements %%g929 and beamline design, we take the MT6
we assign a beam energy spread of
%a beam energy of 
$(2.7\pm0.3)$\% for beam energies between
2 and 4\gev, and $(2.3\pm 0.3)$\% for the range 8--32\gev.
To estimate the intrinsic energy resolution of the prototype, this
energy spread should be quadratically subtracted from the energy
resolution determined.
%%Nige Watson 20161222
%%Section~\ref{section:meanResolutionWstatUncertainties}.
The systematic uncertainty on this procedure arises from the
uncertainty of the intrinsic beam energy spread, taken to be 0.3\%,
and is motivated by the spread and uncertainties of the available
measurements.
\textredSecond{
\item[Fitting range of the energy spectra]~}

\textredSecond{
We determined the fitting range of the energy spectra to $\pm 1.65\,\sigma$. \textblueSecond{
%%%A mean and a $\sigma$ were obtained as conversing results of iterations of the Gaussian fitting to the spectrum.  
The mean and $\sigma$ were obtained using a recursion method of the Gaussian fitting to the spectrum \ie
the mean and $\sigma$ obtained in one iteration step were used in the next iteration to determine the fitting range.
The mean converged in fewer than four iterations.
For smaller fit ranges, the reduced $\chi^2$ does not improve in a significant way. For larger fit ranges of
  $\pm 2.33\,\sigma$ and $\pm 2.58\,\sigma$, the reduced  $\chi^2$ increases by up to factor 3.
 (The reasons for this large reduced $\chi^2$ are
}
%The reduced $\chi^2$ with the fitting range of $\pm\ 1.65\,\sigma$  was no significant difference with fitting range of $\pm\ 1.28\,\sigma$, whereas it increases 
%up to 3 for some beam energies with fitting ranges of  $\pm\ 2.33\,\sigma$ and $2.58\,\sigma$ as the range becomes wider: reason of 
small residual contaminations in the tail of spectrum.)
\textblueSecond{Thus a large fitting range introduces bias to the reconstructed energy mean and its resolution. 
%  To account for the impact this bias may have
To account for the impact these range variations may have,
%%  on linear fitting for the energy response and quadratic fitting for the energy resolution, 
we assigned systematic uncertainties for the reconstructed energy mean
and $\sigma_E/E_\mathrm{reco}$ using differences obtained between
fitting within $\pm\ 1.65\,\sigma$  and smaller. 
These systematic uncertainties are negligible for all energies except $E_\mathrm{beam} = 2$\,GeV, where a systematic contribution of 
0.01\% is added in quadrature  to the ``totalff in Table\,\ref{table:summaryUncertaintyIndividual}.
 %These systematic uncertainties are almost negligible: only energy mean of $E_\mathrm{beam} = 2$\,GeV was quadratically contributed so that 0.01\% negatively shifted  
 % on ``total'' in Table\,\ref{table:summaryUncertaintyIndividual}.
 %, and the second and fourth rightmost column of Table\,5, 
  %whereas the single variation by the change of this fitting range  was $- 0.1$\%.
  % The estimated single variation by the larger fitting range of $\pm\ 2.58\,\sigma$ is less than 0.2\% for all beam energy.
 }  
%Because a large fitting range \textblue{thus} adds  a bias on the reconstructed energy mean\,---\,although still small\,---\,we include the systematic uncertainty 
 %from the differences of reconstructed energy mean and $\sigma_E/E_\mathrm{reco}$ between the fitting range  $\pm\ 1.65\,\sigma$  and smaller.
% This systematic uncertainties are almost negligible: only energy mean of $E_\mathrm{beam} = 2$\,GeV was apparently affected so that 0.01\% increases 
 % on ``total'' in Table\,\label{table:summaryUncertaintyIndividual}.''
}

%%\subsubsection{Residuals of the mean value of deposited energy among runs}
%%\subsubsection{Summary of uncertainties on each beam momentum}\label{section:systematicEachMom}
%%Nige Watson 20161222 \item {\bf Summary of uncertainties on each beam momentum}
\item[Summary of uncertainties on each beam energy]
  \label{section:systematicEachMom}~
  
Table~\ref{table:summaryUncertaintyIndividual} summarises the different systematic uncertainties for 
%% Frank g923 respective
the considered beam energies together with the statistical
uncertainties.  Figure~\ref{meanDepOnTemp_withSysE} shows the same
data as those of Fig.~\ref{meanDepOnTemp}, but with systematic uncertainties
discussed above included.  The systematic uncertainties have a size
comparable with the run-to-run variations, except for the 12, 15, and
20\gev\ cases, where the variation is larger than the estimated
uncertainties.  Those data were acquired early in the test beam period
%%%% 20170511 David: and 
when there were frequent changes made to the beam conditions.
%: we confirmed it from that the center of beam positions were changed. 
This potentially results in changes of the beam energy with changing
beam conditions.
%
%%% NB - change this wording - NKW
%%%%As the recording of these changes in beam conditions were not
%%%%sufficiently well documented, w
We conservatively assign
the SD of the observed run-to-run
%%Nige Watson 20161222 Because we unfortunately have no detail record of beam condition
%%Nige Watson 20161222 changes, the run-to-run variations cannot be explained clearly.
variations as the systematic uncertainties in such cases.
Table~\ref{runDif_totalUncertainties} lists the sum of the
individually estimated uncertainties and the deviations estimated from
the run-to-run variation.  To reduce the impact of double counting of
uncertainties, the larger of the two values is assigned as the final
systematic uncertainty for each individual beam energy.
Table~\ref{table:summaryUncertaintyResoIndividual} lists the energy
resolution at each beam energy after subtraction of the beam
energy spread, together with its systematic %%% 20170511 David: uncertainty 
and statistical uncertainties. % are listed in
% Table~\ref{table:summaryUncertaintyResoIndividual}.
%The dominant contribution from the beam energy spread makes other
%contributions invisible in the sum in quadrature.
%\textred{
The quadrature sum of all systematic effects is completely dominated by the beam energy spread.
%}
\end{description}

%%%%%%%%%%%%%%%%%%%%%%%%%%%%%%%%%%%%%%%%%%%
%%         Fig   runDiff: temperrature vs mean (2)             %%%%%%%%%%%%%%%
%%%%%%%%%%%%%%%%%%%%%%%%%%%%%%%%%%%%%%%%%%%
%%% he:/home/coterra/FNALpaper/temper_mean_20GeV4runs
\begin{figure}[tbhp]				
\captionsetup{width=0.75\textwidth}						
\begin{center}\includegraphics[width=0.6\textwidth]{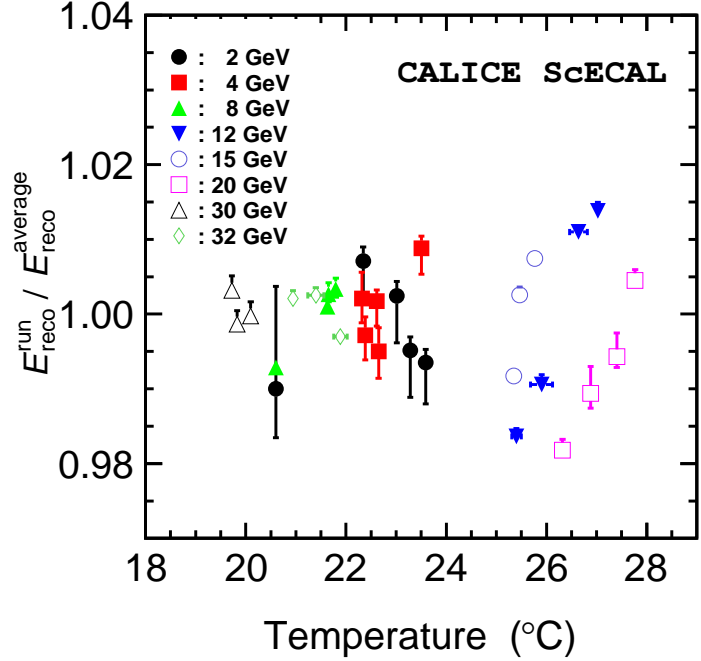}
\caption{\label{meanDepOnTemp_withSysE} \small{ {\textblueSecond{
Ratio of the reconstructed mean of individual runs to the average of
  all runs at a given beam energy (including systematic uncertainties)
  \vs\ temperature during data taking.}
        }}}
\end{center} 
\end{figure}

\begin{table}[tbhp]
\begin{center}
\caption{\label{table:summaryUncertaintyIndividual}\small The
  uncertainties in the mean measured energy 
  (\%) for combined data sets.
 % beam energy, $E_\mathrm{beam}$,
%%%he:/home/coterra/FNALpaper/temper_mean_20GeV4runs/temperVSmeanLargeW.sysErrEachRun.C
}
\begin{tabular}{cccccccc}
\hline \hline
\vspace{-2mm}&&\\
$E_{\mathrm{beam}}$  & range-$x$ & other cuts & \cmiptwenty) &
\cmipslope & \Npix& statistical & total \\
$[$\gev$]$   &   & & & & & \\
\hline %\aline
\vspace{-3mm}\\
2  &$_{-0.45}^{+0.22}$&$_{-0.37}^{+0.09}$&$\pm$0.23&$\pm$0.03&$\pm$0.11&$\pm$0.03&$_{-0.65}^{+0.36}$\\
\vspace{-3mm}\\
4  &$_{-0.25}^{+0.21}$&$_{-0.22}^{+0.07}$&$\pm$0.09&$\pm$0.02&$\pm$0.01&$\pm$0.02&$_{-0.35}^{+0.24}$\\
\vspace{-3mm}\\
8  &$_{-0.08}^{+0.12}$&$_{-0.03}^{+0.06}$&$\pm$0.21&$\pm$0.03&$\pm$0.05&$\pm$0.01&$_{-0.25}^{+0.27}$\\
\vspace{-3mm}\\
12 &$_{-0.02}^{+0.10}$&$_{-0.04}^{+0.04}$&$\pm$0.16&$\pm$0.03&$\pm$0.05&$\pm$0.01&$_{-0.19}^{+0.21}$\\
\vspace{-3mm}\\
15 &$_{-0.06}^{+0.07}$&$_{-0.03}^{+0.04}$&$\pm$0.13&$\pm$0.04&$\pm$0.04&$\pm$0.01&$_{-0.17}^{+0.18}$\\
\vspace{-3mm}\\
20 &$_{-0.04}^{+0.18}$&$_{-0.04}^{+0.06}$&$\pm$0.13&$\pm$0.04&$\pm$0.04&$\pm$0.01&$_{-0.16}^{+0.24}$\\
\vspace{-3mm}\\
30 &$_{-0.01}^{+0.13}$&$_{-0.02}^{+0.12}$&$\pm$0.12&$\pm$0.06&$\pm$0.16&$\pm$0.01&$_{-0.22}^{+0.28}$\\
\vspace{-3mm}\\
32 &$_{-0.00}^{+0.02}$&$_{-0.03}^{+0.09}$&$\pm$0.23&$\pm$0.04&$\pm$0.13&$\pm$0.02&$_{-0.28}^{+0.30}$\\
%\vspace{-3mm}\\
\hline
%%Nige Watson 20161222 \multicolumn{8}{l}{\small $*$ Beam momentum (GeV/$c$).}
\end{tabular}
\end{center}
\end{table}

\begin{table}[tbhp]
\captionsetup{width=0.85\textwidth}	
\begin{center}
\caption{\label{runDif_totalUncertainties}\small %Standard deviations
%  among runs and estimated uncertainties---including statistical
%  uncertainties---of the measured energy resolution, for each
 % beam energy, $E_\mathrm{beam}$, see
%  Section~\ref{section:systematics} for details.
 % The values are relative to the average (\%). The rightmost column
%  shows the values used in Fig.~\ref{fig:resultPlots}.
  Relative uncertainties of \Ereco\ from run-to-run variations (second column) and 
  from all sources of estimated systematic uncertainties summed up in quadrature
   with statistical uncertainties (third column). The final relative uncertainties applied 
   in the further linearity and resolution analysis are shown in the last column.
}

\begin{tabular}{ccrcrc}
\hline \hline
\vspace{-2mm}\\
 $E_\mathrm{beam}$ &\multicolumn{5}{c}{Relative uncertainty (\%)}\\ 
 \cline{2-6}
$[$\gev$]$  &\vspace{-2mm}\\
%  uncertainties}&\multicolumn{2}{c}{Final uncertainties}  \\
%$E_{\mathrm{beam}}$ &Run difference& \multicolumn{2}{c}{Estimated
%  uncertainties}&\multicolumn{2}{c}{Final uncertainties}  \\
 &Run variations& \multicolumn{2}{c}{Estimated  uncertainties}&\multicolumn{2}{c}{Final uncertainties}  \\
\hline %\aline
2   &$\pm$0.58&\hspace{7mm}$-0.65$&\hspace{-12mm}+0.36&\hspace{3mm}$-0.65$&\hspace{-10mm}\textcolor{black}{+0.58}\\
4   &$\pm$0.34&\hspace{7mm}$-0.35$&\hspace{-12mm}+0.24&\hspace{3mm}$-0.35$&\hspace{-10mm}\textcolor{black}{+0.34}\\
8   &$\pm$0.44&\hspace{7mm}$-0.25$&\hspace{-12mm}+0.27&\hspace{3mm}$\textcolor{black}{-0.44}$&\hspace{-10mm}\textcolor{black}{+0.44}\\
12  &$\pm$1.23&\hspace{7mm}$-0.19$&\hspace{-12mm}+0.21&\hspace{3mm}$\textcolor{black}{-1.23}$&\hspace{-10mm}\textcolor{black}{+1.23}\\
15  &$\pm$0.66&\hspace{7mm}$-0.17$&\hspace{-12mm}+0.18&\hspace{3mm}$\textcolor{black}{-0.66}$&\hspace{-10mm}\textcolor{black}{+0.66}\\
20  &$\pm$0.79&\hspace{7mm}$-0.16$&\hspace{-12mm}+0.24&\hspace{3mm}$\textcolor{black}{-0.79}$&\hspace{-10mm}\textcolor{black}{+0.79}\\
30  &$\pm$0.17&\hspace{7mm}$-0.22$&\hspace{-12mm}+0.28&\hspace{3mm}$-0.22$&\hspace{-10mm}+0.28\\
32  &$\pm$0.27&\hspace{7mm}$-0.28$&\hspace{-12mm}+0.30&\hspace{3mm}$-0.28$&\hspace{-10mm}+0.30\\
\hline
%%Nige Watson 20161222 \multicolumn{3}{l}{$*$ Beam momentum (GeV/$c$).}\\
\end{tabular}
\end{center}
\end{table}

\begin{table}[tbhp]
\captionsetup{width=0.8\textwidth}
\begin{center}
\caption{\label{table:summaryUncertaintyResoIndividual}\small
  Measured energy resolutions and their statistical and systematic
  uncertainties, after subtraction of beam energy spread, for each
  beam energy, $E_\mathrm{beam}$.
}
\begin{tabular}{cccc}
\hline \hline
\vspace{-2mm}&&\\
$E_{\mathrm{beam}}$ &energy resolution  &systematic & statistical \\
$[$\gev$]$       & $\sigma_E/E$ (\%)   &            & \\
\hline %\aline
2   &9.06&$\pm$0.34&$\pm$0.038\\
4   &6.25&$\pm$0.35&$\pm$0.028\\
8   &4.48&$\pm$0.33&$\pm$0.016\\
12  &3.72&$\pm$0.32&$\pm$0.018\\
15  &3.55&$\pm$0.31&$\pm$0.015\\
20  &3.04&$\pm$0.33&$\pm$0.030\\
30  &2.59&$\pm$0.34&$\pm$0.018\\
32  &2.52&$\pm$0.33&$\pm$0.022\\
\hline
%%Nige Watson 20161222 \multicolumn{3}{l}{$*$ Beam momentum (\gevc).}\\
%%Nige Watson 20161222 \multicolumn{3}{l}{$\dagger$ Beam momentum spread is subtracted.}\\
%%Nige Watson 20161222 \multicolumn{3}{l}{$\ddagger$ Absolute value: uncertainty of $\sigma_E/p$ (\%).}\\
\end{tabular}
\end{center}
\end{table}

\subsection{Linearity and energy resolution of the ScECAL prototype}\label{section:results_linearity_resolution}

\begin{figure}[htb]		
 \captionsetup{width=0.75\textwidth}								
\begin{center}{\includegraphics[width=0.6\textwidth]{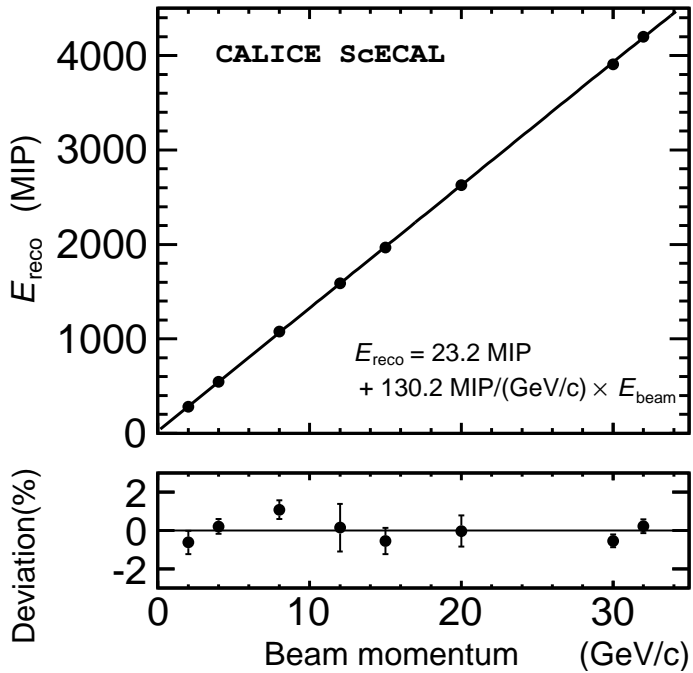}
%%%he:/home/coterra/FNALpaper/ResultForFNALPaper_MultA3800_20GeV4runs_g714
%\includegraphics[width=0.6\textwidth]{./resolutionScenario1_1can16cg714.eps}
}
\caption{\label{fig:resultPlotLinear} \small{Response of the ScECAL
    prototype to 2--32\gev\ electrons ({\it top}), deviation from
    the result of a linear fit divided by the linear fit ({\it bottom}).  
    The error bars show the sum in quadrature
    of the statistical and systematic uncertainties.
  }
}
\end{center} 
\end{figure}

Figure~\ref{fig:resultPlotLinear}
shows the mean reconstructed
energy
(as shown in Table~\ref{table:meanResolution_eachBean},
with uncertainties from Table~\ref{runDif_totalUncertainties}) as a function of the incident beam energy.
The solid line is the result of a linear fit to these measurements.
The \textred{slope  and} offset are ($130.22\pm0.26$)\,MIP/\gev\ and $(23.2\pm1.6)$ MIP, respectively.
The figure also shows the deviation from linearity at each beam energy.
The maximum deviation from linearity is $(1.1\pm0.4)$\%, at 8~\gev.

\begin{figure}[htb]
 \captionsetup{width=0.75\textwidth}									
\begin{center}{%\includegraphics[width=0.6\textwidth]{./DEVIATIONcan16cg714.eps}
%%%he:/home/coterra/FNALpaper/ResultForFNALPaper_MultA3800_20GeV4runs_g714
%%%% data he:/home/coterra/Fnal2009Anaf830start/2009MultiEventsCut_intcalDB2sig/??GeV/All10x10_C2_MultA3800_20GeV4runs/pdg_average.dat
%%%% for NIMA revices he:/home/coterra/FNALpaper/ResultForFNALPaper_MultA3800_20GeV4runs_g714_replyNIM1_h918
%%%% error table was made at he:/home/coterra/FNALpaper/temper_mean_20GeV4runs
%%%% by using calcAddErrorFromFittingRange.C
\includegraphics[width=0.6\textwidth]{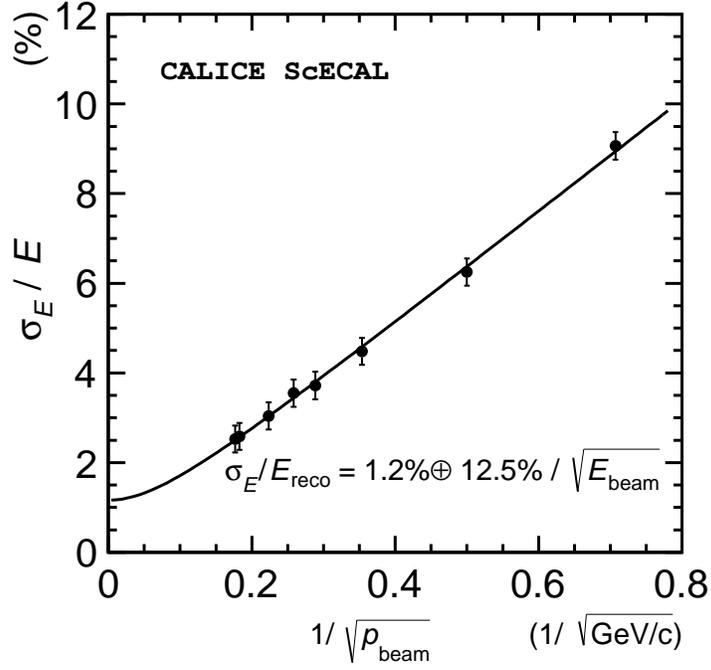}
}
\caption{\label{fig:resultPlotResolution} \small{ Energy
    resolution of the ScECAL as a function of the inverse square root of the beam
    \textred{energy}.  The error bars show the sum in quadrature
    of the statistical and systematic uncertainties.
}}
\end{center} 
\end{figure}

Figure~\ref{fig:resultPlotResolution}
shows the energy resolution
as a function of the inverse of the square root of the beam energy. 
The data points and their uncertainties are taken from Table~\ref{table:summaryUncertaintyResoIndividual}:
 the intrinsic beam energy spread has been subtracted.
%%%% Frank's requirement.
%%The curve shows the result of a fit to the data using a quadratic parametrization of the energy resolution. --->
The curve shows the result of a fit to the data using a %%% frank g929 quadratic 
two-component parametrisation of the energy resolution: 
\begin{equation}
  \label{eq:fittingEq}
\frac{\sigma_{E}}{E_\mathrm{reco}} = \frac {\Cstoch}{\sqrt{E_\mathrm{beam}[\mathrm{\gev}]}} \oplus \Cconst,
\end{equation}
where \Cstoch\ and \Cconst\ are free
to vary in the fit and determined to be $(12.5 \pm0.4)$\% and $(1.2\pm0.4)$\%, respectively.
%%%Difference between $E_\mathrm{beam}$ and  $E_\mathrm{beam}$ is ignorable for electron beam.
The uncertainties include both systematic and statistical contributions.
%%Nige Watson 20161222 because those are results of fit adopting systematic and statistic uncertainties in Table\,\ref{table:summaryUncertaintyResoIndividual}.
%%%%%%%%%%%%%%%%%%%%%%%%%%%%%%%%%%%%%%%%%%%
%%         Fig   runDiff4GeV              %%%%%%%%%%%%%%%%%%%%
%%%%%%%%%%%%%%%%%%%%%%%%%%%%%%%%%%%%%%%%%%%
%Fig. 22 Original data analysls
%	he:~/Fnal2009Anaf830start/2009MultiEventsCut_intcalDB2sig/xxGeV/All10x10_C2_MultA3800_20GeV4runs/
%
%	plot making
%	he:~/FNALpaper/ResultForFNALPaper_MultA3800_20GeV4runs/
%%
%%%%%%%%%%%%%%%%%%%%%%%%%%%%%%%%%%%%%%%%%%%%%

\begin{figure}[htbp]									
  \begin{center}{\includegraphics[width=0.45\textwidth]{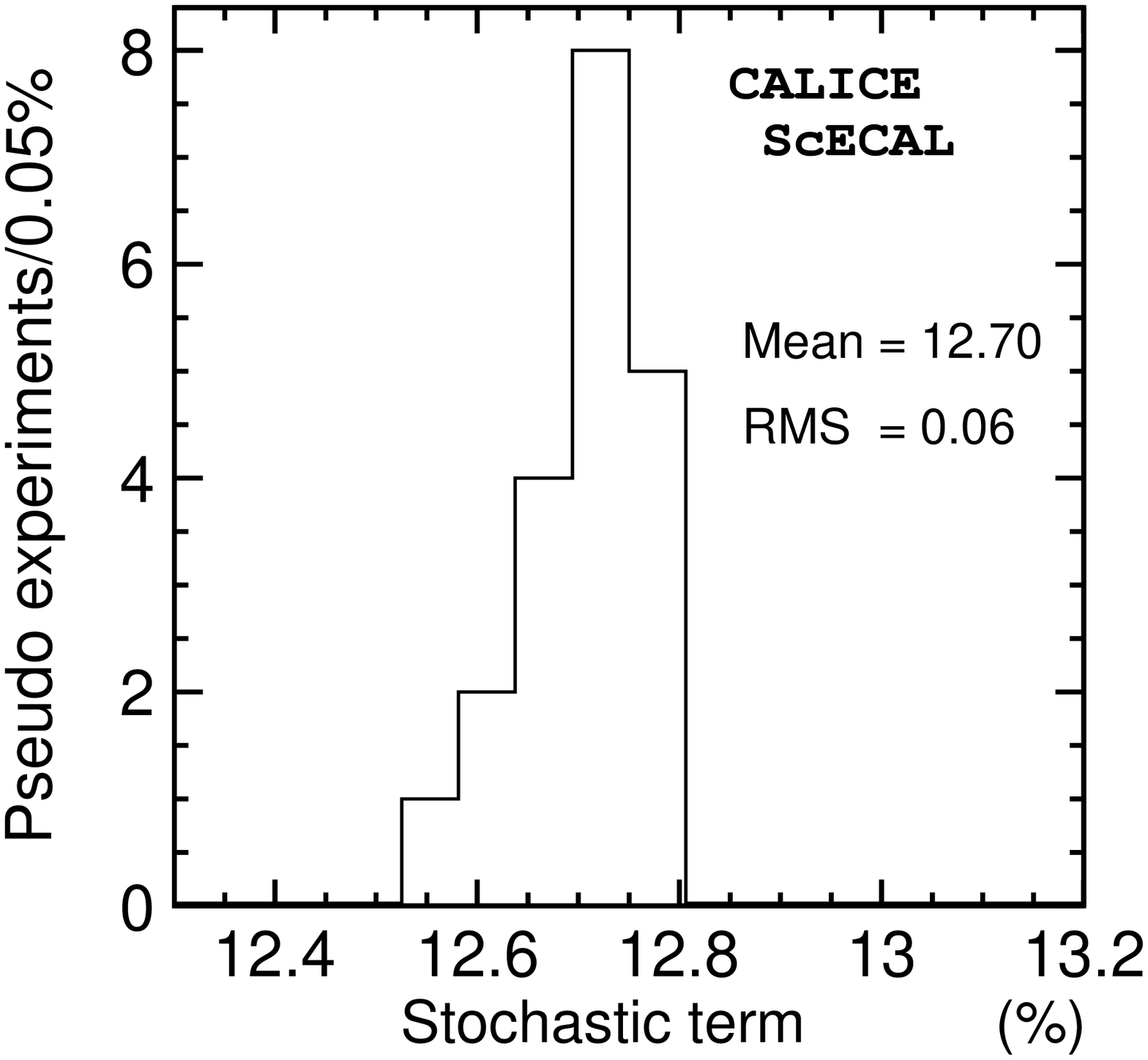}
\includegraphics[width=0.45\textwidth]{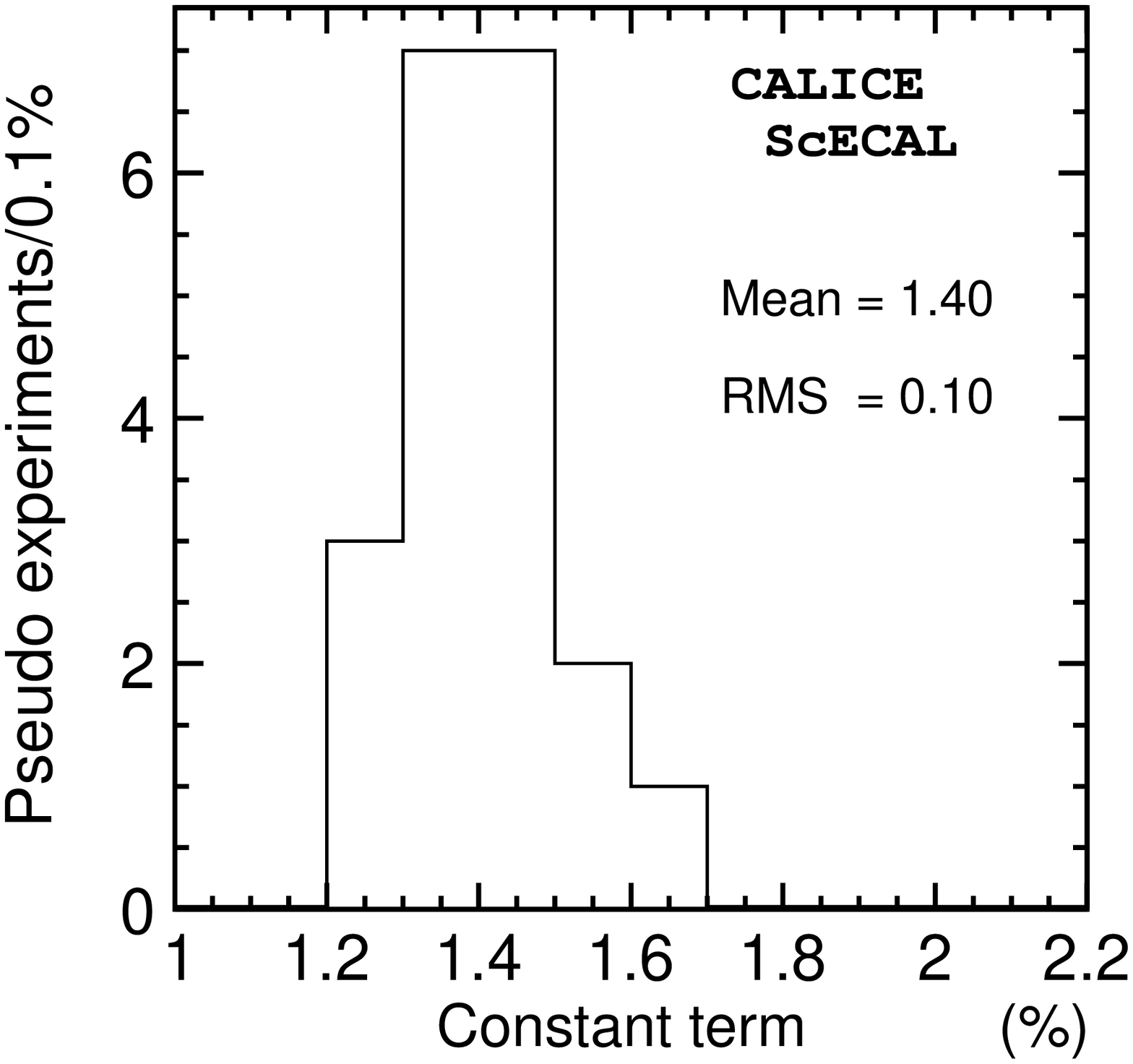}
}
\caption{\label{fig:systematic_cmip_const_on_reso} \small{
    Distribution of the stochastic (${\it left}$) and constant (${\it
      right}$) terms of the energy resolution in 20 pseudo-experiments
    in which \cmiptwenty\ was varied.  }}
\end{center} 
\end{figure}

The systematic uncertainties originating from the three calibration
factors, \cmip, \cphoton, and \cinter\ on the stochastic and constant
terms of the energy resolution were investigated by using a
pseudo-experiment method as discussed
%%Nige Watson 20161222 in term
%%Nige Watson 20161222 \ref{section:ADC-MIPsystematic}--\ref{section:InterCalibSystematic}
in Section~\ref{section:systematics}.  As examples,
Fig.~\ref{fig:systematic_cmip_const_on_reso} shows the distribution of
the stochastic ({\it left}) and constant ({\it right}) terms of the
energy resolution in the pseudo-experiments in which \cmiptwenty\ was
varied.
%%% Katsu adds the following: (2015/09/01)
%%% 20170511 David Those 
The mean values slightly increased from the nominal value, because the random \textredSecond{variations} 
of those constants keep them away from true values.
Therefore, we take RMS values of those for the uncertainty.
The RMS of the energy resolution for each beam energy  
\textredSecond{is} included in the systematic uncertainties in Table~\ref{table:summaryUncertaintyResoIndividual}
as well as the uncertainty of \Npixeff\ and cut variations.
%%%%%%%%%%%%%%%%%%%%%%%%%%%%%%%%%%%%%%%%%%%
%%         Fig  figs:systematic_cmip_const_on_reso               %%%%%%%%%%%%%
%%%%%%%%%%%%%%%%%%%%%%%%%%%%%%%%%%%%%%%%%%%
%%% he:/home/coterra/FNALpaper/AdilsPseudoExp
%%% he:/home/coterra/FNALpaper/AdilsPseudoExp

The statistical uncertainties in the energy resolution and stochastic
terms of Equation~\ref{eq:fittingEq} are determined by fitting to
data, taking into account only statistical contributions from
Table~\ref{table:summaryUncertaintyResoIndividual}.  The central
values of the stochastic term and the constant term are determined by
using both statistical and systematic uncertainties in these fits.
 %the statistical and systematic uncertainties except the uncertainty from 
%the beam momentum spread.
 
 The uncertainty arising from the intrinsic beam energy spread is
 considered to be completely correlated across all beam energies.
 The propagation of these uncertainties into the stochastic and
 constant terms are therefore conservatively estimated %%% 20170511 Nigel: by comparing the
%%% 20170511 Nigel:cases where all assumed fluctuations are coherently increased or
%%% 20170511 Nigel: decreased by 0.3\% with respect to their central values: shift of
%%% 20170511 Nigel: \Cstoch\ and \Cconst\ from the
 %%% 20170511 Nigel:nominal results by 0.3\% increasing or decreasing energy resolution
%%% 20170511 Nigel: of each energy are taken as the systematic uncertainties of those
%%% 20170511 Nigel: terms arising from the uncertainty of the beam energy spread,
%%% 20170511 Nigel: merged with the statistical uncertainty.  
 %%%
 as the change from the nominal result caused by varying  \Cstoch\ and \Cconst\ coherently by $\pm$\,0.3\% at all energies. 
These changes are taken to be the systematic uncertainties associated with these terms due to the beam energy spread, 
combined with the statistical uncertainty.
 %%%
 Therefore, the residuals
 after quadratically subtracting statistical uncertainties from the
 uncertainties determined above, are considered as the systematic
 uncertainties from the beam energy spread.
 % The resulting systematic uncertainties of the constant term from
 % to v05-00 The uncertainty in the intrinsic beam momentum spread is  $-0.7$\%,
 The uncertainty of the constant term from the intrinsic beam energy spread is  $-0.7$\%,
 $+ 0.5$\%, while all other sources combined correspond to $\pm$0.09\%.  The
 uncertainty assuming incoherent fluctuations is negligibly small.

Regarding the stochastic term, the uncertainties estimated above are
much smaller than the case assuming the uncertainties of beam energy
spread do not have coherent behaviour among energy points.  Therefore,
the systematic uncertainty originating from the uncertainties due to beam
energy spread is conservatively adopted from the incoherent case as
0.4\%.

Therefore, the final results of the stochastic term and constant term
can be expressed as:
\begin{eqnarray*}
\Cstoch & = & 12.5\pm0.1(\mathrm{stat.})\pm0.4(\mathrm{syst.})\%\,
\mathrm{\gev}^{1/2}  \\
\Cconst & =  & 1.2\pm0.1(\mathrm{stat.})^{+0.6}_{-0.7}(\mathrm{syst.}) \% \,.
\end{eqnarray*}

%%%%%%%%%%%%%%%%%%%%%%%%%%%%%%%%%%%%%%%%%%%%%%%%%%%%%%%%%%%%%%%%%%%%%%%%%%%
%%%%%%%%%%%%%%%%%%%%%%%%%%%%%%%%%%%%%%%%%%%%%%%%%%%%%%%%%%%%%%%%%%%%%%%%%%%
%%%%%%%%%%%%%%%%%%%%%%%%%%%%%%%%%%%%%%%%%%%%%%%%%%%%%%%%%%%%%%%%%%%%%%%%%%%
\section{Comparison with Monte Carlo simulation}\label{section:simulation}
%%%%%%%%%%%%%%%%%%%%%%%%%%%%%%%%%%%%%%%%%%%%%%%%%%%%%%%%%%%%%%%%%%%%%%%%%%%
%%%%%%%%%%%%%%%%%%%%%%%%%%%%%%%%%%%%%%%%%%%%%%%%%%%%%%%%%%%%%%%%%%%%%%%%%%%
\subsection{ScECAL prototype simulation}\label{section:MC_model}
The test beam setup was simulated using \scmokka\ \cite{mokka}, a
\scgeant\ \cite{Geant4} based detector simulation framework. 
We selected a reference physics list of %% 20170511 Nigel: simulation engine 
QGSP\_BERT in the \scgeant\ version 9.6\,p1.
 The ScECAL simulation model consisted of 30 layers, each being composed of
the %layers; absorber, reflector, scintillator, the second reflector,
%\textred{
absorber, a scintillator  between two reflectors, %} 
readout instrumentation, and an air gap.  The readout instrumentation
layer was simulated as a uniform mixture of polyimide flat cable,
clear fibre, %black 
polyvinyl chloride sheet, glass fibre and air.  The
scintillator layer was segmented in the same way as the prototype, but
the reflectors between strips and the MPPC volumes were not simulated
because the physical properties of those small materials are close to
those of the scintillator.  The absorber layers were made of a mixture
of elements, as discussed in Section~\ref{section:construction}, with
the measured density and mass fraction.

%%%%%% g229 trial
As the first step of simulation, 32\gev\ muon events were generated
corresponding to each real run.  From these simulated events, 
%%% 20170511 David: a deposited energy 
the energy deposited 
by a MIP, $E^\mathrm{dep}_{\mathrm{MIP},i}$/MIP, 
%a ratio of
%deposited energy to a MIP, $E^\mathrm{dep}_{\mathrm{MIP},i}$/MIP, 
was determined as the MPV of the distribution of deposited energy in each
channel.  After determining $E^\mathrm{dep}_{\mathrm{MIP},i}$/MIP,
each energy deposit was converted into the number of % detected photons, 
\textredSecond{photoelectrons},
p.e.$_{i,k}$ using the following:
\begin{eqnarray} \label{xxxxxxxxxxx} 
\mathrm{p.e.}_{i,k} = e^\mathrm{dep}_{i,k}
(E^\mathrm{dep}_{\mathrm{MIP},i}/\mathrm{MIP})^{-1} R_\mathrm{MIP/p.e.}(T_0)\,,
\end{eqnarray}
where $e^\mathrm{dep}_{i,k}$ is %%% 20170511 David: a deposited energy on a channel 
the energy deposited in channel
$i$ in the event under consideration, $k$, and 
$R_\mathrm{MIP/p.e.}(T)
= \cmip_i(T)/(\cphoton_i(T)/c^\mathrm{inter}_i)$ is taken from real
data.
This p.e.$_{i,k}$ was then binomially fluctuated, thereby smearing the
distribution of deposited energy in the number of p.e. for each
channel \cite{ThesisOskar}.  This smearing
method---photon-statistics-smearing---was also applied to all
electron beam events in the simulation.  From the MPV of the smeared
distribution, an averaged $E^\mathrm{dep}_\mathrm{MIP}$/MIP of all
channels was determined.

%From these simulation, a ratio of deposited energy to a MIP, $E^\mathrm{dep}_{\mathrm{MIP},i}$/MIP, was determined as a MPV of the distribution of deposited energy 
%in each channel. 
%Preceding this, the distribution of deposited energy was  smeared binomially according to the number of photons corresponding to the deposited energy.
%The number of photons were determined with  $c^\mathrm{MIP}_i(T_0)/c^\mathrm{p.e.}_i(T_0)$ of each channel from real data.

%With the average MPVs of all channels\,--\,MPV of deposited energy in each channel simulated with 32\,GeV/$c$ muon beam\,--, a ratio of deposited energy to a MIP were determined; %the energy deposits were smeared 
%binomially according to the photon statistics referring the real data, $c^\mathrm{MIP}_i(T_0)/c^\mathrm{p.e.}_i(T_0)$, of each channel.
%%%%With the electron event simulation, a calibration factor--total energy deposit to the incident energy, sampling ratio toward  the absorbers was determined.
With this ratio, $E^\mathrm{dep}_\mathrm{MIP}$/MIP, the digitisation
procedure for each electron event is carried out as follows:
\begin{enumerate}
\item the deposited energy of each channel is converted into the
  equivalent number of MIPs: $n_{i,k}^\mathrm{MIP} =
  e_{i,k}^\mathrm{dep}/(E^\mathrm{dep}_\mathrm{MIP}$/MIP),
%\item each despited energy on a channel converts into the number of mip: $E_i^\mathrm{MIP} = E_i^\mathrm{dep}/(E^\mathrm{dep}_\mathrm{MIP}$/MIP),
\item $n_{i,k}^\mathrm{MIP} $ is converted into the number of photoelectrons,
  $n_{i,k}^\mathrm{p.e.}$, %%%% 20170511 David: being multiplied by
  by multiplying by 
  $R_\mathrm{MIP/p.e.}(T)$,
%\item $E_i^\mathrm{MIP} $ converts into the number of photons, $E_i^\mathrm{p.e.}$, being multiplied by $c^\mathrm{p.e.}_i(T)/c^\mathrm{inter}_i$,
\item MPPC non-linear response is taken into account according to
  Equation~\ref{eq:saturationSimple} with \Npixeff\ yielding $n_{i,k}^\mathrm{sat} $,
%\item $E_i^\mathrm{p.e.} $ saturates according to Equation\,\ref{eq:saturation}; $E_i^\mathrm{sat} $,
\item Binomial fluctuations are applied to $n_{i,k}^\mathrm{sat} $ to
  account for effects of photon statistics and finally this value is
  converted into the ADC counts, %%%% 20170511 David: being multiplied by
    by multiplying by 
  $c^\mathrm{p.e.}_i(T)/c^\mathrm{inter}_i$.
%\item  $E_i^\mathrm{sat} $ binomially fluctuates and finally this value converts into the ADC counts again with  $c^\mathrm{p.e.}_i(T)/c^\mathrm{inter}_i$.
\end{enumerate}  
These digitised simulation data were analysed with the same computer code %processor
as the real data.

%\textred{
In this way, both the photon statistics and %broad of the fluctuation by the saturation 
effects of non-linear response are %%%implemented 
taken into account for each channel of each event.
%}
\textredSecond{The beam energy spread discussed in
Section~\ref{section:systematics} was implemented as a Gaussian
distribution.  %and the geometrical beam spread in $x$ and $y$ were also %implemented 
%taken into account.
%from 
The geometrical beam spread  in $x$ and $y$ were  taken
\textblueSecond{from} the \textblueSecond{observed} energy weighted distribution in data.}
%
%As the first step, 32\,GeV muon events were generated according to the above procedure corresponding to each real run.
%With the average of the MPV of all channels, the factor to convert the energy deposit on a channel in MC to the ADC counts were estimated, and this factor then was used to convert the energy in each channel in the MC of all electron runs to the  ADC counts. 
%After those step, the MC data for every electron runs were analyzed with the same analysis code as the real data, using also  $c^{\mathrm{MIP}}$ for each channel measured from the real data
%
The material in the beam-line upstream of the prototype was simulated
as three plastic scintillator trigger counters and one plastic
scintillator veto counter. %, and four drift chambers.  
The (downstream) AHCAL and TCMT prototypes were not simulated, because they were used to remove
muon and pion contaminations in data, whereas the simulated events did not
include these contaminations, and the electron efficiency is almost
unchanged. %\textcolor{red}{$\leftarrow$ Really!!}

%We used square shaped beam and smeared the beam position with Gaussian shape,
%and we then tuned the beam parameters run by run adjusting the beam shape and position for the real data. 
%It was confirmed that the beam position did not affect the mean energy deposit or its fluctuation within the 
%measured variations of beam position.  
%% g923: Dead channels and detector noise were also implemented according to run-by-run real detector conditions;
%% g923the random trigger data discussed in section \ref{AnalysisFlow} provided the noise signal overlaid onto each channel of each event.
Dead channels and detector noise were also implemented according to
run-by-run detector conditions.  The modelling of noise was carried
out using the random trigger data introduced in
Section~\ref{section:analysis}, allowing a noise signal to be overlaid
onto each channel of each event: the noise signal of each channel in 
data is added to the simulated signal of the channel concerned %at 
in the digitisation procedure.  
%In the reconstruction procedure, signals that are less than half a MIP in size were rejected in both data and the simulation. 
 The number of the random trigger events is 
between 5\,000 and 10\,000
per run.  Therefore, the noise events were
reused cyclically for the simulations of a given run. % in the

%%%%%%%%%%%%%%%%%%%%%%%%%%%%%%%%%%%%%%%%%%%
%%%%%%%%%%%%%%%%%%%%%%%%%%%%%%%%%%%%%%%%%%%
%%%%%%%%%%%%%%%%%%%%%%%%%%%%%%%%%%%%%%%%%%%
%\subsection{Longitudinal profile}%%\label{section:linearity_resolution}
\subsection{Shower profile}\label{sec:ShowerProfile}
%%%%%%%%%%%%%%%%%%%%%%%%%%%%%%%%%%%%%%%%%%%
%%%%%%%%%%%%%%%%%%%%%%%%%%%%%%%%%%%%%%%%%%%

%%% g930.1139
%% structure is changed according to Ffank's suggestion.
%Explain that two versions of the absorber materials were studied, then, refer to the Figure. 25
%%

It is essential for the simulation to accurately model
%model accurately 
the material composition of the detector.  As mentioned in
Section~\ref{section:construction}, the measured density of the
absorber plates is $14.25\pm0.04$g/cm$^3$.  This can be  compared with the
density calculated from the known constituents of the detector and
their properties, giving a density of $14.76\pm0.13$~g/cm$^3$ with
$\rho_\mathrm{WC} = 15.63\pm0.1$g/cm$^3$~\cite{WCDense},
$\rho_\mathrm{Co} = 8.9$~g/cm$^3$, and $\rho_\mathrm{Cr} =
7.19$~g/cm$^3$.  This discrepancy requires a correction of the
composition measured by EDX and X-ray diffraction, because the density by direct measurement
is reliable.

We investigated two models for the correction: %in which
1) weight \textred{ratio} of Co to WC was changed to the \textredSecond{directly} measured density of the plate (``balanced'' model), and 
2) vacancies were uniformly distributed into the plate keeping the composition of materials (``vacancy'' model).
Details are explained in Appendix A.

Figure~\ref{fig:longitudinalComparison} shows comparisons of energy deposits on layers among 
%MCs with two components 
both simulation cases
and data.
The
best agreement is found using the balanced model, which  agrees with data in the mean ratio, 0.98\,$\pm$\,0.04 (SD)
 with a small slope of $-0.00064\,\pm\,0.00003/$layer.
 Therefore, we use the ``balanced'' model in subsequent discussions.  The
systematic uncertainty from the model dependence is negligible;
$-0.16\pm 0.01$ %on response 
on the mean response dMIP/d\ebeam,
$+0.67\pm0.01$~MIP on the offset, $0.05\pm$0.05\% on the constant term
of the energy resolution, and $+0.17\pm0.05$\% on the
stochastic term.

\begin{figure}[tbph]			
\captionsetup{width=0.75\textwidth}						
\begin{center}{\includegraphics[width=90mm]{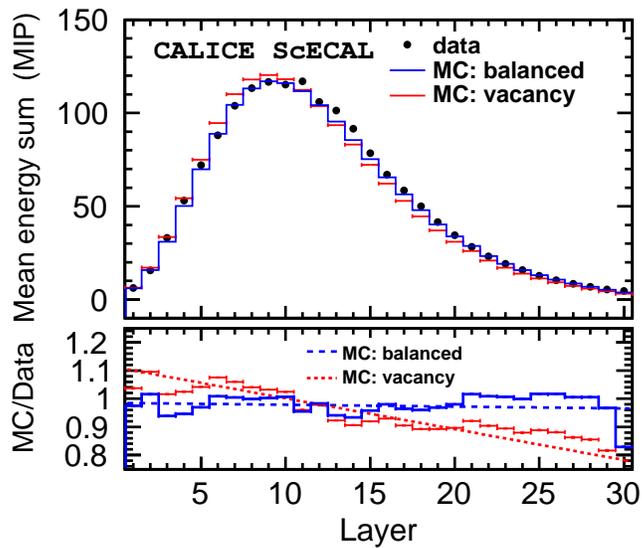}
}
\caption{\label{fig:longitudinalComparison} %\small{ Comparison of the
    %longitudinal profiles of energy deposit--as example of
   % 12\gev.}  
   \small{Comparison between data and Monte Carlo for the longitudinal energy deposition profile, using 12\gev electron beam data.
 %   Ratio of energy deposit in MC to data shows
 % agreement between data and MC especially with the second composition
 % of the absorber plates (balanced) at bottom. 
  MC to data ratio shows up to 10\% discrepancy for the vacancy model.  The balanced model which is the second composition of the absorber plates clearly improves agreement between data and MC. 
  Dotted  and dashed
  lines show linear fitting results.  }}
\end{center} 
\end{figure}
\begin{figure}[tbph]	
\captionsetup{width=0.75\textwidth}								
\begin{center}{\includegraphics[width=0.6\textwidth]{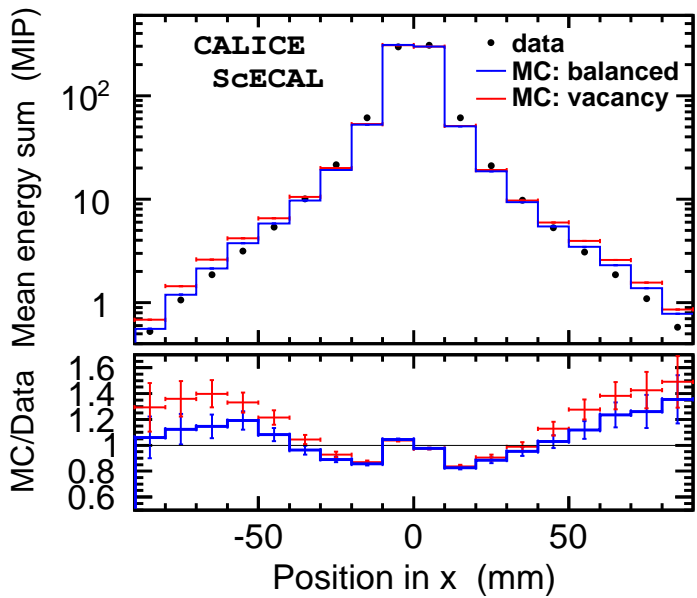}
}
\caption{\label{fig:lateralComparison} \small{ Comparison of the
    lateral energy deposition profile, using 12\gev\ electron beam
    data as an example.}
  An energy sum is a collection of energy in the same lateral position
  on only those layers which have a 10\,mm segmentation in the $x$ direction.  The
  position is the distance from the energy centre of the
  cluster---centre of gravity---for each event.
%MC shows a narrower profile than data at its kernel part.
%The second composition of the absorber plates gives (balanced) closer profile to data than the first composition (vacancy) as well as the longitudinal profile.
}
\end{center} 
\end{figure}

 
Figure~\ref{fig:lateralComparison} shows comparisons of energy deposits projected on 
the $x$  axis in simulation and data.
%%%%%%%%%%%%%%%%%%%%%%%%%%%%%%%%%%%%%%%%%%%
%%         Fig  fig:longitudinalComparison               %%%%%%%%%%%%%%%%%%%%
%%%%%%%%%%%%%%%%%%%%%%%%%%%%%%%%%%%%%%%%%%%
%%% nafhh-ilc02:/afs/desy.de/user/c/coterra/FNALpaper/longitudeProje
%%%
%
%Although 
%\textred{
The simulations predict narrower lateral profiles than those
observed in data %%%at their kernel part\,--\,
in the ``core'' region (within $\pm30$~mm),
whereas the simulations have wider tails than the data.
The origin of this discrepancy is as yet unexplained: 
we %compared 
investigated the results of changing the
%}
detector angle with respect to the beam direction, 
the number of effective pixels, %the thickness of reflector film between scintillators, 
and the physics list to higher precision electromagnetic tracking \cite{ThesisOskar},
none of which was responsible for the effects observed.
The narrower shower cores are not explained  by the uncertainty of absorber composition, because 
we validated  that the radiation length which determines the Moli\`{e}re radius of the detector was correct
by agreement of longitudinal profiles  between data and MC with the balanced absorber composition. 

\if 0
The balanced composition of the absorber plates 
%gives closer profile to data than the first composition as well as the longitudinal profile.
provides a better description of the profile observed in data than
does the ``vacancy'' model, in agreement with 
the observations made for the longitudinal profile.
Although simulations predict narrower lateral profiles than those
observed in data %%%at their kernel part\,--\,
in the ``core'' region (within $\pm30$~mm),
the origin of this discrepancy is as yet unexplained: 
we compared the results of changing detector angle with respect to the beam direction, 
the number of effective pixels, the thickness of reflector film between scintillators, 
and physics list to higher precision electromagnetic tracking \cite{ThesisOskar},
non of which was responsible for the effects observed.
This narrower shower cores is inexplicable also by the uncertainty of absorber composition, because 
we confirmed that the radiation length, which determines the Moli\`{e}re radius of the detector,  was correct
by agreement of longitudinal profiles  between data and MC with the balanced absorber composition. 
\fi 
%%%%%%%%%%%%%%%%%%%%%%%%%%%%%%%%%%%%%%%%%%%
%%%%%%%%%%%%%%%%%%%%%%%%%%%%%%%%%%%%%%%%%%%
%%%%%%%%%%%%%%%%%%%%%%%%%%%%%%%%%%%%%%%%%%%
\subsection{Comparisons of linearity and resolution}\label{section:extructionEffects}
\begin{figure}[htbp]									
\begin{center}
\includegraphics[width=8.0cm]{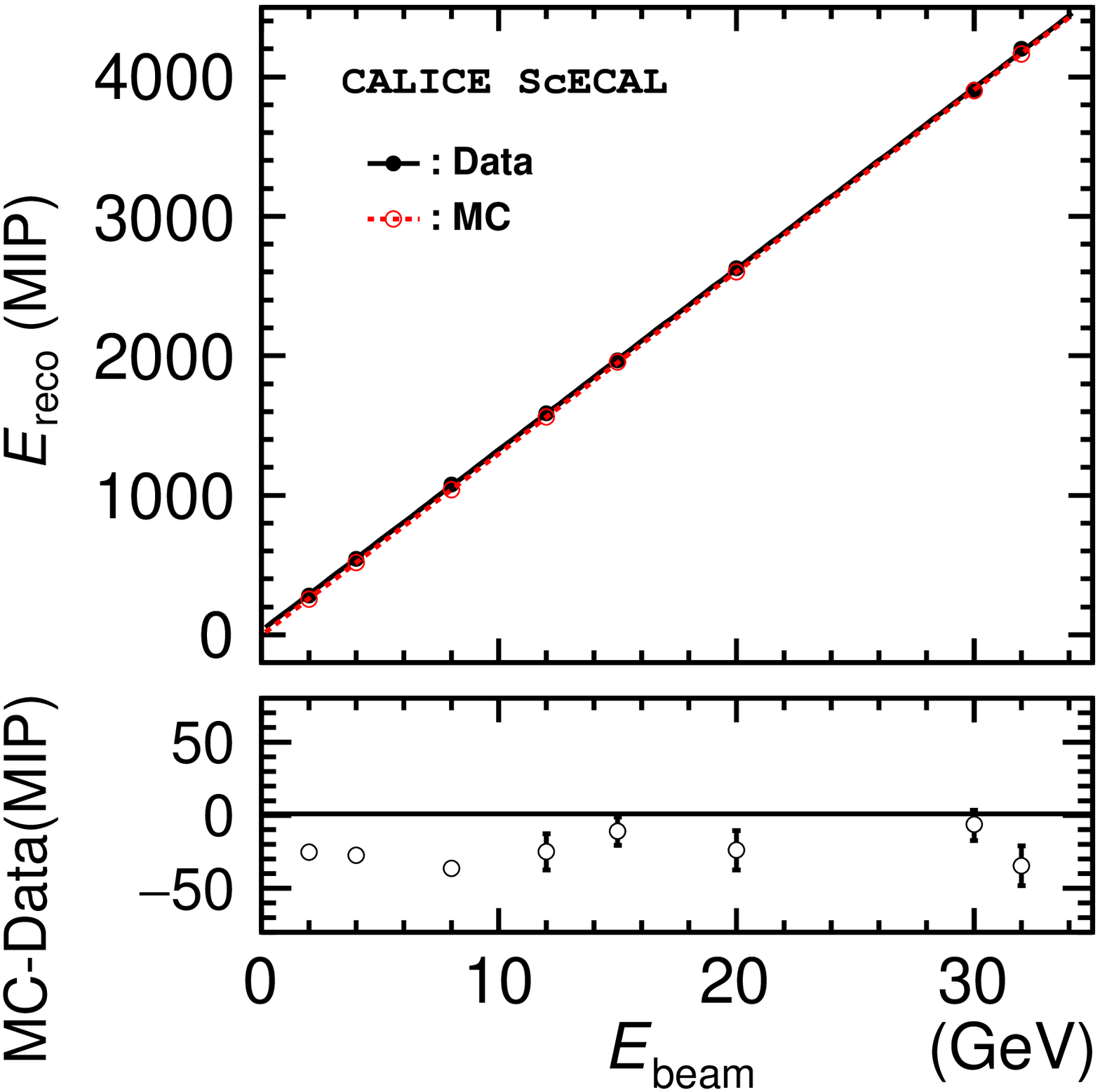}
\includegraphics[width=8.0cm]{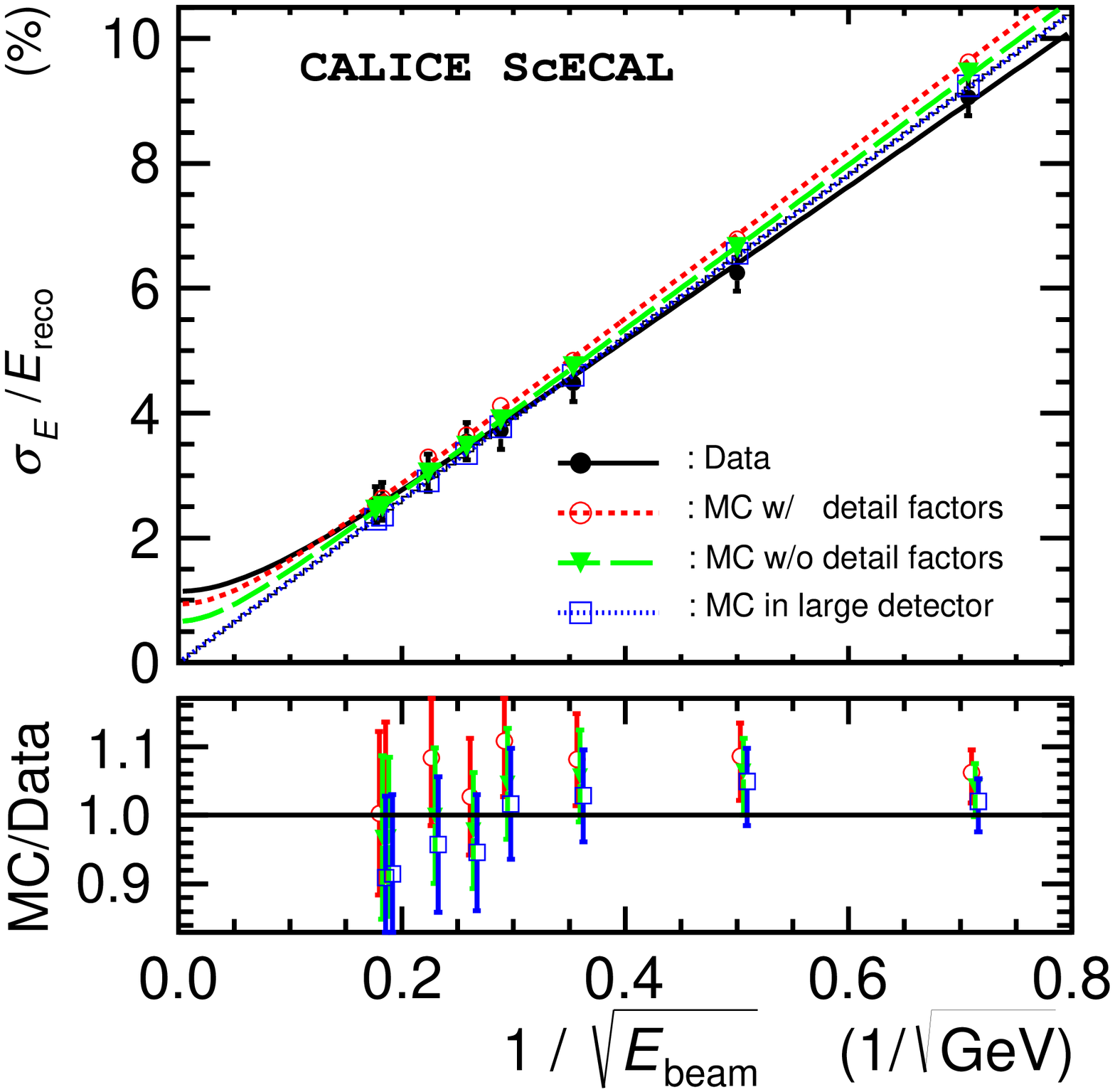}
\caption{\label{fig:mcresults} \small{
{The response ({\it left}) and the energy resolution ({\it right}) of data and the simulated prototype to the electron beams.
``detail factors'' refers to the implementation of fine details,
including photon statistics, effects of the MPPC non-linear response and the
overlaying of noise.
``large detector'' refers  to a simulation in which the dimensions of
the detector have been increased by a factor of three, without the implementation of the
``detail factors'', to study the impact of leakage.}}}
\end{center} 
\end{figure}

%%%%%%%%%%%%%%%%%%%%%%%%%%%%%%%%%%%%%%%%%%%
%%%%%%%%%%%%%%%%%%%%%%%%%%%%%%%%%%%%%%%%%%%
Figure~\ref{fig:mcresults} {\it left} compares the predicted response of the prototype to
electrons with data.
The slope observed in the simulation, $\gevmip= 130.27\pm0.06$~MIP/\gev,
is consistent with that in the data of $130.03\pm0.24$~MIP/\gev, whereas the offset
is $-3.0\pm 0.1$~MIP, some 27~MIP smaller than found in data.
%% g923 With this facts,
This observation is illustrated clearly by the ratio of simulation to
data in Fig.~\ref{fig:mcresults} {\it left, bottom}, suggesting the existence
of
%% g923 coherent 
a constant difference 
for all energies. This %may be due to 
potentially originated by 
a small, residual background contamination
in the data\,% rather %rather 
%than a mis-modelling of the correction of non-linear response,
%because mis-modelling is considered to make energy dependence.
\footnote{\textred{
The mean of \textblueSecond{the} noise in  highly granular calorimeters naturally becomes finite, because of the treatment of the individual detector cells:
\textblueSecond{the} amplitude \textblueSecond{of} each cell is required to be above the threshold which is three times larger than the noise width.
Therefore, there \textblueSecond{are no negative amplitudes contributing to the energy sum by construction, leading to a positive mean of the noise contribution.}}
},
despite %the background 
the detector noise is determined using random trigger events
overlaid on the simulated events.
%
% indicates a small resider of backgrounds in data,
 %as well as the large difference of MC/data from unit at lower energy.
The average difference
of  reconstructed energy between simulation and data is
$-0.18\pm0.20$(RMS)\,\gev.
%%Nige Watson 20161222 in corresponding beam momentum. %, 
%and each response agree within the uncertainty listed in 
%Table\,\ref{table:summaryUncertaintyIndividual}
 
Figures~\ref{fig:mcresults} {\it right} shows the
energy resolution of data and simulation with several different conditions modelled in the simulations.  
%%Simulation explained in Section~\ref{section:simulation}---denoted with ``MC w/ detail factors''---agrees data within one total uncertainty 
%%with maximum of 1.6 times of the total uncertainty at 2\gev.
\textred{The simulation described in Section~\ref{section:simulation}---denoted by gMC w/ detail factors''---agrees with the data, within uncertainties.}
%
%The real prototype has a better resolution at low energy than the simulation, with
%the largest difference amounting to 1.6 times the total uncertainty at 2\gev.
%%whereas less than one uncertainty at other beam momenta.
 %
%%%% g715 coterra The difference is resolved if the momenta increase the amount corresponding to the difference of response  between data and MC in
%Fig.\,\ref{fig:mcresults} {\it left} , 0.18\,$\pm\,$0.20\,(RMS)\,GeV/$c$.
%%% The obtained constant term with this momentum shift is 1.0\%, and  stochastic term is 12.9\%/$\sqrt{\mathrm{GeV}}$: differences from the nominal values are within their uncertainties;
% 1.1$\pm$0.1(stat.)$^{+0.6}_{-0.7}$(syst.)\% and 12.6$\pm$0.1(stat.)$\pm$0.4(syst.)\%/$\sqrt{\mathrm{GeV}}$.
% in the range of their uncertainties. 
The discrepancy persists even if the beam energy spread from 
higher beam energies %were 
is applied to the data recorded at 2 and
4\gev\ ($2.3\pm0.3$\%).
We discuss other MC models in Section~\ref{section:discussion}.

\if 0
Table~\ref{table:compareResolutionMC_DATAl} lists the energy
resolution of  data and simulation, comparing the modelling of several
alternative sets of conditions. 
%% g923 From Figure~\ref{fig:mcresults}, Table\,\ref{table:compareResolutionMC_DATAl},  and other additional simulation studies,
%% the following discusses effects of energy leakage, noise existing, photon statistics, saturation correction, 
%% and the non uniformity of single scintillator response. 
The effect of the energy leakage, detector noise, photon statistics, correction of the MPPC non-linear response, 
and the non-uniformity of a single scintillator response are studied,
based on the results given in Fig.~\ref{fig:mcresults} and
Table~\ref{table:compareResolutionMC_DATAl}.  Additional simulation studies are discussed below.
\fi

%%%%%%%%%%%%%%%%%%%%%%%%%%%%%%%%%%%%%%%%%%%
%%         Fig  sim_results                %%%%%%%%%%%%%%%%%%%%%%%%%%
%%%%%%%%%%%%%%%%%%%%%%%%%%%%%%%%%%%%%%%%%%%
%%% he:/home/coterra/FNALpaper/compaData_MC_paperV02lastBalanced
%%% he:/home/coterra/FNALpaper/compaData_MC_paperV02lastBalanced
%%% he:/home/coterra/FNALpaper/compaData_MC_paperV02lastBalanced:w

%%%%%The following discusses the degrading effects extracted by the simulation:
%

%%%%%%%%%%%%%%%%%%%%%%%%%%%%%%%%%
%%%%%%%%%%%%%%%%%%%%%%%%%%%%%%%%%%%
%%%%%%%%%%%%%%%%%%%%%%%%%%%%%%%%%%
\section{Discussion}\label{section:discussion}
%%%%%%%%%%%%%%%%%%%%%%%%%%%%%%%%%%%
%%%%%%%%%%%%%%%%%%%%%%%%%%%%%%%%%%%
%%%%%%%%%%%%%%%%%%%%%%%%%%%%%%%%%%%%

%%%%%%%%%%%%%%%%%%%%%%%%%%%%%%%%%%%%%%%%%%%
 %%         Fig  fig:tempBeforeAfter                %%%%%%%%%%%%%%%%%%%%%%%%%%
%%%%%%%%%%%%%%%%%%%%%%%%%%%%%%%%%%%%%%%%%%%
 \begin{figure}[h]%tb]									
\captionsetup{width=0.75\textwidth}	
\begin{center}\includegraphics[width=0.6\textwidth]{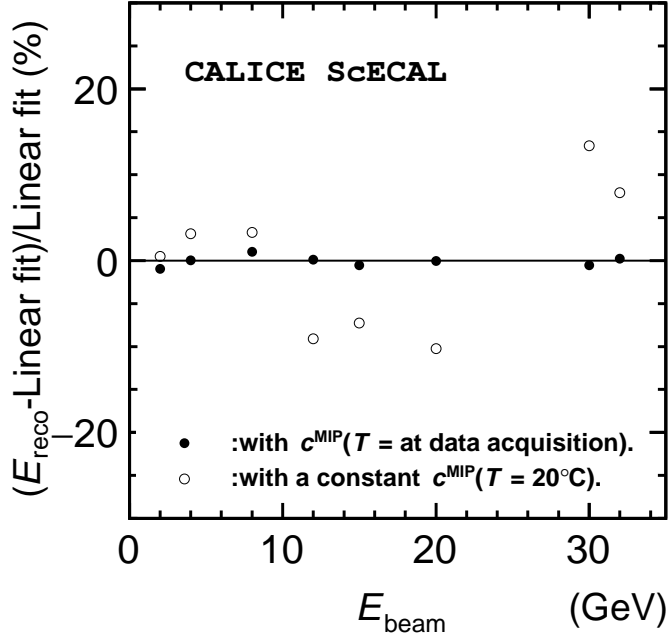}
\caption{\label{fig:tempBeforeAfter} \small{
 Deviations from linear energy response: 
the data were calibrated with the ADC-MIP and ADC--photoelectron conversion factors at 20$^{\circ}$C (circles: only with statistical errors),
and with the ADC-MIP and ADC--photoelectron conversion factors as functions of temperature (black dots: with statistical and systematic errors).
 }}
\end{center} 
\end{figure} 
The ScECAL prototype has shown a linear energy response for electron beam
energies in the range between 2 and 32\gev,
 with a maximum deviation from linearity of $(1.1\pm0.4)$\% at 8\gev.
%%%%%%For further wide range, we are developing the ScECAL with MPPCs having 10,000 pixel on 1\,$\times$\,1\,mm$^2$.
%
Although this experiment was performed in an environment subject to large
variations of the ambient temperature, between 19$^{\circ}$C and 27.5$^{\circ}$C,
the calibration procedure, consisting of temperature-dependent of ADC-MIP and ADC--photoelectron conversion factors for each channel,
successfully controlled the influence of these temperature variations.  
This gives clear evidence that a SiPM-based scintillator tungsten ECAL can be used even in such serious temperature conditions. 
Figure~\ref{fig:tempBeforeAfter} compares the deviations from linear energy response 
when using temperature-independent, without temperature correction, and temperature-dependent conversion factors, with temperature correction.

%%Nige Watson 20161203 \if 0 %g424.
%%Nige Watson 20161203 To demonstrate the efficacy of this procedure, Figure~\ref{fig:tempBeforeAfter} compares the deviations from linear energy response
%%Nige Watson 20161203  when using temperature-independent and -dependent conversion factors.
%%Nige Watson 20161203 %%%%%%%%%%%%%%%%%%%%%%%%%%%%%%%%%%%%%%%%%%%
%%Nige Watson 20161203 %%         Fig  fig:tempBeforeAfter                %%%%%%%%%%%%%%%%%%%%%%%%%%
%%Nige Watson 20161203 %%%%%%%%%%%%%%%%%%%%%%%%%%%%%%%%%%%%%%%%%%%
%%Nige Watson 20161203 \begin{figure}[htb]									
%%Nige Watson 20161203 \begin{center}\includegraphics[width=7cm]{./CompareWtempWOtemp.eps}
%%Nige Watson 20161203 \caption{\label{fig:tempBeforeAfter} \small{
%%Nige Watson 20161203 Deviations from linear energy response: 
%%Nige Watson 20161203 data were calibrated with the ADC-MIP and ADC--photo-electron conversion factors at 20$^{\circ}$C (circles: only with statistical errors),
%%Nige Watson 20161203 and with the ADC-MIP and ADC--photo-electron conversion factors as functions of temperature (black dots: with statistical and systematic errors).
%%Nige Watson 20161203 }}
%%Nige Watson 20161203 \end{center} 
%%Nige Watson 20161203 \end{figure} 
%%Nige Watson 20161203 \fi 

%The standard deviation of the distribution of ADC-MIP conversion factors shown in Fig.\,\ref{fig:adc_mip_distribution} corresponds to 
%19.0\% of its mean.

The variation of the %ADC-MIP conversion factors
\cmip\ is 23\% as shown in Fig.~\ref{fig:adc_mip_distribution}.
This variation is larger than the 
expected value considering the variance in capacitance of the MPPCs
used, as shown in Fig.~\ref{fig:breakdown_capacitance},
and that the over-voltage of every channel was uniformly set to 3~V.
The most probable reason for this variation is a mis-alignment of the WLS fibre and MPPC positions: 
a lateral shift of the WLS fibre to the sensitive area of the MPPC
decreases the photon yield of this scintillator-MPPC unit.
This is caused by difficulties in the precise control of the position and size of the hole when using the extrusion method to 
manufacture the scintillator strips.
Although the performance of the present prototype is sufficient, 
improved MPPC-fibre matching or direct coupling between the MPPC and scintillator have the potential to improve performance \cite{Frank}.
The CALICE Collaboration %ScECAL group
 is currently studying 5~mm-wide scintillator strips directly coupled to MPPCs \cite{IEEE2014}.
 
 The stochastic term in the energy resolution, determined as $(12.5\pm
0.4)\%/\sqrt{E\mathrm{[\gev]}}$ for electron beam energies in the range 2--32\gev,
is significantly better than the requirement of $15\%/\sqrt{E\mathrm{[\gev]}}$.
This fact indicates that we can reduce the sampling ratio by reducing scintillator thickness.
This is one of the advantages of the ScECAL that users can easily optimise the scintillator thickness to achieve a suitable performance.
Actually, the CALICE Collaboration is currently developing the ScECAL with 1.5--2~mm thick scintillator strips \cite{IEEE2014}.

%%Nige Watson 20161203 \if 0
%%Nige Watson 20161203 The sampling fraction of the number of photon by MPPC is less than 5\% as discussed in Section~\ref{section:MC_model}, and this small sampling fraction was not implemented in the 
%%Nige Watson 20161203 simulation model. 
%%Nige Watson 20161203 Despite this, the energy resolution is well modeled, indicating that this small sampling ratio had negligible effect on the 
%%Nige Watson 20161203 energy resolution. 
%%Nige Watson 20161203 This fact is important, since a smaller light yield per MIP may also give satisfactory results, giving a larger effective dynamic range.
%%Nige Watson 20161203 \fi

%\,--\,discussed with Fig\,\ref{fig:uniformity} in Section\,\ref{section:extructionEffects}\,--\,is sufficient for an ScECAL.

%% Moved to Section 2 according to Jerry's suggestion
%Four out of 2160 channels of the present ScECAL prototype were not operational.
%One possible cause is the development of short-circuits between the MPPC electrodes %% g923 occurred 
%caused 
%by the conductive cut edges of the reflector film.
%The CALICE collaboration has another candidate for the reflector
%design that does not have any conductive layer \cite{ThreeM}.

The simulation provided a good description of the prototype
data after the inclusion of a model of photon statistics, effect of the MPPC non-linear response and noise effects.
The largest uncertainty in the input parameters for the  simulation
was %the intrinsic beam energy spread and its uncertainty.  
the uncertainty of the intrinsic beam energy spread.
%%These were estimated using the maximum possible information provided by FTBF. 
The energy resolution of data and simulation are consistent when
all of these uncertainties are taken into consideration.

Regarding the positive offset of the response corresponding to
$0.18\pm0.20$(RMS)\,\gev, [$23\pm26$(RMS)\,MIP], the overlaying of noise on % simulated data onto 
the MC events does not %represent 
reproduce this phenomenon.
%%%The agreement of $\diff E /\diff$MIP between data and simulation,
%%%and the behaviour of MC-data in Fig.~\ref{fig:mcresults} {\it left, bottom},
 %%%indicate that the effect is coherent among the incident energies 
 %%%so that  the offset is not induced by possible deficiencies in the
%%% correction of the MPPC non-linear response: 
%%% effect by deficiencies possibly increases with the deposited energy increase.
\textredSecond{
 The  $\diff E /\diff$MIP of  data and simulation agree with each other. However, the offset of fitting results in data is larger than \textblueSecond{in} 
 MC, and the difference 
between data and MC \textblueSecond{for} each energy point is \textblueSecond{approximately} constant as shown in Fig.~\ref{fig:mcresults} {\it left, bottom}.
These facts indicate that the offset is not induced by possible deficiencies in the correction of the MPPC non-linear response, 
\textblueSecond{because such effects increase with increasing energy.}
}
 
\begin{figure}[htbp]	
\captionsetup{width=0.75\textwidth}									
\begin{center}\includegraphics[width=0.6\textwidth]{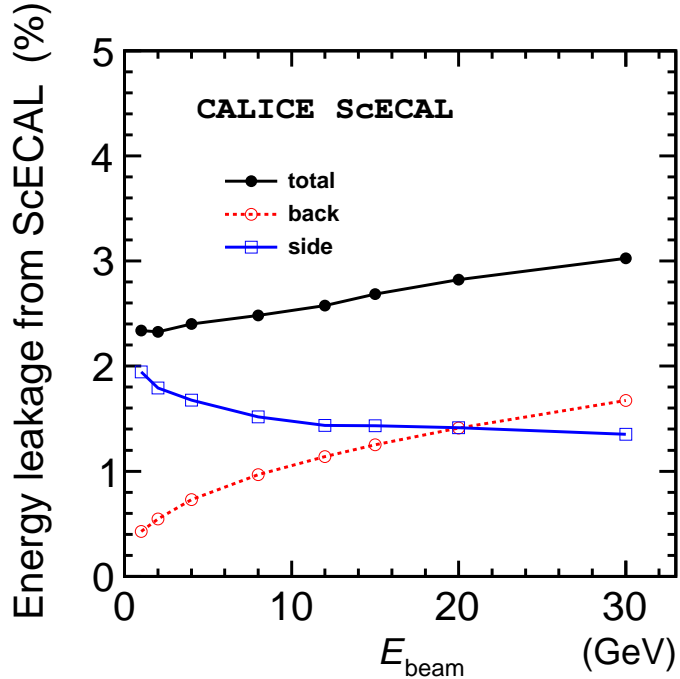}
\caption{\label{fig:leakage} \small{
Relative leakage of the electron energy in the lateral (open boxes)
and longitudinal (open circles) directions. The black markers show the total leakage.}}
\end{center} 
\end{figure}

 We studied what conditions contributed to the energy resolution by comparing 
 data and MC modelling of several alternative sets of conditions.
 To extract the effect of energy leakage
  a study was performed using a simulation in which a detector of linear
dimensions three times greater in each dimension (\scaleLarge).
Figure~\ref{fig:leakage} shows the fraction of energy leakage
perpendicular to the nominal beam direction (lateral leakage) and in
depth (longitudinal leakage) of the ScECAL prototype, estimated
by comparison of deposited energy between large detector and  prototype size.
The total leakage is between 2.3 and 3\% at all measured energies: 
%% g923 There is some compensation between the lateral and longitudinal leakage at the different energies.
the lateral leakage ratio decreases with increasing energy and
dominates below 20\gev, while the longitudinal ratio increases with energy.
\textredSecond{
In \textblueSecond{a future collider}, we can ignore the lateral
leakage because \textblueSecond{the ECAL will have a very large lateral extent. 
Longitudinal leakage will be measured in the hadron calorimeter behind the ECAL and will thus also 
be included in the global energy measurement.}
}
The total deposited energy as the reference does not include the energy leaking out via the
front face of the ScECAL.
%%%%%%%%%%%%%%%%%%%%%%%%%%%%%%%%%%%%%%%%%%%
%%         Fig  fig:leakage                %%%%%%%%%%%%%%%%%%%%%%%%%%
%%%%%%%%%%%%%%%%%%%%%%%%%%%%%%%%%%%%%%%%%%%
%%% he:/home/coterra/FNALpaper/yujiMC

Table~\ref{table:compareResolutionMC_DATAl} lists the energy
resolution of  data and simulation of such a large detector with % 20170511 other 
several other modelings.
Comparison of simulated results of the resolution parameters between large detector and the actual size of the detector
 shows that the leakage alone contributes half  the uncertainty in the constant term; 
  increasing the constant term  
 $\Delta(\sigma_E/E) = +0.66\%$.
The leakage also
  increases the stochastic term by a relative 1.8\%, which
corresponds to a factor of 2.5 for the statistical uncertainty.

\begin{table}[h!]
\begin{center}
%\textcolor{red}{
\caption{\protect\label{table:compareResolutionMC_DATAl}\small The constant
  term and the stochastic term of the energy resolution for data and the various simulations.}
%}
%\vspace{-4mm}
\begin{tabular}{ccccc}
\hline \hline
\vspace{-2mm}&&\\
%data/MC & size &\hspace{-2mm}simulation details$^*$ &constant term\,(\%)$^{\dagger}$&\hspace{-3mm}stochastic term\,(\%)$^{\dagger}$  \\ 
%\hline %\aline
%MC &$0.9\times0.9$m$^2\times90$\,layers&without&$0.00\,\pm0.22$&$13.03\,\pm\,0.04$\\%\textcolor{white}{0}\,\pm0.2\textcolor{white}{0}$&$13.03\,\pm\,0.04$\\
%MC &$0.18\times0.18$m$^2\times30$\,layers&without&$0.66\,\pm0.08$&$13.26\,\pm\,0.08$\\
%MC &$0.18\times0.18$m$^2\times30$\,layers&with&$0.94\,\pm0.03$&$13.58\,\pm\,0.04$\\
%MC &$0.18\times0.18$m$^2\times30$\,layers&with$^{\ddagger}$&$0.78\,\pm0.03$&$13.52\,\pm\,0.03$\\
%data &$0.18\times0.18$m$^2\times30$\,layers&--&$1.1\textcolor{white}{0}\,\pm0.7\textcolor{white}{0}$&$12.6\textcolor{white}{0}\,\pm\,0.4\textcolor{white}{0}$\\
data/MC & size(m;m;layers) &\hspace{-2mm}simulation details$^*$ &constant term\,(\%)$^{\dagger}$&\hspace{-3mm}stochastic term\,(\%)$^{\dagger}$  \\ 
\hline %\aline
MC &$0.9\times0.9\times90$&without&$0.00\,\pm0.22$&$13.03\,\pm\,0.04$\\%\textcolor{white}{0}\,\pm0.2\textcolor{white}{0}$&$13.03\,\pm\,0.04$\\
MC &$0.18\times0.18\times30$\,layers&without&$0.66\,\pm0.08$&$13.26\,\pm\,0.08$\\
MC &$0.18\times0.18\times30$\,layers&with&$0.94\,\pm0.03$&$13.58\,\pm\,0.04$\\
MC &$0.18\times0.18\times30$\,layers&with$^{\ddagger}$&$0.78\,\pm0.03$&$13.52\,\pm\,0.03$\\
data &$0.18\times0.18\times30$\,layers&--&$1.2\textcolor{white}{0}\,\pm0.7\textcolor{white}{0}$&$12.5\textcolor{white}{0}\,\pm\,0.4\textcolor{white}{0}$\\
\hline
\multicolumn{5}{l}{$*$ Includes modelling of finite photon statistics, MPPC non-linear response, beam energy,  position }\\%and noise; }\\ 
\multicolumn{5}{l}{\hspace{4mm}fluctuation and noise; see Section\,\ref{section:MC_model}.}\\
\multicolumn{5}{l}{$\dagger$ including systematic and statistical uncertainties for data; statistic only for MC. }\\
\multicolumn{5}{l}{$\ddagger$ all details modelled as with $*$,
  with the exception of overlaying of detector noise. }
\end{tabular}
\end{center}
\end{table}

%\paragraph{
Photon statistics, correction of the MPPC non-linear response, non-uniformity of single scintillator response and the noise %} %% g923 existing} 
%totally add comparable degradation of energy resolution to the effect from the leakage. 
have a combined contribution to the degradation of the energy
resolution that is comparable to the effect of leakage.
%%% g923 \is{red}{The saturation of MPPC signal and its correction effect on constant term, $\Delta(\sigma_E/p) = +0.55\%$ and on stochastic term, $-1.72\%$.}
%%20170511 For details of this contribution,
%%20170511 comparing simulation with and without the saturation of MPPC signals and its correction procedure after photon statistic smearing indicated that 
%%20170511 these increase the constant term with $\Delta(\sigma_E/E) = +0.55$\%
%%20170511 and decrease the stochastic term by a relative factor of $1.72\%$.
For details of these contributions,
comparing simulation with and without these effects indicated that 
increase of the constant term is $\Delta(\sigma_E/E) = +0.67$\%
whereas decrease of the stochastic term is relatively $2.4\%$.
%Beam momentum fluctuation makes relatively large constant term, thus both of those effects seems almost negligible.
%However, the non-uniform response and the signal saturation by MPPC %degrade 
%change the energy resolution as well as the shower leakage. 

%% g923 Noise increases the constant term of the energy resolution by +0.5\%--assuming quadratic sum of its effect--whereas the effect on the stochastic term is negligible. 
Similarly, a comparison of the impact of overlaying noise on
the simulation indicated that overlaying the noise increases the value of the constant term of the energy resolution by $+0.5$\% whereas the effect on the stochastic term is negligible.

\section{Conclusion}\label{section:summary}
A prototype of a %CALICE 
Scintillator-Tungsten ECAL, designed for a future linear collider experiment, was constructed and tested at Fermilab in May 2009. 
This represents the large scale application of novel SiPM (MPPC) sensors and is
a feasibility study for the realisation of a highly granular calorimeter using this  type of photodetector. 
%%This paper describes the design of the prototype, its operation during the test beam data taking periods, 
%%%the calibration of all calorimeter channels and the reconstruction of events collected in electron beams.

%A combined experimental configuration consisting of the ScECAL, the AHCAL, the TCMT, the trigger system,  
%\v{C}erenkov counter and various devices for particle identification and monitoring of the beam parameters was %%installed and commissioned.
%used for this test beam campaign.

The response of the prototype to electron beams with energies between 2
and 32\gev\ was studied.  Despite the large environmental temperature
variation, 19\celdeg\---27.5\celdeg, a stable, linear response with
a maximum deviation from linearity of 1.1\% was
verified with a standard temperature correction procedure.  The
intrinsic energy resolution performance obtained,
$(12.5\pm0.1(\mathrm{stat.})\pm0.4(\mathrm{syst.}))/\sqrt{E[\mathrm{\gev}]}\oplus
(1.2\pm0.1(\mathrm{stat.})^{+0.6}_{-0.7}$(syst.))\%, is sufficient
for the anticipated requirements of a future linear collider.
Each scintillator strip has sufficient uniformity of response with $(88.3\pm4.3)$\% %%%% NIMA rev1 of the further side of the SiPM. 
 at the further side of the SiPM because of the light lost.
%%Nige Watson 20161203 \if 0
%%Nige Watson 20161203 The temperature dependence and intrinsic non-linearity of PPD (MPPC) devices presented a 
%%Nige Watson 20161203 particular challenge to the calibration of the ScECAL.
%%Nige Watson 20161203 A MIP-based calibration was applied, using data collected from 32 GeV muon beams. 
%%Nige Watson 20161203 The temperature dependence of the prototype's response was measured and understood with these data,
%%Nige Watson 20161203 and then used to remove the influence of temperature fluctuations in electron beam data.
%%Nige Watson 20161203 The saturation behavior of the MPPCs was determined by using a pico second laser system. 
%%Nige Watson 20161203 %%The use of t
%%Nige Watson 20161203 These calibration methods have been established by showing sufficient performance of the ScECAL.
%%Nige Watson 20161203 \fi

Potential systematic uncertainties arising from a number of sources
have been studied, including:
the precision of the beam energy spread; event selection cuts; 
ADC-MIP, ADC--photoelectron and inter-calibration  factors; and
the effective number of MPPC pixels. 
The most important uncertainty in the energy resolution is due to the uncertainty of the beam energy spread, 0.3\%. 

%%Nige Watson 20161203 \if 0
%%Nige Watson 20161203 The intrinsic ScECAL energy resolution, after subtraction of the beam momentum spread, 
%%Nige Watson 20161203 was determined to be 
%%Nige Watson 20161203 12.6\,$\pm$\,0.1(stat.) $\pm$\,0.4(syst.)\% for the stochastic term and 
%%Nige Watson 20161203 1.1$\pm$0.1(stat.)$^{+0.6}_{-0.7}$(syst.)\%  for the constant term.
%%Nige Watson 20161203 The deviations of the energy response from linear behavior are less than 1.1\%.
%%Nige Watson 20161203 \fi

\section{Acknowledgements}

We gratefully acknowledge the Fermilab managements for their support and hospitality, and their accelerator staff for the reliable and efficient beam operation. 
This work was supported 
by the FWO, Belgium; 
by the Natural Sciences and Engineering Research Council of Canada;
by the Ministry of Education, Youth and Sports of the Czech Republic;
by the European Union's Horizon 2020 Research and Innovation programme under Grant Agreement 654168; %% AIDA-2020
by the European Commission within Framework Programme 7 Capacities, Grant Agreement 262025;  %% AIDA (not 2020)
by the Alexander von Humboldt Stiftung (AvH), Germany;
by the Bundesministerium f\"{u}r Bildung und Forschung (BMBF), Germany; 
by the Deutsche Forschungsgemeinschaft (DFG), Germany; 
by the Helmholtz-Gemeinschaft (HGF), Germany; 
by the I-CORE Program of the Planning and Budgeting Committee, Israel;
by the Nella and Leon Benoziyo Center for High Energy Physics, Israel;
by the Israeli Science Foundation, Israel;
by the JSPS KAKENHI Grant-in-Aid for Scientific Research (B) No.\,17340071 and specially promoted research No.\,223000002, Japan;
by the National Research Foundation of Korea;
by the Korea-EU cooperation programme of National Research Foundation of Korea, Grant Agreement 2014K1A3A7A03075053; 
%%% coterra 20170506 added requested by Marina 
by the Russian Ministry of Education and Science contracts 3.2989.2017 and 14.A12.31.0006;
by the Spanish Ministry of Economy and Competitiveness FPA2014-53938-C3-R2 and Grant MDM-2015-0509;
by the Science and Technology Facilities Council, UK;
by the US Department of Energy;
by the National Science Foundation of the United States of America, and 
by the Nuclear Physics, Particle Physics, Astrophysics and Cosmology Initiative, a Laboratory Directed Research, USA. %and Development program at the Pacific Northwest National Laboratory, USA.

\begin{appendices}
\section{Composition of the absorber plate} 
As discussed in \ref{sec:ShowerProfile},
there is a discrepancy of absorber density between its direct measurement and estimation from the
composition of materials, as determined with EDX. 
Two plausible explanations behind this apparent discrepancy are the following:
%\begin{enumerate}
\begin{description}
\item [``balanced'' model:]EDX results have potentially unknown systematic
  uncertainties; the WC material  is too hard to provide sufficiently many samples at various locations
  in a plate, although the two samples used for tests showed no
  evidence of significant differences in their composition;
\item [``vacancy'' model:]
%\textred{
because the WC is a sintered material, produced by compressing a powder,
%}
the absorber plate is not entirely uniform, and has vacancies;
%% g923 photo by 
back-scattered electron imaging %~(BSEI)\footnote{JSM-5310E, JEOL Ltd., at 15\,kV, 700\,pA}
shows that the absorber plate is an aggregate of WC %grits;
grains.
%%% g923 \item EDX results have unknown systematic uncertainties: an example is that sampling areas were biased\,--\,although 
\end{description}

In the ``balanced'' model, 
the ratio of mass of WC to Co and Cr was decreased keeping the ratio of Co and Cr
%the composition of WC was reduced, and Co and Cr were increased keeping their ratio.
In the ``vacancy'' model, 
the absolute mass quantity of material
within the MC model was reduced, so that the relative composition was
maintained and the absorber had the density %%g923
obtained from direct measurement, meaning that the absorber material has vacancies.
Table~\ref{densityTune} lists the composition of the absorber in these two cases.
 \begin{table}[tbhp]
 \captionsetup{width=0.55\textwidth}
\begin{center}
\caption{\label{densityTune}\small Composition of the absorber plate in mass fraction (\%)
  measured with EDX (vacancy) and adjusted components, WC : Co+Cr, to
  have the density measured directly (balanced).  }
\begin{tabular}{ccc}
\hline \hline
\vspace{-3mm}\\
Component &vacancy &balanced  \\ 
\hline %\aline
\vspace{-3mm}\\
W   &81.82\,\plm\,0.31&74.43\,\plm\,0.30\\
\vspace{-3mm}\\
C  &\textcolor{white}{0}5.35\,\plm\,0.02&\textcolor{white}{0}4.86\,\plm\,0.02\\
\vspace{-3mm}\\
Co  &12.39\,\plm\,0.47&19.99\,\plm\,0.45\\
\vspace{-3mm}\\
Cr   &\textcolor{white}{0}0.45\,\plm\,0.47&\textcolor{white}{0}0.72\,\plm\,0.45\\
%\vspace{-3mm}\\
\hline
%\multicolumn{2}{l}{\small $*$ estimated from W component.}\\
\end{tabular}
\end{center}
\end{table}

Although both models agree with data in the mean ratio of longitudinal profile, 
0.98\,$\pm$\,0.04 (SD) for ``balanced'' model and 
0.96\,$\pm$\,0.07 (SD) for ``vacancy'' model as we can see in Fig.\,\ref{fig:longitudinalComparison},
the gradients from a linear fit to the ratios show clearly better
agreement with the ``balanced'' model.  
The slope for the ``balanced'' (``vacancy'') model
is $-0.00064\,\pm\,0.00003/$layer ($-0.01043\,\pm\,0.00003/$layer).

\end{appendices}

% ****************************************************************************
% BIBLIOGRAPHY AREA
% ****************************************************************************

\begin{footnotesize}
% IF YOU DO NOT USE BIBTEX, USE THE FOLLOWING SAMPLE SCHEME FOR THE REFERENCES
% ----------------------------------------------------------------------------

% ----------------------------------------------------------------------------

\end{footnotesize}

 \end{document}

%% file: CALICEMemberList170519.tex
%\documentclass[12pt]{JINST}
%%%%\documentclass[11pt, a4]{article}
%\usepackage{epsfig}
%\usepackage{a4}
%\usepackage{amsbsy}
%%%%%%%%%%%%%%\usepackage{rotating}
%%%%%%%%%%%%%%%%%\usepackage{alphalph}
%\usepackage{pennames} 
%%%%%%%%%%%%%%%%%%\usepackage{cite}
%%%%%%%%%%%%%%%%%%%\parindent 0pt
%%%%%%%%%%%%%%%%%%%%%\parskip 10pt plus 1pt minus 1pt
%

%

% carefull: Hacked version of JINST style to get more than 9 footnote symbols

%\title{Full CALICE Member List {\normalsize (16 November 2016)}}

\renewcommand{\thefootnote}{\alph{footnote}}
\setcounter{footnote}{0}

%\noindent{ \Large \bf The CALICE Collaboration}

{
%\bf{
{\centering 
{\LARGE\bf The CALICE Collaboration}
%}

\if 0
\vspace{4mm}\bf{\centering
J.-J.\,Blaising, 
M.\,Chefdeville, 
C.\,Drancourt, vspace{4mm}\bf
N.\,Geffroy, 
Y.\,Karyotakis, 
G.\,Vouters
\\ \it
Laboratoire d'Annecy-le-Vieux de Physique des Particules, Universit\'{e} de Savoie,
CNRS/IN2P3,
9 Chemin de Bellevue BP110, F-74941 Annecy-le-Vieux CEDEX, France
}
\fi

%Opt in
\vspace{4mm}\bf{\centering
J.\,Repond, 
L.\,Xia
\\ \it
Argonne National Laboratory,
9700 S.\ Cass Avenue,
Argonne, IL 60439-4815,
USA}

\if 0
\vspace{4mm}\bf{\centering
A.\,White,
J.\,Yu
\\ \it
University of Texas at Arlington, Department of Physics, SH108, Arlington, TX 76019, USA
}

\vspace{4mm}\bf{\centering
J.\,Veloso,
C.\,Azevedo,
P.\,Correia 
\\ \it
I3N, Physics Department, University of Aveiro, 3810-193, Aveiro, Portugal
}
\vspace{4mm}\bf{\centering
Y.\,Wang, 
D.\,Han
\\ \it
Tsinghua University, Department of Engineering Physics, Beijing, 100084, P.R. China
}
\fi

%Opt in
\vspace{4mm}\bf{\centering
G.\,Eigen, 
\\ \it
University of Bergen, Inst.\, of Physics, Allegaten 55, N-5007 Bergen, Norway
}

%Opt in
\vspace{4mm}\bf{\centering
T.\,Price,
N.K.\,Watson,
A.\,Winter
\\ \it
University of Birmingham,
School of Physics and Astronomy,
Edgbaston, Birmingham B15 2TT, UK
}

\if 0
\vspace{4mm}\bf{\centering
F.\,Salvatore,
C.\,Chavez Barajas,
T.\,Coates
\\ \it
University of Sussex, 
Physics and Astronomy Department, 
Brighton, Sussex, BN1 9QH, UK
}
\fi

%Opt in
\vspace{4mm}\bf{\centering 
%S.\,Green,
%J.S.\,Marshall,
M.A.\,Thomson,
%D.R.\,Ward
\\ \it
University of Cambridge, Cavendish Laboratory, J J Thomson Avenue, CB3 0HE, UK
}

\if 0
\vspace{4mm}\bf{\centering 
D.\,Benchekroun, 
A.\,Hoummada, 
Y.\,Khoulaki
\\ \it
Universit\'{e} Hassan II de Casablanca, Facult\'{e} des Sciences A\"{\i}n Chock, Casablanca, Morocco
}
\fi
%%\vspace{4mm}\bf{\centering 
%J.\,Apostolakis, 
%D.\,Dannheim, 
%K.\,Elsener,
%G.\,Folger, 
%V.\,Ivantchenko, 
%W.\,Klempt, 
%L.\,Linssen,
%A.A.\,Maier,
%N.\,Nikiforou,
%F.\,Pitters,
%A.\,Ribon,
%A.\,Sailer,
%E.\,Sicking,
%V.\,Uzhinskiy,
%M.A.\,Weber
%\\ \it 
%CERN, 1211 Gen\`{e}ve 23, Switzerland
%}

\if 0
\vspace{4mm}\bf{\centering
D.\,Boumediene,
V.\,Francais
\\ \it
Universit\'e Clermont Auvergne, Universit\'e Blaise Pascal, CNRS/IN2P3, LPC, 4 Av. Blaise Pascal, TSA/CS 60026,
F-63178 Aubi\`ere, France
}
\vspace{4mm}\bf{\centering
J.\,Marques Ferreira dos Santos,
F.D.\,Amaro,
C.M.B.\,Monteiro,
L.M.P.\,Fernandes,
E.D.C.\,Freitas,
X.\,Carvalho,
R.J.C.\,Roque
\\ \it
LIBPhys, Physics Departement, University of Coimbra, 3004-516 Coimbra, Portugal
}
\vspace{4mm}\bf{\centering
A.\,Pingault, 
M.\,Tytgat, 
N.\,Zaganidis
\\ \it
Ghent University, Department of Physics and Astronomy,
Proeftuinstraat 86, B-9000 Gent, Belgium
}
\fi

%% OPT in 
\vspace{4mm}\bf{\centering
G. C.\,Blazey,
A.\,Dyshkant, 
K.\,Francis, 
V.\,Zutshi
\\ \it
NICADD, Northern Illinois University,
Department of Physics,
DeKalb, IL 60115,
USA
}

\if 0
\vspace{4mm}\bf{\centering 
G.\,Cho, 
D-W.\,Kim,
S.\,C.\,Lee, 
W.\,Park, 
S.\,Vallecorsa
\\ \it
Gangneung-Wonju National University
Gangneung 25457, South Korea
}

\vspace{4mm}\bf{\centering 
J.\,Giraud, 
D.\,Grondin, 
J.-Y.\,Hostachy
\\ \it
Laboratoire de Physique Subatomique et de Cosmologie - Universit\'{e}
Grenoble-Alpes, CNRS/IN2P3, Grenoble, France
}
\fi

%Opt in
\vspace{4mm}\bf{\centering 
K.\,Gadow, 
P.\,G\"{o}ttlicher, 
O.\,Hartbrich\footnote{now at University of Hawaii at Manoa, High Energy Physics Group, 2505 Correa Road, HI, Honolulu 96822, USA}, 
K.\,Kotera\footnote{also at Shinshu University, now at Osaka University},
F.\,Krivan, 
K.\,Kr\"{u}ger, 
S.\,Lu, 
B.\,Lutz, 
M.\,Reinecke, 
F.\,Sefkow, 
Y.\,Sudo,
H.L.\,Tran
\\ \it
DESY, Notkestrasse 85,
D-22603 Hamburg, Germany
}

\if 0
\vspace{4mm}\bf{\centering 
P.\,Buhmann,
E.\,Garutti,
S.\,Laurien,
D.\,Lomidze,
M.\,Matysek
\\ \it
Univ. Hamburg,
Physics Department,
Institut f\"ur Experimentalphysik,
Luruper Chaussee 149,
22761 Hamburg, Germany
}
\fi

%Opt in
\vspace{4mm}\bf{\centering 
%K.\,Briggl, 
%P.\,Eckert, 
A.\,Kaplan, %\footnote{now at 
 %Next Thing Co
 %2461 Peralta St
 %Oakland, CA 94607
 %USA}, 
%Y.\,Munwes, 
H.-Ch.\,Schultz-Coulon, 
%W.\,Shen, 
%R.\,Stamen
\\ \it
University of Heidelberg, Fakultat fur Physik und Astronomie,
Albert Uberle Str. 3-5 2.OG Ost,
D-69120 Heidelberg, Germany
}

%Opt in
\vspace{4mm}\bf{\centering 
B.\,Bilki\footnote{Also at Beykent University, Istanbul, Turkey}, 
D.\,Northacker,
Y.\,Onel
\\ \it
University of Iowa, Dept. of Physics and Astronomy,
203 Van Allen Hall, Iowa City, IA 52242-1479, USA
}

%Opt In
\vspace{4mm}\bf{\centering 
G.W.\,Wilson
\\ \it
University of Kansas, Department of Physics and Astronomy,
Malott Hall, 1251 Wescoe Hall Drive, Lawrence, KS 66045-7582, USA
}

%Opt in
\vspace{4mm}\bf{\centering 
K.\,Kawagoe,
I.\,Sekiya,
T.\,Suehara,
H.\,Yamashiro,
T.\,Yoshioka,
\\ \it
Department of Physics and Research Center for Advanced Particle Physics,
Kyushu University, 744 Motooka, Nishi-ku, Fukuoka 819-0395, Japan
}

\if 0
\vspace{4mm}\bf{\centering 
P.\,Dauncey
\\ \it
Imperial College London, Blackett Laboratory,
Department of Physics,
Prince Consort Road,
London SW7 2AZ, UK 
}
\fi
%\vspace{4mm}\bf{\centering 
%M.\,Wing\footnote{Also at DESY}
%\\ \it
%Department of Physics and Astronomy, University College London,
%Gower Street,
%London WC1E 6BT, UK
%}

\if 0
\vspace{4mm}\bf{\centering 
E.\,Cortina Gil, 
S.\,Mannai
\\ \it
Center for Cosmology, Particle Physics and Phenomenology (CP3)
Universit\'{e} catholique de Louvain, Chemin du cyclotron 2,
1348 Louvain-la-Neuve, Belgium
}
\vspace{4mm}\bf{\centering 
P.\,Calabria, 
L.\,Caponetto, 
C.\,Combaret, 
D.\,Delauany, 
F.\,Doizon, 
R.\,Ete, 
A.\,Eynard, 
G.\,Garillot, 
L.\,Germani,
G.\,Grenier, 
J-C.\,Ianigro, 
T. \,Kurca, 
I.\,Laktineh, 
N.\,Lumb, 
H.\,Mathez, 
L.\,Mirabito, 
A.\,Petrukhin,
A.\,Steen,
W.\,Tromeur, 
Y.\,Zoccarato
\\ \it
Univ Lyon, Universit\'{e} Lyon 1, 
CNRS/IN2P3, IPN-Lyon, F-69622 
Villeurbanne, France
}
\fi

%Opt in
\vspace{4mm}\bf{\centering 
E.\,Calvo Alamillo,
M.C.\,Fouz,
J.\,Marin,
J.\,Navarrete,
J.\,Puerta Pelayo,
A.\,Verdugo
\\ \it
CIEMAT, Centro de Investigaciones Energeticas, Medioambientales y Tecnologicas, Madrid, Spain 
}

\if 0
\vspace{4mm}\bf{\centering 
V.\,B\"uscher,
P.\,Chau,
S.\,Krause,
Y.\,Liu,
L.\,Masetti,
U.\,Sch\"afer,
S.\,Tapprogge,
R.\,Wanke,
Q.\,Weitzel
\\ \it
Institut f\"ur Physik, Universit\"at Mainz, Staudinger Weg 7, D-55099 Mainz,
Germany
}
\vspace{4mm}\bf{\centering 
F.\,Corriveau,
B.\,Freund\footnote{Also at Argonne National Laboratory}
\\ \it
Department of Physics, McGill University,
Ernest Rutherford Physics Bldg.,
3600 University Ave.,
Montr\'{e}al, Qu\'{e}bec,
Canada H3A 2T8
}
\fi

%Opt in
\vspace{4mm}\bf{\centering 
M.\,Chadeeva\footnote{Also at MEPhI}, 
M.\,Danilov\footnote{Also at MEPhI and at Moscow Institute of Physics and Technology}, 
%A.\,Drutskoy\footnote{Also at MEPhI}, 
%N.\,Kirikova,
%V.\,Kozlov, 
%R.\,Mizuk\footnote{Also at MEPhI}
\\ \it
P.\,N.\, Lebedev Physical Institute,
Russian Academy of Sciences,
117924 GSP-1 Moscow, B-333, Russia
}

\if 0
\vspace{4mm}\bf{\centering
D.\,Besson,
B.\,Bobchenko, 
P.\,Buzhan, 
A.\,Ilyin, 
O.\,Markin, 
D.\,Mironov\footnote{Also at Moscow Institute of Physics and Technology}, 
D.\,Nikolaev, 
E.\,Popova, 
V.\,Rusinov,
E.\,Tarkovsky
\\ \it
National Research Nuclear University 
MEPhI (Moscow Engineering Physics Institute)
31, Kashirskoye shosse,
115409 Moscow, Russia
}
\vspace{4mm}\bf{\centering 
N.\,Baranova,
E.\,Boos, 
L.\,Gladilin,
D.\,Karmanov, 
M.\,Korolev, 
M.\,Merkin,
A.\,Voronin
\\ \it
M.V.Lomonosov Moscow State University, D.V.Skobeltsyn Institute of Nuclear
Physics (SINP MSU),
1/2 Leninskiye Gory, Moscow, 119991, Russia
}
\fi

%% OPT IN
\vspace{4mm}\bf{\centering 
M.\,Gabriel,
P.\,Goecke,
C.\,Graf,
Y.\,Israeli,
N.\,van der Kolk,
F.\,Simon,
M.\,Szalay,
H.\,Windel
\\ \it
Max-Planck-Institut f\"ur Physik,
F\"ohringer Ring 6,
D-80805 Munich, Germany
}

\if 0
\vspace{4mm}\bf{\centering 
J.E.\,Augustin,
J.\,David,
P.\,Ghislain, 
D.\,Lacour,
L.\,Lavergne,
J.M.\, Parraud
\\ \it
Laboratoire de Physique Nucl\'eaire et de Hautes Energies (LPNHE),
UPMC, UPD, CNRS/IN2P3, 4 Place Jussieu, 75005 Paris, France 
}
\fi

%Opt in
\vspace{4mm}\bf{\centering 
S.\,Bilokin, 
J.\,Bonis, 
R.\,P\"oschl, 
A.\,Thiebault, 
F.\,Richard, 
D.\,Zerwas
\\ \it
Laboratoire de l'Acc\'elerateur Lin\'eaire,
CNRS/IN2P3 et Universit\'e de Paris-Sud XI,
Centre Scientifique d'Orsay B\^atiment 200, 
BP 34,
F-91898 Orsay CEDEX, France
}

%Opt in
\vspace{4mm}\bf{\centering 
%M.\,Anduze,
V.\,Balagura,
%E.\,Becheva,
V.\,Boudry,
J-C.\,Brient, 
R.\,Cornat,
%E.\,Edy,
%M.\,Frotin\footnote{ Now at GEPI, Meudon.}, 
%F.\,Gastaldi, 
%Y.\,Haddad\footnote{ Now at Imperial College, London.}, 
%B.\,Li, 
%F.\,Magniette,
%J.\,Nanni,
% M.\,Ruan\footnote{ Now at IHEP, Beijing.},
%M.\,Rubio-Roy,
%K.\,Shpak, 
%T.H.\,Tran, 
%H.\,Videau,
%D.\,Yu\footnote{ Also at at IHEP, Beijing.} 
\\ \it
Laboratoire Leprince-Ringuet (LLR) -- \'{E}cole Polytechnique,
CNRS/IN2P3,
Palaiseau, F-91128 France
}

\if 0
\vspace{4mm}\bf{\centering 
S.\,Callier,
F.\,Dulucq, 
Ch.\,de la Taille, 
G.\,Martin-Chassard,
L.\,Raux, 
N.\,Seguin-Moreau
\\ \it
Laboratoire OMEGA -- \'{E}cole Polytechnique-CNRS/IN2P3, 
Palaiseau, F-91128 France
}
\fi

%Opt in
\vspace{4mm}\bf{\centering 
J.\,Cvach, 
M.\,Janata, 
M.\,Kovalcuk, 
J.\,Kvasnicka\footnote{also at DESY}, 
I.\,Polak, 
J.\,Smolik, 
V.\,Vrba, 
J.\,Zalesak, 
J.\,Zuklin
\\ \it
Institute of Physics, The Czech Academy of Sciences,
Na Slovance 2, CZ-18221 Prague 8, Czech Republic
}

\if 0
\vspace{4mm}\bf{\centering 
M.\,Havranek, 
M.\,Hejtmanek, 
Z.\,Janoska, 
V.\,Kafka, 
O.\,Korchak, 
M.\,Marcisovsky, 
G.\,Neue,
R.\,Novotny, 
L.\,Tomasek
\\ \it
Faculty of Nuclear Sciences and Physical Engineering,
Czech Technical University in Prague,
Brehova 7, CZ-11519 Prague 1, Czech Republic 
}
\vspace{4mm}\bf{\centering 
S.\,Bressler,
L.\,Arazi,
A.\,Breskin,
L.\,Moler,
D.\,Shaked-Renous,
A.\,Coimbra,
A.\,Roy,
P.\,Bhattacharya,
\\ \it
Weizmann Institute of Science, 234 Herzl Street, Rehovot 7610001 Israel
}
\vspace{4mm}\bf{\centering 
J.\,Strube
\\ \it
Pacific Northwest National Laboratory,
902 Battelle Boulevard,
Richland, WA, USA
}
\fi

%Opt in
\vspace{4mm}\bf{\centering 
%Y.\,Fujita, 
%H.\,Itoh, 
%I.\,Kanzaki, 
W.\,Choi, 
K.\,Kotera\footnote{also at DESY, now at Osaka University},
M.\,Nishiyama,
T.\,Sakuma,
T.\,Takeshita, 
S.\,Tozuka,
T.\,Tsubokawa,
S.\,Uozumi\footnote{now at Okayama University},
%R.\,Terada 
\\ \it
Shinshu Univ.\,,
Dept. of Physics,
3-1-1, Asahi,
Matsumoto-shi, Nagano 390-8621,
Japan
}

%Opt in
\vspace{4mm}\bf{\centering 
D.\,Jeans\footnote{now at KEK},
W.\,Ootani, 
%T.\,Mori, 
%S.\,Yamashita, 
L.\,Liu, 
%N.\,Matsuzawa, 
%M.\,Usami
\\ \it
ICEPP, The University of Tokyo, 7-3-1 Hongo, Bunkyo-ku, Tokyo
113-0033, Japan
\if 0
}
\vspace{4mm}\bf{\centering 
S.\,Komamiya, 
D.\,Jeans\footnote{now at KEK}, 
Y.\,Kamiya, 
H.\,Nakanishi\\ \it
Department of Physics, The University of Tokyo, 7-3-1 Hongo, Bunkyo-ku, Tokyo
113-0033, Japan
}
\fi

%Opt in
\vspace{4mm}\bf{\centering 
S.\,Chang, 
A.\,Khan\footnote{now at Islamia College University, Peshawar, Pakistan}, 
D.H.\,Kim, 
D.J.\,Kong, 
Y.D.\,Oh 
\\ \it
Department of Physics, Kyungpook National University, Daegu, 702-701,
Republic of Korea
}

%Opt in
\vspace{4mm}\bf{\centering 
T.\,Ikuno, 
Y.\,Sudo, 
Y.\,Takahashi 
\\ \it
Division of Physics, Faculty of Pure and Applied Sciences, University of Tsukuba, 
Tennoudai 1-1-1, Tsukuba-shi, Ibaraki-ken 305-8571, Japan
}

\if 0
\vspace{4mm}\bf{\centering 
A.\,van den Brink, 
G.-J.\,Nooren, T.\,Peitzmann,
T.\,Richert,
M.\,van Leeuwen,
H.\,Wang,
C.\,Zhang
\\ \it
Institute for Subatomic Physics, Utrecht University/Nikhef, 3584CC Utrecht, The Netherlands
}
\fi

%Opt in
\vspace{4mm}\bf{\centering 
%A.\,Elkhalii,
M.\,G\"{o}tze,
%C.\,Zeitnitz
\\ \it
Bergische Universit\"{a}t Wuppertal
Fakult\"at 4 / Physik,
Gaussstrasse 20,
D-42097 Wuppertal, Germany
}

}
}